\documentclass[rmp,aps,amsfonts,amssymb,nofootinbib,floats,twocolumn]{revtex4}
\usepackage{graphics}
\usepackage{epsfig}
\usepackage{bm}

\begin{document}
\title{Inelastic Light Scattering from Correlated Electrons}

\author{Thomas P. Devereaux}
\email{tpd@lorax.uwaterloo.ca} \affiliation{Department of Physics,
University of Waterloo, Waterloo, ON, Canada N2L 3G1}
\author{Rudi Hackl}
\email{hackl@wmi.badw.de}
\affiliation{Walther Meissner Institute, Bavarian Academy of
Sciences, 85748 Garching, Germany }

\begin{abstract}
Inelastic light scattering is an intensively used tool in the
study of electronic properties of solids. Triggered by the
discovery of high temperature superconductivity in the cuprates
and by new developments in instrumentation, light scattering both
in the visible (Raman effect) and the X-ray part of the
electromagnetic spectrum has become a method complementary to
optical (infrared) spectroscopy while providing additional and
relevant information. The main purpose of the review is to
position Raman scattering with regard to single-particle methods
like angle-resolved photoemission spectroscopy (ARPES), and other
transport and thermodynamic measurements in correlated materials.
Particular focus will be placed on photon polarizations and the
role of symmetry to elucidate the dynamics of electrons in
different regions of the Brillouin zone. This advantage over
conventional transport (usually measuring averaged properties)
indeed provides new insights into anisotropic and complex
many-body behavior of electrons in various systems. We review
recent developments in the theory of electronic Raman scattering
in correlated systems and experimental results in paradigmatic
materials such as the A15 superconductors, magnetic and
paramagnetic insulators, compounds with competing orders, as well
as the cuprates with high superconducting transition temperatures.
We present an overview of the manifestations of complexity in the Raman response due to the impact of correlations and developing competing orders. In a variety of materials we discuss which observations may be understood and summarize important open questions that pave the way to a detailed understanding of correlated electron systems.
\end{abstract}

\date{July 16, 2006}
\maketitle
\tableofcontents

\section{INTRODUCTION}
\label{sec:intro}

\subsection{Overview}

Raman scattering is a photon-in photon-out process with energy
transfered to a target material. Most of the light is elastically
scattered from the sample, a fraction is color-shifted and
collected at the detector.

\begin{figure*}[floatfix]
\begin{center}
\includegraphics[width=11.5cm,angle=0]{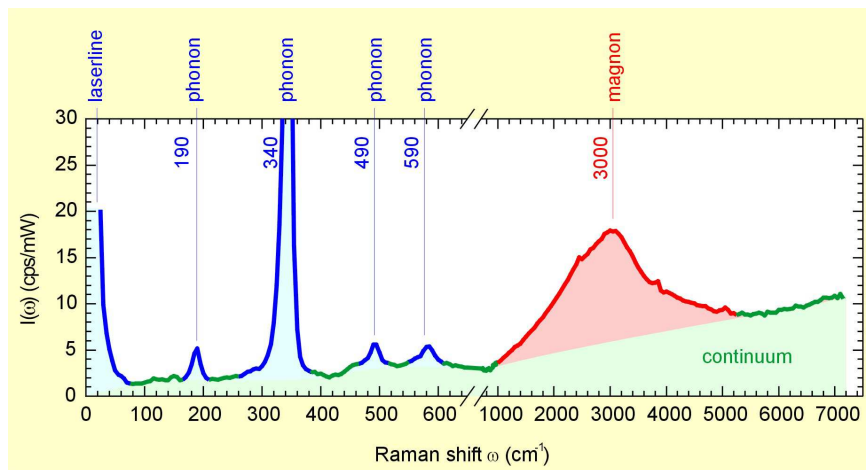}
\end{center}
\caption{Characteristic Raman scattering  spectrum taken on $\rm
YBa_2Cu_3O_{6.5}$ showing light scattering from phonons (blue
lines), magnons (red), and electrons (green). Courtesy of Matthias Opel.
}\label{Fig:gen_idea}
\end{figure*}

Light couples to electronic  charge in solids and can scatter
inelastically from many types of excitations in a sample, as
shown schematically for $\rm YBa_2Cu_3O_{6.5}$ in Figure
\ref{Fig:gen_idea}. Optical phonons produce sharp peaks at well
known positions and orientations of the incoming and outgoing
photon  polarizations, while a large broad feature centered at
much higher energies due to two-magnon scattering occurs in
compounds with antiferromagnetic correlations. This review article largely places
emphasis on the electronic Raman scattering continuum upon which
the phonons and magnons are superimposed.

Light scatters off of electrons by creating variations of
electronic charge density in the illuminated region of a sample.
By observing the frequency shift and polarization change of the
outgoing photon compared to the incoming photon, the properties of
charge density relaxation is measured. However, measuring the
Raman effect of photons scattering from electrons is a difficult
proposal to carry forward. It is made difficult precisely because
of the coupling of the real photon vector potential and the
exchange of virtual photons which mediate the Coulomb forces
between electrons. In simple metals, variations of the charge
density will be largely screened by the mobile electrons, and the
system of electrons responds collectively at a characteristic
plasma frequency of several electron volts. In semiconductors or
well-developed band insulators, the creation of charge density
fluctuations occurs only via the population of excited states
across a band gap - again on the scale of several electron volts.
Therefore it is difficult to investigate the behavior of electrons
at low energies. In fact, hardly {\it any} measurements of
electronic Raman scattering in simple metals exist precisely for
this reason, and focus on semiconductors is usually placed on
plasma excitations. Since the dynamics of electrons lying near the
Fermi surface govern the behavior of transport in most systems,
this would give the impression that Raman would have but little to
offer in simple metals and insulators.

Yet the Raman effect is extremely well suited to study electrons
in systems with non-trivial electron dynamics. First well-studied
in the context of breaking Cooper pairs in superconductors in the
period from 1980 to 1990, the field of Raman scattering from
electronic excitations has grown tremendously over the past few
decades to study the evolution of electron correlations in a
variety of systems in which many-body interactions are essential
to the physics of new materials and their potential device
applications.

Raman spectroscopy has become an indispensable tool in the arsenal
for understanding many-body physics. One of the most celebrated
achievements of electronic Raman scattering has been the ability
to focus on the nature of electron dynamics in different regions
of the Brillouin zone. This distinguishes Raman scattering from
most other transport and thermodynamic measurements, allowing the
study of the development of correlations in projected regions of
the Brillouin zone. By simply aligning the polarization
orientations of the incoming and outgoing photons, charge
excitations can be selectively mapped and analyzed using
group-theoretical symmetry arguments. The search for conventional
as well as exotic excitations in strongly correlated matter has
been greatly enhanced. Raman spectroscopy has provided new and
valuable insights into unconventional superconductivity and
collective modes, excitations in charge, spin and/or orbitally
ordered systems as well as the competition between the various
ordered phases. In addition, new insights into electron dynamics
of metal-insulator transitions, quantum phase transitions, and the
concomitant quantum critical behavior could be obtained. The
purpose of this article is to review the essence of these new
developments in a ``snapshot'' of the current state of
investigation.

The overall agenda of the paper is to provide a
vehicle to sort through the extensive literature, learn about the
outstanding problems, and become aware of the level of consensus.
In order to present a detailed picture of the current status of
electronic light scattering, other types of excitations, such as phonons and magnons, are mainly
ignored. There have been many reviews on inelastic light
scattering from phonons and magnons. The reader is referred to
earlier reviews by \textcite{Klein:1982b} for a fundamental
treatment of scattering from phonons, while studies of phonons in high
temperature superconductors are summarized by
\textcite{Thomsen:1989} and \textcite{Sherman:2003}. Recently
\textcite{Lemmens:2003} and \textcite{Gozar:2005b} reviewed
magnetic light scattering in low-dimensional quantum spin systems
and cuprates. Due to space limitations we cannot give adequate
commentary on these exciting and developing fields.

The outline of our review is as follows. After a brief historical
summary, the fundamental experimental aspects and theoretical
developments for electronic Raman scattering are presented in the
first part of the article. A general treatise on the theory of
electronic Raman scattering is given in Section \ref{Section:II}
with a view toward the formalism for both weak and strong
correlations. Results from model-specific calculations can be
found in Section~\ref{Section:charge_relaxation}. Readers who are
more interested in summaries of experimental work may want to skim
these sections and skip to Section \ref{Section:III}, where a
review of Raman scattering measurements in a variety of correlated
materials is given with a view toward common features manifest
from strong correlations. The presentation is generally organized in systems
with increasing complexity of correlations and competing orders.

In this framework, the canon of work on the high temperature
superconductors in the last part of our review is presented in
Section~\ref{sec:HTSC}. This detailed part of the review is
organized in conceptual issues of correlations, superconductivity,
normal state properties, and the propensity toward
charge and spin ordering in various families of the cuprates. In
all subsections in this part, data on a variety of cuprate
materials are summarized.

The review closes with a general  discussion of open questions for
both experimental and theoretical developments in Raman
scattering, and points out new directions in which our
understanding of electronic correlations may be further enhanced.

\subsection{Historical Review}

Inelastic scattering of light was discovered independently in
organic liquids by \textcite{Raman:1928} and in quartz by
\textcite{Landsberg:1928} who properly explained the observed
effect: The energy of the incoming photon is split between the
scattered one and an elementary excitation in the solid. Shortly
thereafter in 1930 C.~V. Raman was awarded the Nobel prize, and
his name was associated with the effect
\cite{Pleijel:1930,Fabelinskii:1998,Ginzburg:1998}. Although the
phenomenological description by \textcite{Smekal:1923} in terms of
a periodically modulated polarizability qualitatively captures the
relevant physics including the selection rules, the effect is
genuinely quantum mechanical as described first and ahead of the
experimental discovery by \textcite{Kramers:1925} in context of
the dispersion in dielectrics.

Soon after the observation of light scattering from vibrational
excitations, \textcite{Verkin:1948} and \textcite{Khaikin:1956}
attempted  to use the new technique for studying electronic
excitations. They picked one of the most ambitious subjects, i.e.,
light scattering from superconducting gap excitations in
conventional metals. It is not at all surprising that they could
not succeed. In a seminal paper \textcite{Abrikosov:1961} not only
calculated the Raman response of a typical elemental
superconductor but also demonstrated that the sensitivity in the
early experiments was by approximately 5 or 6 orders of magnitude
too low. In 1980 light was finally scattered successfully from
superconducting electrons in ${2H-\rm NbSe_2}$
\cite{Sooryakumar:1980}. \textcite{Balseiro:1980} and
\textcite{Littlewood:1981,Littlewood:1982} argued that the
superconducting excitations in this system become Raman active
mainly via their coupling to a charge-density wave mode (CDW).
After the observation of gap excitations in the A15 compounds
${\rm Nb_3Sn}$ and ${\rm V_3Si}$
\cite{Klein:1982a,Hackl:1982,Dierker:1983,Hackl:1983} it was clear
that light can be scattered directly by Cooper pairs
\cite{Dierker:1983,Klein:1984}. \textcite{Tutto:1992} demonstrated
that both types of coupling contribute.

Collective excitations of normal electrons were first observed in
semiconductors \cite{Mooradian:1966} following theoretical studies
by \textcite{Pines:1963}, \textcite{Platzman:1964}. As a function
of doping the plasmon peak moves across the phonon energies
leading to strong electron-lattice interactions. In heavily doped
silicon, with the plasma energy well beyond the vibration spectrum,
the evolution of the phonon line shape
\cite{Cerdeira:1973,Fano:1961} clearly demonstrated the existence
of an electron continuum. In 1977, fluctuations of electrons
between pockets of the Fermi surface of silicon were observed by
\textcite{Chandrasekhar:1977} and explained subsequently by
\textcite{Ipatova:1981}. In magnetic fields, transitions between
Landau levels were found \cite{Worlock:1981}\footnote{The subject
has been reviewed in detail by \textcite{Abstreiter:1984}}. Strong
phonon renormalization effects also occur in metallic alloys with
A15 structure \cite{Wipf:1978,Schicktanz:1980}. The origin of the
broad continuum, which interacts with phonons and is
redistributed below the superconducting transition
\cite{Klein:1984}, is certainly electronic but as of today is still
not fully understood.

The full power of the method became apparent after the discovery
of copper-oxygen compounds \cite{Bednorz:1986} with
superconducting transition temperatures above 100~K. It turned out
that, in contrast to infrared spectroscopy, momentum dependent
transport properties can be measured with Raman spectroscopy, since
different regions of the Brillouin zone can be projected out
independently by appropriately selecting the polarizations of the
incident and scattered photons \cite{Devereaux:1994a}. New
theoretical ideas were not only applied to the superconducting but
also to the normal state. The spectra extend over energy ranges as
large as electron Volts (eV)
\cite{Bozovic:1987,Kirillov:1988,Cooper:1988a,Cooper:1988b} and
are similar to those in the A15s or in rare earth elements
\cite{Klein:1991}.  Both elastic \cite{Zawadowski:1990} and
inelastic \cite{Itai:1992,Kostur:1991,Virosztek:1992} relaxation
of electrons indeed produces light scattering over such a broad
range of energies. It soon became clear that a continuum extending
over an eV cannot originate from elastic scattering, but only from
inelastic processes or interband transitions. However, it has not
been straightforward to pin down the types of interactions.

At very low energies, spin- \cite{Yoon:2000} and charge-ordering
fluctuations were reported in manganites and, respectively, in
ladder compounds \cite{Blumberg:2002} and in the cuprates
\cite{Venturini:2002b}. The response should not be confused with
that of an ordered spin/charge-density-wave (SDW/CDW) state
\cite{Klein:1982c,Benfatto:2000,Zeyher:2002}. Since a characteristic energy
decreases rather than increases upon cooling
\cite{Caprara:2002,Venturini:2002b,Blumberg:2002,Caprara:2005,Yoon:2000},
which is in clear contrast to typical order parameter behavior.

Along with the early studies of charge excitations, Fleury and
coworkers observed Raman scattering from spin waves in
antiferromagnetically ordered ${\rm FeF_2}$, ${\rm MnF_2}$, and
${\rm K_2NiF_4}$ \cite{Fleury:1966,Fleury:1967,Fleury:1970}.
\textcite{Elliot:1963} and \textcite{Fleury:1968} presented a
detailed theoretical description which allowed them to
semi-quantitatively understand the spectral shape and the cross
section. With the discovery of the cuprates by
\textcite{Bednorz:1986} also this field experienced a renaissance
in particular for the study at low doping close to the
antiferromagnetic N\'eel state
\cite{Lyons:1988,Sugai:1988,Sulewski:1991,Gozar:2004,Gozar:2005b}.
Raman scattering is probably the most precise method for
determining the exchange coupling $J$ though the theoretical
understanding is still incomplete. In this context, spin-Peierls
systems \cite{Loosdrecht:1996} and ladder compounds
\cite{Abrashev:1997} shifted very much into the focus of interest
\cite{Dagotto:1999}.

Very recently, light scattering from ``orbitons'', i.e. from a
propagating reorientation of orbitals, has been proposed to
explain new modes in the the Raman spectra \cite{Saitoh:2001}.
However, there is no agreement yet on whether or not orbitons can
be observed independent of phonons or other excitations
\cite{Grueninger:2002,Choi:2005,Krueger:2004}.

\subsection{What Can One Learn from Electronic Raman Scattering?}
\label{sec:learn}

We give now a qualitative  introduction into the relationship
between Raman spectroscopy and other experimental techniques. We
begin by drawing the distinction between one-particle and
many-particle properties.

Typically, electronic states in solids are characterized by  their
energy dispersions as well as the characteristic lifetime of an
electron placed into such a state. This state is characterized by
the single particle propagator or Green's function for the
electron,
\begin{equation}
G({\bf k},\omega)=\frac{1} {\omega-\xi_{\bf k} -
\Sigma({\bf k},\omega)}. \label{Eq:Green}
\end{equation}
Here $\xi_{\bf k}$ denotes the bare energy band dispersion
calculated from a solvable model.
$\Sigma$ represents the
electron self-energy which encompasses all the information
pertaining to interactions of the single electron in state ${\bf
k}$ to all other excitations of the system. Usually the self
energy can only be obtained via approximate methods. Some of these
approximations are quite good - such as electron-phonon
interactions in metals (known as Migdal's approach
\cite{Migdal:1958}) for example, while others are more difficult -
such as the Coulomb interaction between other electrons. The self
energy is a complex function, $\Sigma=\Sigma^{\prime} +
i\Sigma^{\prime\prime} $, which, in general, depends on
temperature, momentum and energy. The real part of the self energy
determines how the energy dispersion $\xi_{\bf k}$ is renormalized
by the interactions while the imaginary part determines the
lifetime of the quasiparticle placed into the state {\bf k}.

The spectral function is directly related to the analytically
continued electron's Green's function for frequencies on the real
axis via the replacement $i\omega\rightarrow\omega+i\delta$
\begin{equation}
A({\bf k},\omega)= -\frac{1}{\pi}\lim_{\delta\rightarrow
0}G^{\prime\prime}({\bf k},\omega+i\delta)
\end{equation}
which is measurable via modern angle-resolved photoemission
(ARPES) techniques and has provided an immense amount of
information in strongly correlated systems
\cite{Campuzano:2002,Damascelli:2003}. For non-interacting
electrons, $A({\bf k},\omega)$ is a $\delta$-function peaked at the
pole of the propagator when the frequency $\omega$ equals the bare
band energy $\xi_{\bf k}$. Interactions broaden the spectral
function and give it non-trivial temperature and frequency
dependences as well as non-trivial anisotropies in momentum-space
if the interactions among electrons are anisotropic. The spectral
function describes real electrons, hence integrals over all
energies must obey sum rules, such as (i) $\int d\omega A({\bf
k},\omega)=1$ and (ii) $\int d\omega f(\omega)A({\bf
k},\omega)=n({\bf k})$ with the Fermi Dirac distribution
$f(\omega)$ and the momentum distribution function $n({\bf k})$.

If the electronic interactions are weak, one usually uses the
nomenclature of Landau and refers to dressed quasiparticles
replacing the electron as the fundamental excitation in the solid.
These interactions may be characterized by the residue of the pole
(usually denoted by $Z_{\bf k}$) and the quasiparticle effective
mass $m^{\ast}/m_{\rm b}=(Z_{\bf k})^{-1}$ with $m_{\rm b}$ the
bare band mass for quasiparticles lying near the Fermi surface.
$Z_{\bf k}$ is related to the real part of the self energy
$\Sigma^{\prime}$ which can be expanded for electrons near the
Fermi surface as $\Sigma^{\prime}({\bf k},\omega) \simeq
\Sigma_{0}^{\prime}({\bf k}) + \omega
\partial\Sigma^{\prime}({\bf k},\omega=0)/
\partial\omega$. According to Luttinger's theorem,
the Fermi surface average of $\Sigma_{0}^{\prime}({\bf k})$
vanishes.  The enhancement of the quasiparticle mass over the band
mass can be written as $m^{\ast}/m_{\rm b} = (1-\partial
\Sigma^{\prime}/\partial\omega)$. One often defines $m^{\ast}/m_{\rm
b} = 1+\lambda$ with the dimensionless coupling constant $\lambda
\ge 0$ (see, e.g., \cite{Ashcroft:1976}). $Z_{\bf k} = (1-\partial
\Sigma^{\prime}/\partial\omega)^{-1}$ is always smaller than 1,
reflecting the fact that even for $\omega=0$ and $T=0$ only a
fraction $Z_{\bf k}$ of the spectral weight (coherent part) is in
the pole of $A({\bf k},\omega)$ while $1-Z_{\bf k}$ (incoherent
part) is distributed over larger energy scales. Equivalently,
$Z_{\bf k} \le 1$ is the discontinuity at ${\bf k}_F$ of the
zero-temperature momentum distribution function $n({\bf k})$. If
$Z_{\bf k}$ approaches zero ($\propto 1/\ln\omega$) the system is
referred to as a marginal Fermi liquid \cite{Varma:1989a}, and sum
rule (i) is exhausted only at energies much larger than $\xi_{\bf
k}$.
Thus, knowledge of the self energy is an important requisite to
understanding many-body interactions.

For this reason, single particle methods such as ARPES, electron
tunneling and specific heat measurements have been applied
extensively to study correlated electron systems. Very much
stimulated by the discovery of superconductivity in the cuprates
\cite{Bednorz:1986} ARPES and tunneling spectroscopy have
developed more rapidly than any other method in the last decade.
ARPES data has given unprecedented insight into momentum-resolved
single electron properties and their many body effects
\cite{Campuzano:2002,Damascelli:2003}, while tunneling measurements have
provided information on pairing
\cite{Mandrus:1991,Renner:1995,Zasadzinski:2002} and have recently
elucidated many issues of nanoscale inhomogeneities in the
cuprates and their connection to superconductivity
\cite{Vershinin:2004,McElroy:2003,Hanaguri:2004,Howald:2003,McElroy:2005,Fang:2006,Hoffman:2002,Kivelson:2003}. Detailed knowledge of
phase transitions in the cuprates has been obtained from specific
heat studies \cite{Moler:1994,Roulin:1998,Loram:2001}. Due to
space limitations we only have listed some of the later references
of important experimental papers which we hope serve as an entry
point for the reader to search backwards in time to follow the
developments.

Yet knowledge of the spectral function and single-particle
excitation spectra do not yield information about how the
electrons may transport heat, current, entropy, or energy. For
this one needs two-particle correlation functions for charge or
spin which can be measured by, e.g., ordinary and heat transport,
optical spectroscopy, neutron and light scattering. As an example
for such a correlation function we consider a standard expression
for the generalized Kubo susceptibility $\chi_{a,b}^{\prime
\prime}(\Omega)$ of weakly interacting, isotropic normal electrons
(see, e.g., \textcite{Mahan:2000}),
\begin{eqnarray}
\medskip&& \chi_{a,b}^{\prime \prime}(\Omega) =
\frac{2}{V}\displaystyle \sum_{{\bf k}} a_{{\bf
k}}b_{{\bf k}} \int_{-\infty}^{\infty} \frac{d\omega}{\pi}
G^{\prime \prime}({\bf
k},\omega) G^{\prime\prime}({\bf k},\omega+\Omega) \nonumber\\
&&\times \left[f(\omega)-f(\omega+\Omega) \right].
\label{Eq:Kubo}
\end{eqnarray}
Here $V$ is the volume, $a_{{\bf k}},b_{\bf k}$ are the bare
vertices representing quasiparticle charge $(a_{{\bf k}}=1)$ or
current $(a_{{\bf k}}=j_{{\bf k}}=e{\bf k})$ correlation
functions, and the factor 2 accounts for spin degeneracy. The
absorptive part of the conductivity $\sigma^{\prime}=\chi_{\rm
j,j}^{\prime \prime}(\Omega)/\Omega$ measures essentially a
convolution of occupied and unoccupied states. For electrons
weakly interacting with impurities the conductivity can readily be
calculated to exhibit a Lorentzian dependence on $\Omega$
represented by
\begin{equation}
\sigma^{\prime}(\Omega) = \sigma_0\frac{1}{1+(\Omega\tau)^{2}}
\label{eq:Drude}
\end{equation}
where $\sigma_0$ is the dc ($\Omega=0$) conductivity, and the
relaxation time $\tau=-\hbar(2\Sigma^{\prime\prime})^{-1}$
controls both the width of the spectral function and the
conductivity as a function of frequency.\footnote{Note that
Eq.~(\ref{Eq:Kubo}) does not return the proper transport lifetime
$\tau_t$ which differs by a factor of typically $1-\cos\theta$
with the scattering angle $\theta$ since events with $\theta
\approx 0$ do not contribute to the resistivity. This deficit must
be taken care of by vertex corrections \cite{Mahan:2000}.} A very
similar expression is found for light scattering\footnote{The case of
Raman scattering is described in detail in reference
\cite{Opel:2000} and touched upon briefly in
section~\ref{sec:ordering}.}. Thus in the case non-interacting
electrons, single and two-particle correlation
functions give similar results.

This is also true by and large if weak but essentially isotropic
interactions lead to an energy dependent
$\Sigma^{\prime\prime}(\omega)$ and, for causality, to a finite
$\Sigma^{\prime}(\omega)$. \textcite{Goetze:1972} and, more
phenomenologically, \textcite{Allen:1977} (for a more recent
reference see \cite{Basov:2005}) discuss how this generalization
modifies the response given by Eq.~(\ref{eq:Drude}), which is then
often referred to as the extended Drude model. However, interacting systems require
some care. For example, in superconductors, both the single and the
two-particle responses yield the energy gap. Yet two-particle
correlation functions also have coherence factors which can be
crucially important to determine the gap symmetry in
unconventional systems. Generally, collective modes (such as the
plasmon or excitons) appear directly in two-particle correlation functions
but only indirectly in the spectral function.

Sometimes the results from single- and two-particle measurements
can be qualitatively different, even for non-interacting
electrons. As an illustration, we consider first the
metal-insulator transition occurring in a system of otherwise
non-interacting electrons in a disordered environment (Anderson
transition). Here, backscattering of electrons from impurities
leads to destructive phase interference, and the electrons become
localized once a critical concentration of impurities is in place
in three dimensions. Thus while the conductivity is critical
and vanishes at the metal-insulator transition, the spectral
function, or equivalently the density of states, is uncritical. This
distinction becomes even more pronounced if the electron
interactions are strong and anisotropic, and the bare vertices
along with the Green's functions entering into Eq.~(\ref{Eq:Kubo})
must be renormalized by the strong interactions.

As a second example, in the spinless Falicov-Kimball model light
$d-$electrons strongly interact with localized $f-$electrons and
are characterized by the Hamiltonian \cite{Falikov:1969}
\begin{eqnarray}
&& H=-\frac{t^*}{2\sqrt{D}}\sum_{\langle
i,j\rangle}(c^\dagger_ic_j+c^\dagger_jc_i) + E_{f}\sum_{i}w_{i}
\nonumber\\
&&-\mu\sum_{i}(c^{\dagger}_{i}c_{i}+w_{i})
+U\sum_ic^\dagger_ic_iw_i \label{Eq:FK}
\end{eqnarray}
where $c^\dagger_i$ ($c_i$) create (destroy) a conduction electron
at site $i$, $w_i$ is a classical variable (representing the
localized electron number at site $i$) that equals 0 or 1,
$t^{\ast}$ is a renormalized hopping matrix element that is
nonzero between nearest neighbors on a hypercubic lattice in
$D$-dimensions, and $U$ is the local screened Coulomb interaction
between conduction and localized electrons. $\langle i,j\rangle$
denotes a sum over sites $i$ and nearest neighbors $j$. $E_{\rm
f}$ and $\mu$ are adjusted to set the average filling of
conduction and localized electrons. This model has been solved
exactly for electrons on a hypercubic lattice in the limit of
large coordination number \cite{Freericks:2003b}. The system
undergoes a metal-insulator transition (MIT) at half-filling (one
electron per site) if the interaction $U$ is beyond a critical
value $U_{\rm c}$. On either side of the metal-insulator
transition, the density of states is temperature independent
\cite{vanDongen:1992}, while the conductivity has a strong
temperature dependence \cite{Pruschke:1995} showing the
development of the MIT.

In systems with strong and anisotropic interactions, the
differences between single and two-particle properties are
inescapable. This is been borne out in the cuprates by the large amount of work
using optical \cite{Homes:2004} and thermal conductivities
\cite{Sutherland:2005}, resistivities \cite{Ando:2004}, nuclear
magnetic resonance (NMR) \cite{Alloul:1989} and electron spin
resonance (ESR) \cite{Janossy:2003}. These experiments have
revealed basic properties of strongly correlated systems and have
emerged as key elements to characterize the complex behavior of
the high-$T_c$ cuprates.

Yet these two-particle measurements are largely insensitive of
anisotropies, as they measure Brillouin zone averaged quantities.
As a result they reveal the behavior of quasiparticles having the
highest velocities which, in the cuprates, are the quasiparticles
near the nodal regions of the Brillouin zone. As far as carriers
are concerned, the momentum dependence of neutron scattering serves
mainly to measure spin dynamics in different regions of the
Brillouin zone.

In this review, we illustrate that Raman spectroscopy gives
complementary information to all of these measurement techniques,
and also may provide detailed information of charge and spin
dynamics of electrons in different regions of the Brillouin zone.
This is due to the polarization selection rules. As with phonon
scattering \cite{Hayes:2005}, simple group theoretic symmetry
arguments can be used to focus on electron dynamics in different
regions of the Brillouin zone. For  Raman scattering $(a_{{\bf
k}}b_{{\bf k}})$ is replaced with $\gamma_{\bf k}^{2}$ which, in
certain limits, may be represented by $\gamma_{\bf k}=
\sum_{\mu,\nu}e^{i}_{\mu}e^{s}_{\nu}
\partial^{2}\xi_{\bf k}/\hbar^{2} \partial k_{\mu}\partial k_{\nu}$,
with ${\bf e}^{i,s}$ the incident, scattered light polarization
vectors.\footnote{For a microscopic derivation see
section~\ref{Section:theory}.} Apart from energy-independent
scaling factors and vertices with different ${\bf k}$ dependences
there is an extra factor $1/\Omega$ between Raman and infrared
response. It has been shown first by \textcite{Shastry:1990} and
explicitly demonstrated within Dynamical Mean Field Theory (DMFT)
by \textcite{Freericks:2001a} that, under certain restrictions,
there is a simple correspondence between conductivity and Raman
response,
\begin{equation}
  \Omega \sigma^{\prime}(\Omega) \propto \chi_{\gamma
  \gamma}^{\prime \prime}(\Omega), \label{Eq:SS}
\end{equation}
highlighting that electronic Raman scattering measures transport
properties. However, even the simple form of the vertices given
above shows that a coincidence can be expected only for an
isotropic material. In anisotropic systems, light scattering can
sample parts of the Fermi surface which are unaccessible for
infrared spectroscopy.

\subsection{State of the Art Experimental Technique}

Over decades Raman scattering was predominantly used for the study
of molecular and lattice vibrations which produce isolated and
typically narrow lines in the spectra (see
Fig.~\ref{Fig:gen_idea}). The lines are used as probes which
sensitively react to changes in the environment of the vibrating
atoms. Similar considerations are at the heart of magnetic
resonance techniques such as NMR and ESR.

If light is scattered from electrons in solids, the spectra are
usually continuous (Fig.~\ref{Fig:gen_idea}). To study their
evolution as a function of a control
parameter, such as temperature, doping, magnetic field or pressure,
is rather involved since the overall shape and not the position of
well-defined lines matters. In addition, typical cross sections
per unit solid angle and energy interval are smaller by several
orders of magnitude than those of vibrations. Electronic Raman
scattering in a metal typically produces one energy-shifted photon
per s, meV, and sr (unit solid angle steradian) out of $10^{13}$
incoming ones. The low efficiency is particularly demanding in
studies at high pressure since additional losses and complications
such as fluorescence and birefringence arise from the windows, which are
typically diamond anvils. Although there were successful early
experiments \cite{Zhou:1996} the availability of synthetic
diamonds brought substantial advances \cite{Goncharov:2003}.

There are three inventions which finally produced the required
sensitivity: (i) the laser as an intense light source providing
lines of high spectral purity in a wide energy range, (ii) as an
early application of the laser, holographically fabricated
gratings without secondary images (ghosts) and an extremely low
level of diffusely scattered light and, finally, (iii) the
invention of charge-coupled devices (CCD) as a location-sensitive
detector with an efficiency at the quantum limit and negligible
dark count rate.

Gratings have an extremely well defined number of lines per unit
length (cm in the cgs system). This is the origin of the energy
unit $\rm cm^{-1}$. The following conversions are frequently used:
\begin{eqnarray*}
1~{\rm meV} &=& 11.604~{\rm K}\\
1~{\rm meV} &=& 8.0655~{\rm cm^{-1}}\\
k_{\rm B} &=& 0.69504~\frac{\rm cm^{-1}}{\rm  K}
\end{eqnarray*}

Since the CCD has a high spatial resolution down to a few $\mu$m
it is a superior replacement of the photographic plate.  It
facilitates recording complete spectra in a single exposure with
energy ranges from meV up to approximately 1~eV, depending on the
desired resolution.

\begin{figure}[floatfix]
\centerline{\epsfig{figure=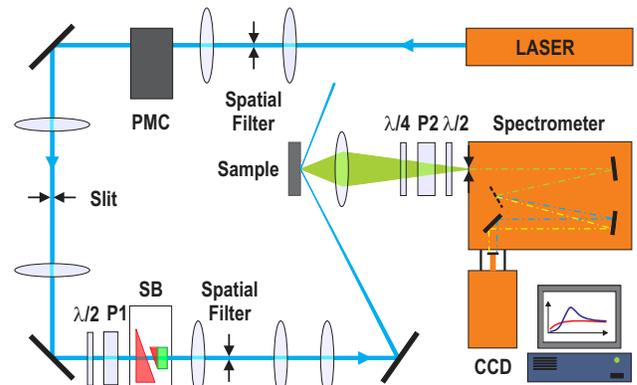,width=1\linewidth,clip=}}
\caption{Schematic drawing of the light path. The laser light with
energy $\hbar \omega_{i}$ is first spatially filtered. A prism
monochromator (PMC) is to suppress the plasma lines of the laser
medium, only the coherent one at $\hbar \omega_{i}$ passes the
slit. The polarization is prepared with polarizer P1 and the
Soleil-Babinet compensator (SB). The $\lambda/2$ retarder in front
of P1 allows one to adjust the power. Before hitting the sample
the light is once again spatially filtered to maintain an
approximately Gaussian intensity profile in the spot. If the angle
of incidence is not close to zero phase shift effects at the
sample surface must be taken into account since the polarization
inside the sample is important. High speed optics collects the scattered light. The polarization state is
selected by a $\lambda/4$ retarder and polarizer P2. The
$\lambda/2$ retarder in front of the entrance slit rotates the light
polarization for maximal transmission of the spectrometer (here
single stage). Except for the compensator most retarders and
polarizers work also for light propagating at small angles (up to
approximately $\pm 5^{\circ}$) with respect to the optical axis.
The configuration shown here is usually referred to as
back-scattering geometry since incoming and outgoing photons have
essentially opposite momenta in the sample.} \label{Fig:set_up}
\vskip -0.5cm
\end{figure}

The essentials of a setup for inelastic light scattering with
polarized photons are shown schematically in
Fig.~\ref{Fig:set_up}. The coherent light at energy
$\hbar\omega_{i}$ from the laser (Ar$^{+}$ and Kr$^{+}$ gas lasers
are still very popular) is spatially filtered. A prism monochromator
(PMC) selects the desired frequency and suppresses incoherent
photons from the laser medium. A combination of a $\lambda/2$
retarder and a polarizer (P1) facilitates the preparation of a
photon flux of a well-defined polarization state and intensity.
For excitation, the polarization inside the sample counts. The
same holds for the selection of the proper polarization for the
scattered photons at $\hbar\omega_{s}$. The best results are
obtained by using a crystal polarizer (P1, e.g. of Glan-Thompson
type) and a Soleil-Babinet compensator for the incoming light and
an achromatic $\lambda/4$ retarder and another crystal polarizer
(P2) for the scattered light. In this way all states, including
circularly polarized ones, can be prepared. The $\lambda/2$
retarder in front of the entrance slit of the spectrometer
rotates the polarization into the direction of highest sensitivity.

Since we wish to discriminate between the $10^{10 - 15}$
elastically scattered photons at $\hbar\omega_{i}$ and the few
Raman photons at $\hbar\omega_{s}$ at very small shifts
$\hbar\Omega = |\hbar\omega_{i} - \hbar\omega_{s}| < 1$~meV, a
single monochromator is insufficient. A modern instrument for
Raman scattering in metallic samples has three stages consisting
of essentially independent grating monochromators. The first two
are usually subtractively coupled and select a band from the
spectrum of inelastically scattered photons. The third stage
disperses the band transmitted through the two stages of the
premonochromator into a spectrum which is recorded by the CCD. In
this configuration, the dispersion is given only by the third stage
while the first two discriminate the elastically scattered laser
light. If all stages are coupled additively, the resolution is
improved by a factor of 3. Because of losses at the mirrors and gratings,
only 15 to 20~\% of the photons entering the spectrometer arrive
at the detector.

Since very interesting physics is going on at energies even below
1~meV (see, e.g., section~\ref{sec:fluctuations}) the
discrimination is a cardinal point. Out of the two options only
the premonochromator gives satisfactory results below 10~meV. The
price one has to pay is a loss of intensity of approximately
60~\%. Alternatively, for energy shifts above 10~meV an
interferometric notch filter can be used. The latter device is
widely used for commercial applications which develop rapidly
since the introduction of the CCD. The main fields are quality
control and analytics.

At finite temperatures, $T>0$, inelastically scattered light is
found on either side of $\hbar \omega_{i}$. As a consequence of
time-reversal symmetry and for phase space arguments the energy
gain (Anti-Stokes) and loss (Stokes) spectra are related by the
principle of detailed balance (equivalent to the
fluctuation-dissipation theorem) \cite{Placzek:1934,Landau:1960}
\begin{equation}
\frac{\dot{N}_{\rm AS}}{\dot{N}_{\rm ST}}=\left(\frac{\omega_{i}+
\Omega}{\omega_{i}-
\Omega}\right)^{2}\exp\left(-\frac{\hbar\Omega}{k_{\rm B}T}\right)
\label{Eq:ST-AS}
\end{equation}
with $\dot{N}_{\rm ST(AS)}$ and $\hbar\Omega$  the rate of photons
per unit time collected on the Stokes (Anti-Stokes) side and the
energy transferred to the system, respectively.
Equation~(\ref{Eq:ST-AS}) can be used to determine the temperature
of the laser spot.

If spectra are measured in large energy ranges, the sensitivity of
the instrument has to be taken into account. To this end, the
spectral response of the whole system, including all optical
elements between the sample and the entrance slit of the
spectrometer, the spectrometer itself, and the detector, must be
calibrated. This is best done by replacing the sample with a
continuous light source of the same size as the laser spot with a
known spectral emissivity. A continuous source is of crucial
importance for including the energy dependence of the dispersion
in addition to the bare transmission. In addition, the frequency
dependence of the sample's index of refraction,
$\sqrt{\varepsilon}=n+ik$, requires attention in order to get the
internal cross section.

The main limitations of present commercial systems come from
geometrical aberrations of the spectrometer optics and from the
relatively low total reflectivity of the large number of mirrors.
It is a matter of resources to improve these caveats. Recently, an
improved type of triple spectrometer with aspherical optics has
been described by \textcite{Schulz:2005}. CCDs and gratings are
close to the theoretical limits.

For many studies, light sources with continuously adjustable lines
in an extended energy range would be desirable. This holds
particularly true for organic materials (e.g. carbon nanotubes or
proteins) which have relatively sharp resonances in the visible
and the ultraviolet. The synchrotron, free electron ``lasers'', as
well as dye and solid-state lasers are developing rapidly and will
gain influence on the field of Raman spectroscopy in the near
future. The same holds for near-field techniques
\cite{Hartschuh:2003} which are capable of improving spatial
resolution by at least an order of magnitude below the diffraction
limit.

\section{THEORY OF ELECTRONIC RAMAN SCATTERING}
\label{Section:II}

\subsection{Electronic Coupling to Light}
\label{Section:theory}

The aim of this section is to formulate the theoretical treatments
for inelastic light scattering in general. Much has been done in
the development of theories for Raman scattering, particularly in
semiconductors and superconductors. Excellent and early reviews
of electronic Raman scattering have been given by
\textcite{Klein:1983} and \textcite{Abstreiter:1984} focusing on
semiconductors. More recent reviews by \textcite{Devereaux:1997}
and \textcite{Sherman:2003} have focused on theory in
superconductors with applications towards the cuprates. We outline
the general formalism for treating systems with weak and strong
correlations and return to a discussion of various theoretical
models in connection with materials in the following section.

\subsubsection{General Approach}

We first consider a Hamiltonian for $N$ electrons coupled to the
electromagnetic fields \cite{Pines:1966,Blum:1970}:
\begin{equation}
H=\sum_{i}^{N} \frac{[{\bf \hat p}_{i}+(e/c){\bf \hat A}({\bf
r}_{i})]^{2}}{2m} + H_{\rm Coulomb} +H_{\rm fields},
\end{equation}
where ${\bf \hat p}=-i\hbar{\bm \nabla}$ is the momentum operator,
$e$ is the magnitude of the elementary charge (the electronic
charge is $q_e=-e$) and $c$ the speed of light. ${\bf \hat
A}({\bf} r_{i})$ is the vector potential of the field at
space-time point ${\bf r_{i}}$ and $m$ the electron mass. $H_{\rm
Coulomb}$ represents the Coulomb interaction and $H_{\rm fields}$
the free electromagnetic part. We use the symbol $\hat A$ to
denote operators. We expand the kinetic energy to obtain
\begin{eqnarray}
H&=&H^{\prime}+\frac{e}{2mc}\sum_{i}\left[{\bf \hat p}_{i}\cdot
{\bf \hat A}({\bf r}_{i})
+ {\bf \hat A}({\bf r}_{i})\cdot {\bf \hat p}_{i}\right]\nonumber\\
&&+\frac{e^{2}}{2mc^{2}}\sum_{i}{\bf \hat A}({\bf r}_{i})\cdot
{\bf \hat A}({\bf r}_{i}), \label{Eq:Hamiltonian}
\end{eqnarray}
with $H^{\prime}=H_{\rm 0}+H_{\rm fields}$ and $H_{\rm 0}$~=
$(1/2m)\sum_{i}{\bf \hat p}_{i}^{2}$~+ $H_{\rm Coulomb}$.
Generally we choose $|\alpha\rangle$
to denote eigenstates of $H_{\rm 0}$ with eigenvalues
$E_{\alpha}$: $H_{\rm 0}|\alpha\rangle=E_{\alpha}|\alpha\rangle$.
The eigenstate is labeled by all the relevant quantum numbers for
the state, such as combinations of band index, wavevector, orbital
and/or spin quantum numbers, for example. The eigenstates may be
considered to be Bloch electrons when the electron-ion interaction
is included in $H_{\rm 0}$, as plane-wave states if it is
neglected, or may represent Hubbard states if $H_{\rm Coulomb}$ is
taken to include short-range Hubbard-like interactions between
electrons.

The electromagnetic vector potential can be expanded into Fourier
modes ${\bf \hat A}({\bf r}_{i})=\sum_{{\bf q}} e^{i{\bf
q}\cdot{\bf r}_{i}} {\bf  \hat A}_{\bf q}$. In second quantized
notation, the electromagnetic field operator takes the
form \cite{Mahan:2000}
\begin{equation}
{\bf \hat A}_{\bf q}=\sqrt{\frac{hc^{2}}{\omega_{{\bf q}}V}}[{\bf
\hat e}_{\bf q}a_{-{\bf q}}+{\bf \hat e}_{\bf q}^{*}a_{{\bf
q}}^{\dagger}],
\end{equation}
with $V$ the volume and $a_{\bf q}^{\dagger},a_{\bf q}$ are the
creation, annihilation operators of transversal photons with
energy $\hbar\omega_{{\bf q}}= \hbar c |{\bf q}|$ having a
polarization direction denoted by the complex unit vector ${\bf \hat e}_{\bf
q}$.

Electronic Raman scattering measures the total cross section for
scattering from all the electrons illuminated by the incident
light. The differential cross section is determined by the
probability that an incident photon $\omega_{i}$ is scattered into
a solid-angle interval between $\Omega$ and $\Omega+d\Omega$ and a
frequency window between $\omega_{s}$ and $\omega_{\rm
s}+d\omega_{s}$. A general expression for the differential light
scattering cross-section is given via the transition rate $R$ of
scattering an incident $({\bf q}_{i},\omega_{i},{\bf \hat e}_{\bf
q}^{(i)})$ photon into a outgoing state $({\bf
q}_{s},\omega_{s},{\bf \hat e}_{\bf q}^{(s)})$,
\begin{equation}
\frac{\partial^{2}\sigma}{\partial\Omega\partial\omega_{s}}=\hbar
r_{0}^{2} \frac{\omega_{s}}{\omega_{i}}R. \label{Eq:cross-section}
\end{equation}
Here, $r_{0}=e^{2}/mc^{2}$ is the Thompson radius, and $R$ is
determined via Fermi's Golden Rule,
\begin{equation}
R=\frac{1}{\cal Z}\sum_{I,F}e^{-\beta E_{I}}\mid M_{F,I}
\mid^{2}\delta(E_{F}-E_{I}-\hbar\Omega), \label{Eq:Golden-Rule}
\end{equation}
with $\beta=1/k_{\rm B}T$, $\cal Z$ the partition function, and
$M_{F,I}=\langle F| M| I\rangle$ where $M$ is the effective light
scattering operator. The sum represents a thermodynamic average
over possible initial and over final states with {\bf k}~vectors
in the solid angle element $d\Omega$ of the many-electron system
having energies $E_{I}$, $E_{F}$, respectively. Here
$\Omega=\omega_{i}-\omega_{s}$ is the transferred frequency and we
denote ${\bf q}={\bf q}_{i}-{\bf q}_{s}$ the net momentum
transfered by the photon. Multiplying Eq. (\ref{Eq:cross-section})
by the incident photon flux gives the number of scattered photons
per second into the solid angle increment $d\Omega$ within the
frequency range $d\omega_{s}$, while multiplying Eq.
(\ref{Eq:cross-section}) by $\omega_{s}/\omega_{i}$ gives the
power scattering cross section \cite{Klein:1983}.\footnote{We note
that Eqs.~(\ref{Eq:cross-section}) and (\ref{Eq:Golden-Rule})
describe scattering inside the material. Trivial
(Fresnel-formulas) and non-trivial ($q_z$~integration)
\cite{Abrikosov:1961,Falkovskii:1990,Falkovskii:1991}
transformations, which we do not discuss here, are required to
fully describe the cross section outside. The $q_z$ integration
originates from the lack of momentum conservation perpendicular to
the surface of an absorbing medium and can change the spectra
qualitatively. From here on, $\hbar\Omega$ is always the energy
transferred to the system.}

\begin{figure}[b!]
\centerline{\epsfig{figure=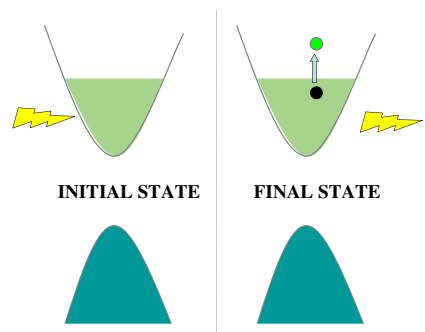,width=6.0cm,clip=}}
\vspace{0.1cm} \caption{Cartoon showing light scattering via
non-resonant intraband scattering.}
\label{Fig:ppt_nonres}\end{figure}

From here on we consider the case relevant to Raman scattering in
the visible range with photon energies of typically 2~eV. Since
the momentum transferred to the electrons, $q \sim 1/\delta$, with
$\delta$ the skin depth at the light energies
\cite{Abrikosov:1961}, is much less than the relevant momentum
scale of order $k_{\rm F}$, the Fermi momentum in metallic
systems, the limit $q\rightarrow 0$ is a good approximation in
practically all cases.\footnote{The applicability of the $q=0$
limit is discussed in more detail at the beginning of
section~\ref{Section:III}.} However, finite $q$ should be
considered if incident light from frequency-doubled or synchrotron
radiation is used where transitions between initial and final
states at finite $q$ can be probed.  Then the structure of the
Landau particle-hole continuum in weakly correlated systems or
transitions across a finite $q$ Mott gap in strongly correlated
insulators can be studied.\footnote{For a review of relevant work
in this regard, the reader is referred to references
\cite{Platzman:1998,Kotani:2001,Devereaux:2003a,Devereaux:2003c}.}

\begin{figure}[b!]
\centerline{\epsfig{figure=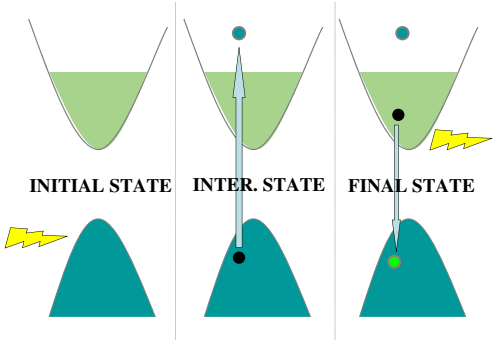,width=6.5cm,clip=}}
\vspace{0.1cm} \caption{Cartoon showing light scattering via
interband transitions.} \label{Fig:ppt_res}\end{figure}

\begin{figure}[b!]
\centerline{\epsfig{figure=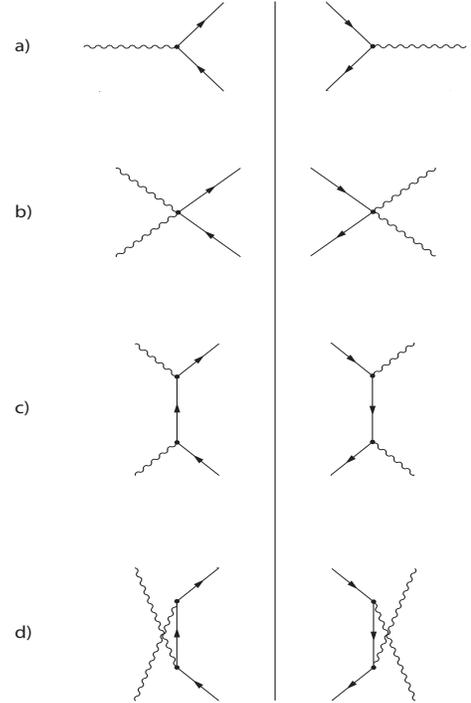,width=6.5cm,height=10.0cm,clip=}}
\vspace{0.1cm} \caption{Feynman diagrams contributing  to the
effective light scattering operator $M$. The top diagram
represents single-photon absorption, the second two photon
scattering, while photon emission (absorption) followed by
absorption (emission) is shown in the 3$^{rd}$ (4$^{th}$) diagram
from top. The panels on the right are the time-reversed partners
of the left diagrams. } \label{Fig:matrix}\end{figure}

$M_{F,I}$ has contributions from either of the last three terms in
Eq. (\ref{Eq:Hamiltonian}): the first two terms coupling the electron's
current to a single photon and the third term coupling the
electron's charge to two photons. This is shown in the schematic
cartoon in Figures \ref{Fig:ppt_nonres} and \ref{Fig:ppt_res}.
Here we consider two bands - one partially filled and the other
completely filled - in which the incident photon excites an
electron from either the partially or the completely filled band,
shown in Figures~\ref{Fig:ppt_nonres} and~\ref{Fig:ppt_res},
respectively. In the non-resonant intraband case, the photon gives up part
of its energy to leave behind a particle-hole pair, while in the
interband case, an intermediate state is involved, which decays via
a particle from the partially filled band into the hole left
behind in the filled band. The latter scattering may be resonant
if the incident or emitted photon energy corresponds to that of
the energy gap separation, otherwise it is non-resonant. In this
simple cartoon, one can see that excitations lying near the Fermi
surface are predominantly probed by non-resonant intraband
scattering while excitations involving transition between
different bands - such as the lower and upper Hubbard band for
example - are probed by intermediate state scattering. The Feynman
diagrams representing these contributions to $M_{F,I}$ are shown
in Figure~\ref{Fig:matrix}.

Referring to Eq.~(\ref{Eq:Hamiltonian}), the current coupling has
odd spatial symmetry and involves single photon emission or
absorption, while the second term is even in parity and involves
two photon scattering of emission followed by absorption and
vice-versa. The cross-section or the transition rate is thus
determined via Fermi's Golden rule by the square of the matrix
elements shown in Figure~\ref{Fig:matrix}.

\begin{figure}[b!]
\centerline{\epsfig{figure=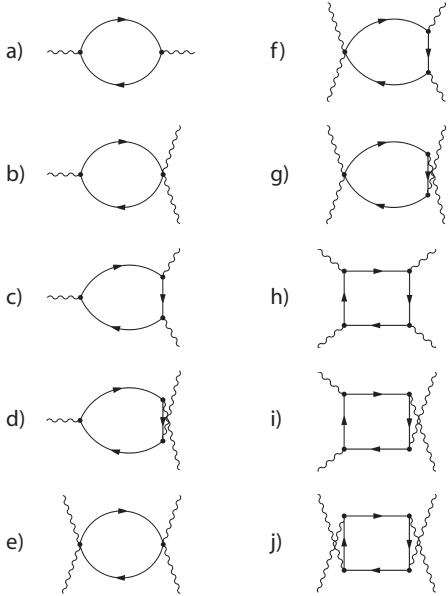,width=6.5cm,clip=}}
\vspace{0.1cm} \caption{Feynman diagrams contained in the
cross-section. (a) gives a renormalized photon propagator in the
solid, while (b)-(d) vanish due to parity and thus (a)-(d) do not
contribute to inelastic light scattering. Of the remaining
contributions, (e) refers to intra-band (sometimes
``non-resonant'') scattering within a single band, (f) and (g) are
referred to as ``mixed'' contributions while (h)-(j) describe
transitions within a single or between different bands via
intermediate band states. If the light energy is equal or close to
the energy difference of the states involved resonance effects
with a strong enhancement of the cross section occur. Therefore
the contributions themselves are sometimes referred to as
``resonant''.} \label{Fig:matrix_squared}\end{figure}

The resulting Feynman diagrams of the contributions to the cross section are
shown in Figure~\ref{Fig:matrix_squared}. However, not all of them
give rise to inelastic light scattering. Some of these terms
vanish either because they represent contributions to the
renormalized photon propagator
(Figure~\ref{Fig:matrix_squared}~(a)), or due to parity arguments
(Figures~\ref{Fig:matrix_squared}~(b)-(d)) in the limit of small
$q$ scattering. The remaining terms can be classified as
non-resonant (Figure~\ref{Fig:matrix_squared}~(e)), resonant
(Figure~\ref{Fig:matrix_squared}~(h)-(j) and mixed terms
(Figure~\ref{Fig:matrix_squared}~(f)-(g)), since in the former case
the initial and final states must share a large sub-set of quantum
numbers, while the other terms can involve transitions through
intermediate states well separated in energy and distinct from the
initial and final states. However, we remark that the response is
only truly resonant if the photon energies are tuned to the energy
gap between intermediate and initial or final states.

To obtain a general expression for the matrix element $M_{F,I}$
for Raman scattering, we use second quantized notation for the
fermions in which the single-particle wave function and it's
conjugate are given by $\psi({\bf
r})=\sum_{\alpha}c_{\alpha}\varphi_{\alpha}({\bf r})$ and
$\psi^{\dagger}({\bf
r})=\sum_{\alpha}c_{\alpha}^{\dagger}\varphi_{\alpha}^{*}({\bf
r})$, with $\varphi,\varphi^{*}$ the eigenstates of the
Hamiltonian $H_{\rm 0}$. Electron states $\alpha,\beta$ are
created, annihilated by $c^{\dagger}_{\alpha}, c_{\beta}$
respectively, and the indices refer to the quantum number
associated with the state, such as the momenta and/or spin states.
The matrix element $M_{F,I}$ can thus be written as
\begin{eqnarray}
   M_{F,I}=&&{\bf e}_{i}\cdot{\bf e}_{s}\sum_{\alpha,\beta}
   \rho_{\alpha,\beta}({\bf q})
   \langle F\left| c^{\dagger}_{\alpha}c_{\beta}\right| I \rangle\nonumber\\
   &&+\frac{1}{m}\sum_{\nu}\sum_{\alpha,\alpha^{\prime},\beta,\beta^{\prime}}
   p_{\alpha,\alpha^{\prime}}({\bf q}_{s})
   p_{\beta,\beta^{\prime}}({\bf q}_{i}) \nonumber\\
   &&\times \biggl( \frac{\langle F\left|
   c_{\alpha}^{\dagger}c_{\alpha^{\prime}}\right| \nu \rangle \langle
   \nu\left| c_{\beta}^{\dagger}c_{\beta^{\prime}}\right| I \rangle}
   {E_{I}-E_{\nu}+\hbar \omega_{i}}\nonumber\\
   &&+ \frac{\langle F\left|
   c_{\beta}^{\dagger}c_{\beta^{\prime}}\right| \nu \rangle \langle
   \nu\left| c_{\alpha}^{\dagger}c_{\alpha^{\prime}}\right| I
   \rangle} {E_{\rm I}-E_{\nu}-\hbar \omega_{\rm s}}\biggr).
\label{Eq:gen_matrix}
\end{eqnarray}
Here $|I\rangle,\mid F\rangle,|\nu\rangle$ represent the initial,
final and intermediate many-electron states having energies
$E_{I,F,\nu}$, respectively. The many-electron states could be
labeled by band index and momentum as, for example, for Bloch
electrons. They may also consist of core and valence electrons on
selected atoms for x-ray scattering, or may represent states of
the many-band Hubbard model for correlated electrons.
$\rho_{\alpha,\beta}({\bf q})=\int d^{3}r\varphi^{*}_{\alpha}({\bf
r}) e^{i {\bf q}\cdot {\bf r}}\varphi_{\beta}({\bf
r})=\langle\alpha\left|e^{i {\bf q}\cdot {\bf r}}\right|
\beta\rangle$ is the matrix element for single-particle density
fluctuations involving states $\alpha,\beta$. The momentum density
matrix element is given by $p_{\alpha,\beta}({\bf
q}_{i,s})=\langle\alpha\left| {\bf p} \cdot {\bf e}_{i,s} e^{\pm
i{\bf q}_{i,s}\cdot {\bf r}}\right| \beta\rangle$. The first term
in the expression arises from the two-photon scattering term in
Eq.~(\ref{Eq:Hamiltonian}) in first order perturbation theory. The
remaining terms arise from the single-photon scattering term in
Eq.~(\ref{Eq:Hamiltonian}) in second order via intermediate states
$\nu$ and involve different time orderings of photon absorption
and emission. The ${\bf p\cdot A}$ coupling does not enter to
first order since the average of the momentum operator is zero.

\subsubsection{Importance of Light Polarization}

At this point, little progress can be made in evaluating the matrix
elements for Raman scattering without specifying the quantum
numbers of the electronic many-body states. Yet, from
Eq.~(\ref{Eq:gen_matrix}), one can apply symmetry arguments to view
what types of excitations can be created by the incident photons.
In this subsection, we employ a general set of symmetry
classifications and put specific emphasis on models in later
subsections.

The first term in Eq.~(\ref{Eq:gen_matrix}) only arises if the
incident and scattering polarization light vectors are not
orthogonal, as electronic charge density fluctuations are created
and destroyed along the polarization directions of the incident
and scattered photons. Thus, for instance, this term does not probe
electron dynamics in which the charge density relaxes in a
direction orthogonal to the incident polarization direction.

In the limit of small momentum transfer ${\bf q} \rightarrow 0$,
the matrix element simplifies  to $\rho_{\alpha,\beta}({\bf
q}\rightarrow 0)=\delta_{\alpha,\beta}$ and thus this term gives
rise to scattering from fluctuations of the isotropic electronic
number density.

The remaining terms in Eq.~(\ref{Eq:gen_matrix})  have
contributions regardless of photon polarization directions.
However, one can further classify scattering contributions by
separating the sum over intermediate states $|\nu\rangle$ into
states which share some quantum numbers with the initial many-body
state, such as band index, and states which do not. Since the
photon momenta are much smaller than the relevant electron
momenta, contributions of the terms where the intermediate states
include the initial states are roughly a factor of $v_{\rm F}/c$
smaller than the first term in Eq.~(\ref{Eq:gen_matrix}) and can
be neglected \cite{Wolff:1966,Pines:1966}. However contributions
where $|\nu\rangle$ includes higher bands cannot in general be
neglected, particularly if the energy of the incident or
scattering light lies near the energy of a transition from the
initial state $E_{I}$ to an intermediate state $E_{\nu}$. These
terms thus give rise to ``mixed'' and ``resonant'' Raman
scattering.

The polarization dependence of Raman scattering can be generally
classified using arguments of group theory. In essence, the charge
density fluctuations brought about by light scattering are
modulated in directions determined by the polarizations of the
incident and scattered photons. These density fluctuations thus
have the symmetry imposed on them by the way in which the light is
oriented, and the charge density fluctuations obey the symmetry
rules governing the scattering geometry. This is manifest in
the dependence of the Raman matrix elements on the initial and
final fermion states. In general, the Raman matrix element
$M_{F,I}=M_{I,F}^{\alpha,\beta}e_{i}^{\alpha}e_{s}^{\beta}$ can be
decomposed into basis functions of the irreducible point group of
the crystal $\Phi_{\mu}$
\cite{Klein:1984,Monien:1990,Shastry:1991,Devereaux:1992,Hayes:2005}
\begin{equation}
M_{F,I}({\bf q}\rightarrow 0)=\sum_{\mu}M_{\mu}\Phi_{\mu},
\label{Eq:decomposition}
\end{equation}
with $\mu$ representing an irreducible representation of the point
group of the crystal. Which set of $\mu$ contributes to the sum is
determined by the orientation of incident and scattered
polarization directions. As an example, if we consider the $D_{\rm
4h}$ group of the tetragonal lattice, as in the cuprates, the
decomposition can be written as
\begin{eqnarray}
M_{F,I}&=&\frac{1}{2} O_{A_{1g}^{(1)}}(e_{i}^{x}e_{s}^{x}+e_{i}^{y}e_{s}^{y})\nonumber\\
&+&\frac{1}{2}O_{A_{1g}^{(2)}}(e_{i}^{z}e_{s}^{z})\nonumber\\
&+&\frac{1}{2}O_{B_{1g}}(e_{i}^{x}e_{s}^{x}-e_{i}^{y}e_{s}^{y})\nonumber\\
&+&\frac{1}{2}O_{B_{2g}}(e_{i}^{x}e_{s}^{y}+e_{i}^{y}e_{s}^{x})\nonumber\\
&+&\frac{1}{2}O_{A_{2g}}(e_{i}^{x}e_{s}^{y}-e_{i}^{y}e_{s}^{x})\nonumber\\
&+&\frac{1}{2}O_{E_{g}^{(1)}}(e_{i}^{x}e_{s}^{z}+e_{i}^{z}e_{s}^{x})\nonumber\\
&+&\frac{1}{2}O_{E_{g}^{(2)}}(e_{i}^{y}e_{s}^{z}+e_{i}^{z}e_{s}^{y}),
\label{Eq:group_theory}
\end{eqnarray}
with $O_{\mu}$ the corresponding projected operators and
$e_{i,s}^{\alpha}$ the light polarizations. This classification
demonstrates that there is no mixing of representations for $q=0$,
i.e. the correlation functions read $R\sim\langle
O_{\mu}^{\dagger}O_{\mu^{\prime}}\rangle=R_{\mu}\delta_{\mu,\mu^{\prime}}$
and there are 6 independent correlation functions, each selected
by combinations of polarization orientations.

Following \textcite{Shastry:1990}, we list in Table~\ref{Table_1}
some common experimentally used polarization geometries in
relation to the elements of the transition rate $R$ selected.  One
observes that a complete characterization of $M$ can be made from
a subset of the polarization orientations listed in the table.
However, additional polarization orientations can be useful to
calibrate data and compare symmetry decompositions from different
combinations of orientations. For geometries with polarizations in
the $(a-b)$~plane in $D_{4h}$ crystals, the irreducible
representations cannot be accessed individually and must be
separated by proper subtraction procedures. As a minimum set, four independent configurations are required, while additional polarizations may be used for consistency checks.

\begin{center}
\begin{table*}[floatfix]
\begin{tabular}{|c|c|c|c|r|}
\hline
Geometry & $\hat e_{i}$ & $\hat e_{s}$ & $R$ & Basis Functions $\Phi_{\mu}(\bf k)$\\
\hline\hline $xx$, $yy$    & $\hat x, \hat y$ &$\hat x, \hat y$&
$R_{A_{1g}}+R_{B_{1g}}$
&$\frac{1}{2}[\cos(k_{x}a)+\cos(k_{y}a)]\pm\frac{1}{2}[\cos(k_{x}a)-\cos(k_{y}a)]$ \\
\hline
$x^{\prime}x^{\prime}$    & $\frac{1}{\sqrt{2}}(\hat x+\hat y)$ &$\frac{1}{\sqrt{2}}(\hat x+\hat y)$& $R_{A_{1g}}+R_{B_{2g}}$ & $\frac{1}{2}[\cos(k_{x}a)+\cos(k_{y}a)]+\sin(k_{x}a)\sin(k_{y}a)$   \\
\hline
$x^{\prime}y^{\prime}$    & $\frac{1}{\sqrt{2}}(\hat x+\hat y)$ &$\frac{1}{\sqrt{2}}(\hat x-\hat y)$& $R_{B_{1g}}+R_{A_{2g}}$ & $\frac{1}{2}[\cos(k_{x}a)-\cos(k_{y}a)][1+\sin(k_{x}a)\sin(k_{y}a)]$ \\
\hline
$xy$    & $\hat x$ & $\hat y$ & $R_{B_{2g}}+R_{A_{2g}}$ & $\sin(k_{x}a)\sin(k_{y}a)\left\{1+\frac{1}{2}[\cos(k_{x}a)-\cos(k_{y}a)]\right\}$  \\
\hline
\cal{L}\cal{R} & $\frac{1}{\sqrt{2}}(\hat x+i\hat y)$& $\frac{1}{\sqrt{2}}(\hat x+i\hat y)$& $R_{B_{1g}}+R_{B_{2g}}$ & $\frac{1}{2}[\cos(k_{x}a)+\cos(k_{y}a)]+\sin(k_{x}a)\sin(k_{y}a)$  \\
\hline
\cal{L}\cal{L}    & $\frac{1}{\sqrt{2}}(\hat x+i\hat y)$& $\frac{1}{\sqrt{2}}(\hat x-i\hat y)$& $R_{A_{1g}}+R_{A_{2g}}$ & $\frac{1}{2}\left\{\cos(k_{x}a)+\cos(k_{y}a)+[\cos(k_{x}a)-\cos(k_{y}a)]\sin(k_{x}a)\sin(k_{y}a)\right\}$\\
\hline
$xz$    & $\hat x$& $\hat z$& $R_{E_{1g}}$  & $\sin(k_{x}a)\sin(k_{z}c)$ \\
\hline
$yz$    & $\hat y$& $\hat z$& $R_{E_{1g}}$   & $\sin(k_{y}a)\sin(k_{z}c)$\\
\hline
$zz$    & $\hat z$& $\hat z$& $R_{A_{1g}^{(2)}}$ & $\cos(k_{z}c)$  \\
\hline
\end{tabular}
\caption{Elements of the transition rate $R$ for experimentally
useful configurations of polarization orientations (given in Porto
notation)  along with the symmetry projections for the $D_{\rm
4h}$ point group relevant for the cuprates. Here we use notations
in which $x$ and $y$ point in directions along the Cu-O bonds in
tetragonal cuprates, while $x^{\prime}$ and $y^{\prime}$ are
directions rotated by $45^{\circ}$. \cal{L} and \cal{R} denote
left and right circularly polarized light, respectively. In our
convention left circular light has positive helicity. (In a
right-handed system the polarization rotates from $x$ to $y$ while
the wavefront travels into positive $z$ direction by $\lambda/4$.)
Note that in back-scattering configuration (see
Fig.~\ref{Fig:set_up}) with $\hat{e}_{i,s}$ pinned to the
coordinate system of the crystal axes the representation for
incoming and outgoing photons with circular polarizations change
sign in order to maintain the proper helicity.} \label{Table_1}
\end{table*}
\end{center}

In addition, we have listed in Table~\ref{Table_1} the
representative basis functions $\Phi_{\mu}(\bf k)$ taken from the
complete set of Brillouin zone (BZ) harmonics for $D_{4h}$ space
group \cite{Allen:1976}. This directly points out the connection
between polarizations and the coupling of light to electrons. By
virtue of the ${\bf k}$-dependence of the light scattering
transition rate, excitations on certain regions of the BZ can be
correspondingly projected out by orienting the incident and
scattered light polarization vectors. Thus, Raman is one of the
few spectroscopic multi-particle probes (the other being inelastic
X-ray scattering) able to examine charge excitations in different
regions of the BZ.

\begin{figure}[floatfix]
\centerline{\epsfig{figure=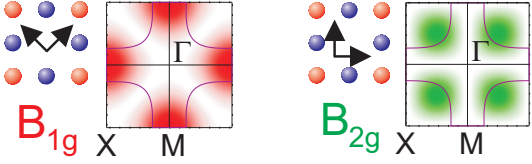,width=7.0cm,clip=}}
\vspace{0.1cm} \caption{Schematic weighting of the light
scattering transition for polarization orientations transforming
as $B_{1g}$ and $B_{2g}$ for a $D_{4h}$ crystal. High symmetry
points are indicated. Here a typical Fermi surface for optimally
doped cuprates is represented by the solid line, and the
orientations of the incident and scattered polarization light
vectors are shown with respect the to copper-oxygen bond
directions.}
\label{Fig:project}\end{figure}

For example, as demonstrated in Figure~\ref{Fig:project}, for
crossed polarizations transforming as $B_{1g}$, light couples to
charge excitations along the BZ axes ($k_{x}$ or $k_{y}=0$), while
for $B_{2g}$, excitations along the BZ diagonals ($k_{x}=\pm
k_{y}$) are projected accordingly. Operators like $O_{A_{2g}}$
cannot be accessed independently by linear polarizations alone.
Only sums including circular polarizations allow the isolation of
$A_{2g}$ components. Light scattering in this orientation can be
coupled to chiral excitations. These important symmetry
classifications have been extremely useful to point out
anisotropic electron dynamics in correlated insulators
\cite{Shastry:1990,Devereaux:2003a,Devereaux:2003c},
superconductors \cite{Devereaux:1994a}, disordered
\cite{Zawadowski:1990,Devereaux:1992} and correlated metals
\cite{Freericks:2001a,Freericks:2001b,Einzel:2004}. More recently
they have been related to sum rules referring to BZ-projected
potential energies in correlated systems \cite{Freericks:2005}. In
the remaining part of this review, such symmetry classifications
will be featured prominently.

\subsection{Formalism: Single-Particle Excitations and Weak Correlations}
\label{SectionC} We now consider specific cases where
simplifications can be made to Eq. (\ref{Eq:gen_matrix}). First we
assume that the intermediate many-particle states only differ from
the initial and final states by single electron excitations. This
is exact in the limit of non-interacting electrons, yet ignores
the effects of many-body correlations, and specifically the role
of Coulomb interactions on the reorganization of the initial state
into the intermediate states by creating many-particle
excitations. We now focus on weakly interacting systems, where
single-particle excitations are relatively well defined, and
discuss strongly correlated systems in later sections.

\subsubsection{Particle-hole Excitations}

Eq.~(\ref{Eq:gen_matrix}) can be simplified by replacing $E_{\nu}$
in the denominators by $E_{I}-E_{\beta^{\prime}}+E_{\beta}$ and
$E_{I}-E_{\alpha^{\prime}}+E_{\alpha}$ in the first and second
terms, respectively, and use the closure relation
$\sum_{\nu}|\nu\rangle\langle\nu|=1$. Commutator algebra
eliminates the four fermion matrix element,
\begin{equation}
   M_{F,I}=\sum_{\alpha,\beta}\gamma_{\alpha,\beta}\langle F\mid
   c^{\dagger}_{\alpha}c_{\beta}\mid I\rangle,
   \label{eq:ni_matrix}
\end{equation}
where
\begin{eqnarray}
   \gamma_{\alpha,\beta}=\rho_{\alpha,\beta}({\bf q}){\bf\hat e}_{i}
   \cdot{\bf\hat e}_{s}&+&\frac{1}{m}\sum_{\beta^{\prime}}
   \biggl(\frac{p^{s}_{\alpha\beta^{\prime}} p^{i}_{\beta^{\prime}\beta}}
   {E_{\beta}-E_{\beta^{\prime}}+\hbar\omega_{i}}\nonumber\\
   &&+\frac{p^{i}_{\alpha,\beta^{\prime}}p^{s}_{\beta^{\prime},\beta}}
   {E_{\beta}-E_{\beta^{\prime}}-\hbar\omega_{s}}\biggr).
\end{eqnarray}
Specifying to states $\alpha,\beta$ indexed by momentum quantum
numbers (such as Bloch electrons), from Eq. (\ref{Eq:cross-section}),
the Raman response simplifies to a correlation function $\tilde S$ of an effective charge
density $\tilde\rho$,
\begin{equation}
   \frac{\partial^{2}\sigma}{\partial \Omega\partial\omega_{\rm
   s}}=\hbar r_{\rm 0}^{2} \frac{\omega_{s}}{\omega_{i}}\tilde S({\bf
   q},i\Omega\rightarrow \Omega +i0).
\end{equation}
Here the Raman effective density-density correlation function is
\begin{equation}
  \tilde S({\bf q},i\Omega)=\sum_{I}\frac{e^{-\beta E_{\rm I}}}{\cal
   Z} \!\int\! d\tau e^{i\Omega\tau}\langle I\mid
   T_{\tau}\tilde\rho({\bf q},\tau)\tilde\rho({\bf -q},0)\mid
   I\rangle,
\label{Eq:S-effective}
\end{equation}
   $T_{\tau}$ is the complex time $\tau$ ordering operator and
\begin{equation}
   \tilde\rho({\bf q})=\sum_{{\bf k},\sigma}\gamma({\bf
   k,q})c^{\dagger}_{{\bf k+q},\sigma}c_{{\bf k},\sigma}.
\label{Eq:effective}
\end{equation}
The scattering amplitude $\gamma$ is determined from the Raman
matrix elements and incident/scattered light polarization vectors
as
\begin{equation}
   \gamma({\bf k,q})=\sum_{\alpha,\beta}\gamma_{\alpha,\beta}({\bf
   k,q}) e_{i}^{\alpha}e_{s}^{\beta},\label{eq:gamma-k-q}
\end{equation}
with
\begin{eqnarray}
   \gamma_{\alpha,\beta}({\bf
   k,q})=\delta_{\alpha,\beta}&+&\frac{1}{m}\sum_{{\bf
   k}_{\nu}}\biggl[ \frac{\langle{\bf k+q}\mid p^{\beta}_{s}\mid{\bf
   k_{\nu}}\rangle\langle{\bf
   k_{\nu}}\mid p_{i}^{\alpha}\mid{\bf k}\rangle}{E_{{\bf k}}-E_{{\bf k_{\nu}}}+\hbar\omega_{i}}\nonumber\\
   &+&\frac{\langle{\bf k+q}\mid p_{i}^{\alpha}\mid{\bf
   k_{\nu}}\rangle\langle{\bf k_{\nu}}\mid p_{s}^{\beta}\mid{\bf
   k}\rangle}{E_{{\bf k+q}}-E_{{\bf k_{\nu}}}-\hbar\omega_{\rm
   s}}\biggr]. \label{Eq:ni_gamma}
\end{eqnarray}
Here, $p_{i,s}^{\alpha}= p^{\alpha} e^{\pm i {\bf q}_{i,s}\cdot
{\bf r}}$. The effective dynamical density-density correlation
function or Raman response $\tilde{S}$ can be written in terms of
a dynamical effective density susceptibility $\tilde\chi$ via the
fluctuation-dissipation theorem, $$\tilde S({\bf
q},\Omega)=-\pi^{-1}\{1+n(\Omega,T)\}\tilde\chi^{\prime\prime}({\bf
q},\Omega),$$ with $n(\Omega,T)$ the Bose-Einstein distribution and
\begin{equation}
   \tilde\chi({\bf q},\Omega)=\langle [\tilde\rho({\bf
   q}),\tilde\rho({\bf -q})]\rangle_{\Omega}. \label{Eq:commutator}
\end{equation}
Thus, for non-interacting
electrons, the Raman response is given as a two-particle effective
density correlation function and can be calculated easily using,
e.g., diagrammatic techniques or via the kinetic equation
\cite{Devereaux:1995a}. This reduces to evaluating the bubble
diagram depicted in Figure~\ref{Fig:matrix_squared} with vertices
$\gamma$ depending upon the incident and scattered photon
frequencies.\footnote{We remark that technically these
expression must be modified if a resonant condition is satisfied.
In that case one needs to expand to higher terms in the vector
potential to capture resonance effects as perturbation theory
breaks down. Yet for low energy Raman scattering, in most cases this is not crucially important
since in real materials the intermediate states reached via a
direct resonance are quite broadened by interactions and the
resonant terms are not orders of magnitude larger than the
non-resonant terms. Thus this treatment may be of more general
utility.}

The vertex $\gamma$ depends on polarization, but does not depend
sensitively on ${\bf q}$ for $q<<k_{\rm F}$. This can be made more
obvious if we consider the sum over intermediate states ${\bf
k}_{\nu}$ in Eq.~(\ref{Eq:ni_gamma}). The sum over intermediate
states includes both the band index of the states created from the
initial state (i.e. the conduction band) as well as intermediate
states separated from the conduction band. The matrix elements of
the former are proportional to the momentum transferred by the
photon, which in the limit $q<<k_{\rm F}$ are smaller by a factor
of $(v_{\rm F}/c)^{2}$ than the other terms, with $v_{\rm F}~(c)$ the Fermi (light) velocity, and can be neglected
\cite{Pines:1966}. For the remaining sum over the intermediate
states separated from the conduction band, we assume that
$\omega_{i,s} << \mid \!E_{{\bf k}_{\nu}}-E_{\bf k}\!\mid$ and
recover the widely used effective mass approximation\footnote{This
is derived in Appendix E of \textcite{Ashcroft:1976}.}
\begin{equation}
\gamma_{\alpha,\beta}({\bf k},q\rightarrow 0)=\frac{1}{\hbar^{2}}
\frac{\partial^{2} E_{\bf k}}{\partial k_{\alpha}k_{\beta}}.
\label{Eq:effective_mass}
\end{equation}
The symmetry classifications listed in Table~\ref{Table_1} thus
can connect excitations created by certain polarization
orientations to properties of the band structure. Yet it must be
kept in mind that this connection can only be made in the limit of
small $\omega_{i,s}$.

\subsubsection{Im$(1/\epsilon)$ and Sum Rules}

We consider first the case when the incident polarization is
parallel to the scattered polarization and the band $E_{\bf k}$ is
isotropic and parabolic, as for the electron gas or lightly-filled
isotropic band metal. The scattering amplitude $\gamma$ is then
independent of $\bf k$ and the effective density $\tilde\rho({\bf
q})$ is simply proportional to the pure charge density $\rho$.
Using the definition of the complex dielectric function
\cite{Pines:1966,Mahan:2000}
\begin{equation}
\epsilon({\bf q},\Omega)=1+v_{q}\chi_{sc}({\bf q},\Omega)
\label{Eq:epsilon}
\end{equation}
is obtained with $v_{q}$ the bare Coulomb interaction, and
$\chi_{sc}$ is the screened or irreducible polarizability. It is
determined from the full polarizability $\chi$ via
\begin{equation}
\chi_{sc}({\bf q},\Omega)=\frac{\chi({\bf q},\Omega)}{1-v_{\rm
q}\chi({\bf q},\Omega)}, \label{Eq:chi}
\end{equation}
but is most easily identified diagrammatically as all
contributions to the polarizability which are irreducible with
respect to the interaction. The dynamical density-density
correlation function (or structure factor) follows as
\begin{equation}
S({\bf q},\Omega)=\frac{1}{\pi v_{q}}[1+n(\Omega)]~
Im\left[\frac{1}{\epsilon({\bf q},\Omega)}\right].
\label{Eq:density}
\end{equation}
The Raman response $\tilde S$ is proportional to $S$ with a
constant of proportionality determined by the ratio of effective
to free electron mass.

Inelastic light scattering occurs via the creation of charge
fluctuations inside the unit cell which are coupled via the
Coulomb interaction to charge fluctuations in other unit cells.
These intercell excitations are therefore well screened by the
Coulomb interaction and reduce the scattering cross section at
small $q$. In particular, for small $q$, $\chi(q,\Omega)\sim
N_{\rm F}v^{2}_{\rm F}q^{2}/\Omega^{2}$, and the response is
governed by the plasma frequency. The Raman response thus obeys
the longitudinal sum rule resulting from particle-number
conservation \cite{Pines:1966},
\begin{equation}
\int_{0}^{\infty}d\Omega~\Omega ~ Im\left[\frac{1}{\epsilon({\bf
q},\Omega)}\right]=\frac{\pi}{2}\Omega_{\rm
pl}^{2}=\frac{2\pi^{2}Ne^{2}}{m}, \label{Eq:f_sum_rule}
\end{equation}
where $\Omega_{pl}$ is the plasma frequency and $N$ is the number
of electrons of mass $m$ and charge $-e$ in the system. Thus the
only contribution for $q\rightarrow 0$ comes from exciting the
plasmon being the only charge excitation available for light
scattering at small $q$ in a free-electron gas.

\subsubsection{Intra- vs. Inter-cell Charge Fluctuations}

In the more general case of light scattering in solids, the Raman
response may have other contributions coming from intra-cell
charge fluctuations, provided the band structure is non-parabolic,
as pointed out by \textcite{Platzman:1965} and
\textcite{Wolff:1968}. Then light can create anisotropic charge
fluctuations which are zero on average inside the unit cell and
thus are not screened via the long-range Coulomb interaction, as
pointed out by \textcite{Abrikosov:1973}. The general expression
for the screened Raman response function
$\chi_{\gamma,\gamma}^{sc}$ can be written as \cite{Devereaux:1995a,Monien:1990}
\begin{equation}
\chi_{\gamma,\gamma}^{sc}=\chi_{\gamma,\gamma}-\frac{\chi_{\gamma,1}\chi_{\rm
1,\gamma}}{\chi_{1,1}}+\frac{\chi_{\gamma,1}\chi_{\rm
1,\gamma}}{\chi_{1,1}^{2}}\chi_{sc}, \label{Eq:gen_screened}
\end{equation}
where $\chi_{sc}=\chi_{1,1}(1-v_{q}\chi_{1,1})^{-1}$. This is an
exact expression, where the subscript $\gamma$ denotes the
effective Raman density and 1 denotes the pure charge density,
obtained when the usually momentum dependent vertex $\gamma$ is
replaced by a constant. The respective $\chi$-s describe the full
density-density, density-Raman density, and Raman density-Raman
density susceptibilities which are again each irreducible with
respect to the interaction.

The first term in Eq.~(\ref{Eq:gen_screened}) is the bare response
for a neutral system, and the other terms represent the backflow
needed to enforce particle number conservation of charge density
fluctuations and gauge invariance. These terms are important for
light scattering configurations which transform according to the
symmetry of the lattice, such as $A_{1g}$ in $D_{\rm 4h}$ crystals.
In particular, if we consider scattering from pure charge density
fluctuations where $\gamma$ is a constant independent of momentum,
which is an $A_{1g}$ representation, the first two terms in
Eq.~(\ref{Eq:gen_screened}) cancel and $\chi_{sc}$ and
Eq.~(\ref{Eq:density}) are recovered. This is inescapable for
$q=0$, since then the scattering operator is given in terms of the
total density of electrons, which commutes with the bare
Hamiltonian and therefore cannot give inelastic scattering
channels to light. On the other hand, if we consider the
scattering vertex $\gamma$ to depend on wavevector, the backflow
terms are not capable of completely canceling the bare Raman
response.

Momentum dependence of the vertex $\gamma$ is quite general for
electrons in solids. In particular, for crossed light polarizations
projecting out representations of lower symmetry than that of the
lattice $\chi_{\gamma,1}$ is identically zero by symmetry for
$q=0$ and the backflow terms make no correction to the Raman
cross-section. This occurs for $B_{1g}, B_{2g}$, and $E_{g}$
scattering geometries in $D_{\rm 4h}$ systems such as the cuprates,
for example. While there is no conservation law for light
scattering from the excitations created by crossed polarizations
\cite{Kosztin:1991}, there are sum rules which relate the Raman
intensity to model-dependent potential energies projected in
different regions of the BZ \cite{Freericks:2005}.

\subsection{Formalism: Strong Correlations}
\label{Section:Formalism_correlations}

\subsubsection{General Approach to Treating Correlations}

Section~\ref{SectionC} outlined a general approach to inelastic
light scattering when the intermediate states differ from the
initial and final states only by individual single electron
energies. This holds in the limit of weakly interacting electrons.
In this section we show that the formalism is also valid in the
Heisenberg limit of the Hubbard model including the manifold of
zero and 1 doubly occupied sites. These are two limits in which
either the kinetic energy or the potential energy of the electrons
is dominant. However, the more general case of interest to most
systems is tackling the problem when both kinetic and potential
energies are roughly equal. The interactions are sufficient to
give broad spectral functions as measured by ARPES, where the
incoherent part of the spectral function is manifest from the
strong many-body interactions mixing individual electron states.

In this situation, one must resort back to
Eq.~(\ref{Eq:gen_matrix}), and correlation functions involving
two, three and four particles are needed, as depicted in
Figure~\ref{Fig:matrix_squared}. Yet usually one is not interested
in treating many-body correlations over various bands and puts
focus on a few bands close to the Fermi level having strong
correlations. For example in the cuprates, one usually takes
downfolded Hamiltonians involving only Cu $3d_{x^{2}-y^{2}}$ and O
$2p_{x,y}$ orbitals, with short-range Coulomb interactions for two
electrons in on-site or neighboring orbital states. The
downfolding procedure, such as that described by
\textcite{Lowdin:1951}, removes all other bands (effectively
moving them infinitely far away in energy from the focus bands),
such as the apical oxygen, Cu $4s$ and other Cu $d$ orbitals, and
treats the electrons in the bands of interest as having
renormalized energy dispersion. In the cuprates the downfolding
results in either a few bands from local density approximation
(LDA) approaches in one case or cell perturbation theory in the
strongly interacting case. Thus we consider a downfolded
tight-binding Hamiltonian
\begin{equation}
H_{\rm 0}=\sum_{\langle i,j\rangle,\sigma} t_{i,j}
c^{\dagger}_{i,\sigma}c_{j,\sigma} + H_{\rm int},
\label{Eq:tight-binding}
\end{equation}
where $t_{i,j}$ are the effective hopping integrals resulting from
the downfolding procedure, and $H_{\rm int}$ describes the
relevant interactions. For example, we may consider a square
lattice of electrons with strong on-site repulsion $U$ much
greater than the electron hopping $t$ in the Hubbard model,
\begin{equation}
H=-t\sum_{\langle i,j\rangle, \sigma} c_{i,\sigma}^{\dagger}
c_{j,\sigma}+U\sum_{i} n_{i,\uparrow}n_{i,\downarrow},
\label{Eq:Hubbard}
\end{equation}
where $\langle i,j\rangle$ denotes a sum over nearest neighbors.

The electronic eigenstates fall into two bands for large $U$ at 1/2 filling -
e.g., the occupied lower and unoccupied upper Hubbard bands -
separated by an energy $U$ for double occupancies. Away from 1/2 filling, quasiparticles
develop. The microscopic
Hamiltonians can be viewed as families of models related to the
Hubbard model, such as the Falicov-Kimball model, Anderson model,
or Anderson-Fano model, without loss of generality.

In essence, the sums over intermediate states in
Eq.~(\ref{Eq:gen_matrix}) are separated into groups of bands lying
far away in energy from the initial and final states (i.e. the
bands projected out) and bands lying nearby the initial and final
states (considered bands with correlations). The former grouping
of intermediate states are considered as in Section~\ref{SectionC}
to be approximated by the effective mass contribution,
Eq.~(\ref{Eq:effective_mass}). The remaining terms involve matrix
elements of the current operator between the remaining band of
interest\footnote{We remark that this is not exact as it inaccurately treats
resonant scattering processes occurring within the conduction band.
Since these processes are usually taken to be damped, even though this approach
is strictly speaking not exact, it provides a
good starting point for considering Raman scattering in correlated
single-band systems, and as such, has been widely used
\cite{Shastry:1991}.}.

\begin{figure}[b!]
\centerline{\epsfig{figure=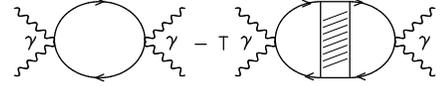,width=6.0cm,clip=}}
\vspace{0.1cm} \caption{Feynman diagrams for non-resonant Raman
scattering. The wavy and the solid lines denote photon and
electron propagators, respectively.  The cross-hatched rectangle
is the \textit{reducible} charge vertex. The symbol $\gamma$
denotes the stress-tensor vertex of the corresponding
electron-photon interaction. From \textcite{Shvaika:2005}.}
\label{Fig:non_res}\end{figure}

\begin{figure}[b!]
\centerline{\epsfig{figure=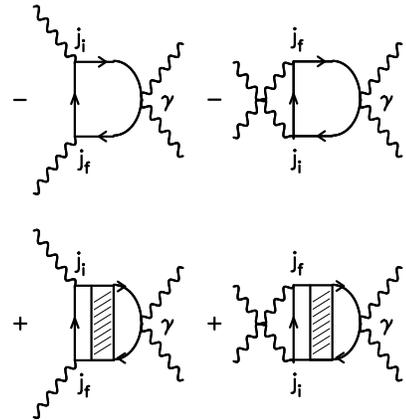,width=6.0cm,clip=}}
\vspace{0.1cm} \caption{Feynman diagrams for the mixed
contributions to Raman scattering. The symbols $j_f$ and $j_i$
remind us to include the relevant vertex factors from the current
operator in the electron-photon interaction. From
\textcite{Shvaika:2005}.} \label{Fig:mixed}\end{figure}

\begin{figure}[b!]
\centerline{\epsfig{figure=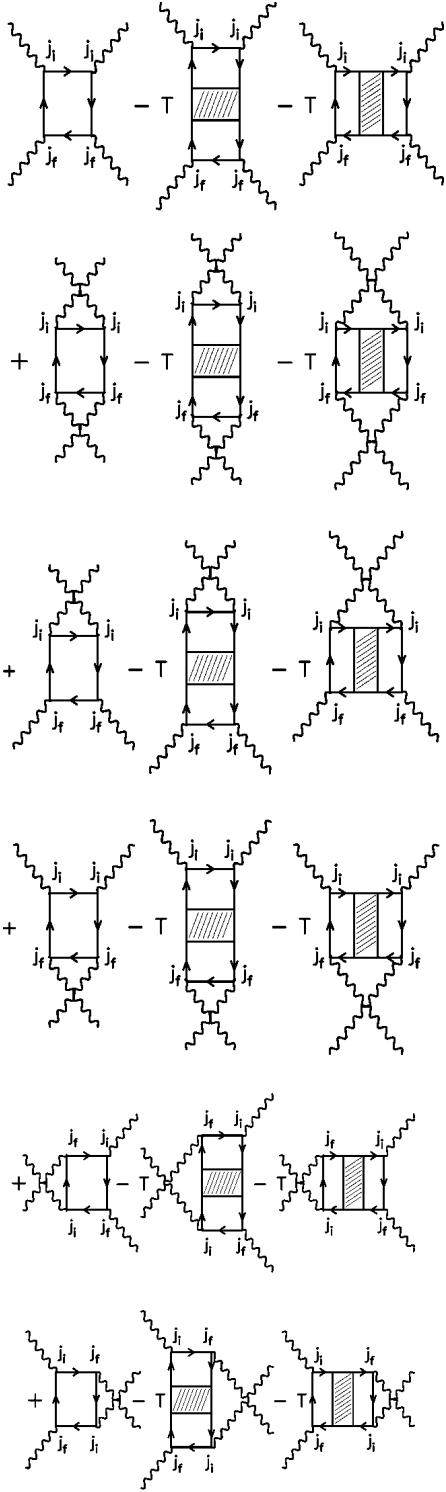,height=20.5cm,clip=}}
\vspace{0.1cm} \caption{Feynman diagrams for the resonant
contributions to Raman scattering. From \textcite{Shvaika:2005}.}
\label{Fig:resonant}\end{figure}

The interaction of light with these downfolded electrons can be
treated via the Peierls construction, in which the creation and
annihilation operators develop a phase,
\begin{equation}
c_{i,\sigma}\rightarrow c_{i,\sigma}e^{-i\frac{e}{\hbar
c}\int_{-\infty}^{{\bf r}_{i}} {\bf A}\cdot d{\bf \ell}}.
\end{equation}
The resulting scattering Hamiltonian obtained by expanding in
powers of ${\bf A}$ reads
\begin{equation}
\label{Hint}
    H_{\text{int}} = \frac{e}{\hbar c} \hat{\bf j} \cdot {\bf A}
    + \frac{e^2}{2\hbar^2 c^2}  \sum_{\alpha\beta}A_\alpha
    \hat\gamma_{\alpha\beta} A_\beta,
\end{equation}
where
\begin{equation}\label{Hint1}
    \hat j_\alpha({\bf q}) = \sum_{\bf k} \frac{\partial\varepsilon({\bf k})}{\partial k_\alpha}
    c_\sigma^\dagger({{\bf k} + {\bf q}}/2) c_\sigma({{\bf k} - {\bf q}}/2 ),
\end{equation}
is a component of the current operator $\hat{\bf j}$ and
\begin{equation}\label{Hint2}
    \hat\gamma_{\alpha\beta}({\bf q}) = \sum_{\bf k}
    \frac{\partial^2\varepsilon({\bf k})}{\partial k_\alpha\partial k_\beta}
    c_\sigma^\dagger({{\bf k} +{\bf q}}/2) c_\sigma({{\bf k} -{\bf q}}/2)
\end{equation}
is the stress tensor operator. Both operators are thus formed from
the energy dispersion of the downfolded band structure. The matrix
element can be written in compact form:
\begin{eqnarray}
\label{M_oper}
&&    M_{F,I}({\bf q})=\sum_{\alpha,\beta}e_{\alpha}^{i}e_{\beta}^{s} M^{\alpha,\beta}({\bf q}),\nonumber\\
&&     M^{\alpha\beta}({\bf q})
    =
    \left\langle F \left| \hat\gamma_{\alpha,\beta}({\bf q}) \right| I \right\rangle
    \\
    \nonumber
 &&   + \sum_{\nu} \left(
    \frac{\left\langle F \left| \hat j_{\beta}({\bf q}_{s}) \right| \nu \right\rangle
    \left\langle \nu \left| \hat j_{\alpha}({\bf q_i}) \right| I \right\rangle}
    {E_{\nu} - E_{I} - \hbar\omega_{i}}
    \right.
    \\
    \nonumber
    &&+
    \left.
    \frac{\left\langle F \left| \hat j_{\alpha}({\bf q}_{i}) \right| \nu \right\rangle
    \left\langle \nu \left| \hat j_{\beta}({\bf q}_{s}) \right| I
    \right\rangle}
    {E_{\nu} - E_{I} + \hbar\omega_s}
    \right),
\end{eqnarray}
with the sum over intermediate states $\nu$ of the Hamiltonian
Eq.~(\ref{Eq:tight-binding}). The Raman cross section can be
separated into non-resonant, mixed, and resonant contributions:
\begin{equation}\label{Raman_gen}
    R(\Omega) = R_{N}(\Omega) + R_{M}(\Omega) +
                      R_{R}(\Omega),
\end{equation}
where the nonresonant contribution is
\begin{equation}\label{Raman_N}
    R_{N}(\Omega)=
    \sum_{I,F} \frac{\exp(-\beta E_{I})}{\mathcal{Z}} \;
    \tilde \gamma^{i,s}_{I,F} \;
    \tilde \gamma^{s,i}_{F,I} \;
    \delta(E_{F} - E_{I} - \hbar\Omega),
\end{equation}
the mixed contribution is
\begin{eqnarray}\label{Raman_M}
    R_{M}(\Omega) &=& \sum_{I,F,\nu}
\frac{\exp(-\beta E_{I})}{\mathcal{Z}}\nonumber
    \\
                &\times&
    \Biggr[
    \tilde \gamma^{i,s}_{I,F}
    \Biggr(
    \frac{
    \tilde j^{(s)}_{{F},\nu}
    \tilde j^{(i)}_{\nu,I}
    }
    {E_{\nu} - E_{I} - \hbar\omega_{i}}
    +
    \frac{
    \tilde j^{(i)}_{F,\nu}
    \tilde j^{(s)}_{\nu,I}
    }
    {E_{\nu} - E_{I} + \hbar\omega_{s}}
    \Biggr)\nonumber
    \\
    & +&
    \Biggr(
    \frac{
    \tilde j^{(i)}_{I,\nu}
    \tilde j^{(s)}_{\nu,F}
    }
    {E_{\nu} - E_{I} - \hbar\omega_{i}}
    +
    \frac{
    \tilde j^{(s)}_{{I},\nu}
    \tilde j^{(i)}_{\nu,{F}}
    }
    {E_{\nu} - E_{I} + \hbar\omega_{s}}
    \Biggr)
    \tilde \gamma^{s,i}_{F,I}
    \Biggr]\nonumber
    \\
    &\times &
    \delta(E_{F} - E_{\nu} - \hbar\Omega),
\end{eqnarray}
and the resonant contribution is
\begin{eqnarray}\label{Raman_R}
    R_{R}(\Omega) &=& \sum_{I,F,\nu\nu^{\prime}}
    \frac{\exp(-\beta E_{I})}{\mathcal{Z}}\nonumber
    \\
    &\times&\Biggr(
    \frac{
    \tilde j^{(i)}_{I,\nu}
    \tilde j^{(s)}_{\nu,F}
    }
    {E_{\nu} - E_I - \hbar\omega_i} +
    \frac{
    \tilde j^{(s)}_{I,\nu}
    \tilde j^{(i)}_{\nu,F}
    }
    {E_{\nu} - E_I + \hbar\omega_s}
    \Biggr)\nonumber
    \\
    &\times&
    \Biggr(
    \frac{
    \tilde j^{(s)}_{F,\nu^{\prime}}
    \tilde j^{(i)}_{\nu^{\prime},I}
    }
    {E_{\nu^{\prime}} - E_I - \hbar\omega_i} +
    \frac{
    \tilde j^{(i)}_{F,\nu^{\prime}}
    \tilde j^{(s)}_{\nu^{\prime},I}
    }
    {E_{\nu^{\prime}} - E_I + \hbar\omega_s}
    \Biggr)\nonumber
    \\
    &\times&
    \delta(E_{F} - E_I - \hbar\Omega).
\end{eqnarray}
We have introduced the symbols
\begin{equation}\label{Raman_not}
    \tilde \gamma^{i,s} = \sum_{\alpha\beta} e_\alpha^i \hat \gamma_{\alpha\beta}({\bf q})
    e_\beta^s, \quad\quad
    \tilde j^{(i,s)} = \sum_\alpha e_\alpha^{i,s} \hat j_{\alpha}({\bf q}_{i,s}),
\end{equation}
and denote $\hat O_{\kappa,\lambda}$ as the matrix element
$\langle \kappa | \hat O |\lambda\rangle$. Some of the
contributions are depicted diagrammatically in
Figures~\ref{Fig:non_res}, \ref{Fig:mixed}, and \ref{Fig:resonant}
for the non-resonant, the mixed, and the resonant terms,
respectively. In the limit $D\rightarrow\infty$ these are the main
diagrams to consider. Additional diagrams  involving
multi-particle vertex renormalizations generally contribute for
finite dimensions (not shown).

\begin{figure}[b!]
\centerline{\epsfig{figure=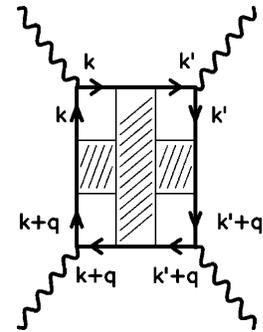,width=4.5cm,clip=}}
\vspace{0.1cm} \caption{Feynman diagrams for a typical
parquet-like renormalization. This resonant diagram has a
simultaneous horizontal and vertical renormalization by the
two-particle reducible charge vertex. From
\textcite{Shvaika:2005}.} \label{Fig:4_part}\end{figure}

In general, the matrix elements that enter into
Eqs.~(\ref{Raman_N}--\ref{Raman_R}) are not easy to calculate for
an interacting system, so the summations are problematic to
evaluate. In particular, one needs to evaluate the irreducible
stress and current vertices, depicted in Figures
\ref{Fig:non_res}-\ref{Fig:resonant} by the hatched symbols.
Contributions to these vertex dressings include many-particle
renormalizations. A particularly complicated one is shown in
Figure \ref{Fig:4_part} which represents 4-particle vertex
corrections. Moreover, analytic continuation must be performed to
obtain the Raman response $R(\Omega)$ on the real axis from the
imaginary axis. While this is relatively straightforward for the
non-resonant case, mixed and resonant cases are problematic
because of the complicated dependences on each of the frequencies
$\omega_{i,s},\Omega$ which have to be analytically continued.
While the continuation has been worked out recently for the
general case, evaluating these diagrams for general interactions
has proved elusive.

The overall complexity of the problem limits the evaluation of the
light scattering cross section to generic interacting systems.
Only recently, these diagrams have been evaluated exactly in a
DMFT treatment of the Falicov-Kimball model
\cite{Shvaika:2004,Shvaika:2005}.

\subsubsection{Correlated Insulators - Heisenberg Limit}
\label{SectionD}

To emphasize the generality of Eq.~(\ref{Eq:gen_matrix}) we now
consider the large $U$ limit for the insulating half-filled
two-dimensional Hubbard model. Following \textcite{Shastry:1991}
we consider a system with $N$ interacting electrons in which the
manifold of states can be classified by the number $n$ of doubly
occupied sites. The well-known Heisenberg Hamiltonian emerges from
projecting the Hubbard model down onto the reduced Hilbert space
containing no double occupancies:
\begin{equation}
   H_{\rm Heisenberg}=J\sum_{i,\delta} {\bf S}_{i}\cdot {\bf
   S}_{i+\delta}, \label{Eq:Heisenberg}
\end{equation}
with $J=4t^{2}/U$ the Heisenberg exchange constant. Higher
manifolds containing empty holes and doubly occupied states can be
labeled according to the net spin configuration $\{\sigma\}$ of
the $N-2n$ singly occupied sites, as well as the locations $\{{\bf
R}\}$ of the empty ${\bf r}_{\rm hole}$ and doubly occupied ${\bf
r}_{\rm double}$ sites. We denote these states as $\mid \!n;
\{\sigma\},\{{\bf R}\}\rangle$. These states are connected to each
other via
\begin{equation}
   \mid \!1; \{\sigma^{\prime}\};\{{\bf R}\}\rangle =
   c_{\sigma}^{\dagger}({\bf r}_{\rm double})c_{\sigma}({\bf r}_{\rm
   hole})\!\mid \!0; \{\sigma\}\rangle, \label{Eq:manifold}
\end{equation}
where $\mid \!\!0;\{\sigma\}\rangle =\Pi_{\bf
r}c_{\sigma_{r}}^{\dagger}({\bf r})\!\!\mid \!\!{\rm vac}\rangle$
and $\mid \!\!{\rm vac}\rangle$ denotes the vacuum.

Light scattering thus occurs via transitions out of the manifold
of singly-occupied states. To leading order for large $U$, only the
$n=0$ and $n=1$ manifold of states contributes to light scattering
via Eq.~(\ref{Eq:gen_matrix}), with $n=0$ denoting the ground
state and $n=1$ the manifold of intermediate states having one
doubly occupied and one empty site. The first term containing
$m_{\alpha,\beta}$ cannot contribute for the half-filled lattice,
and thus only interband scattering between the upper and lower
Hubbard bands occurs via the ${\bf p}\cdot {\bf A}$ term. The
energy difference between these excitations is $U$ to lowest order
in $t/U$, allowing us to write the matrix element
Eq.~(\ref{Eq:gen_matrix}) in the form
\begin{widetext}
\begin{equation}
   M_{F,I}=\sum_{\nu,\bm{r},\bm{r^{\prime}},\bm{\delta},\,\bm{\delta^{\prime}}}
   \langle 0,\{\sigma_{I}\}\mid \hat j_{s}({\bf r})\,
   {\bf \hat e}_{s}\cdot \bm{\delta}\mid 1;\{\sigma_{\nu}\}; {\bf R}_{\nu}\rangle
   \langle 1;\{\sigma_{\nu}\}; {\bf R}_{\nu}\mid
   {\bf \hat e}_{i}\cdot \bm{\delta^{\prime}} \hat j_{i}({\bf r}^{\prime})\mid 0,\{\sigma_{F}\}\rangle
   \left[\frac{1}{U-\hbar\omega_{i}}+\frac{1}{U+\hbar\omega_{s}}\right],
\label{Eq:SS_matrix}
\end{equation}
with the current operator defined as
\begin{equation}
    \hat j_{i,s}({\bf r})=i~\!t[c_{\sigma}^{\dagger}({\bf r}
    +\bm{\delta} a \cdot {\bf \hat e}_{i,s})c_{\sigma}({\bf r})
    -c_{\sigma}^{\dagger}({\bf r})c_{\sigma}({\bf r}
    +\bm{\delta} a\cdot {\bf \hat e}_{i,s})].\nonumber
\end{equation}
Here $\bm{\delta}$ is a unit vector connecting a site with its nearest
neighbors. The intermediate states $\nu$ represent a sum over spin
configurations and locations of both the doubly occupied and the
hole sites. Substituting Eq.~(\ref{Eq:manifold}) into Eq.
(\ref{Eq:SS_matrix}) collapses the intermediate state sum, leaving
four terms connecting initial and final states. Using the identity
$1/2 - 2{\bf S}_{i}\cdot{\bf S}_{j} = c^{\dagger}({\bf r}+\bm{\delta}
a)c({\bf r}) c^{\dagger}({\bf r})c({\bf r}+\bm{\delta} a)$ valid in
the manifold of singly occupied states, one obtains the light
scattering Hamiltonian of \textcite{Elliot:1963} and
\textcite{Fleury:1968},
\begin{equation}
   H_{\rm EFL}=\sum_{{\bf r},\bm{\delta}}
   {\bf S_{\bf r}}\cdot{\bf S_{{\bf r}+\bm{\delta}a}}
   ({\bf \hat e_{s}}\cdot\bm{\delta})({\bf \hat e_{i}}\cdot\bm{\delta})
   \left[\frac{1}{U-\hbar\omega_{i}}+\frac{1}{U+\hbar\omega_{s}}\right].
\label{Eq:ELF}
\end{equation}
\end{widetext}
We note that the polarization dependence is crucial as well. For $xx+yy$ polarizations projecting the fully symmetric components, the light scattering Hamiltonian
commutes with the nearest neighbor Heisenberg Hamiltonian and thus
does not give inelastic scattering in the $A_{1g}$ channel.
Moreover, $B_{2g}$ ($xy$) is also identically zero. As a result, a large signal appears only in the $B_{1g}$ channel ($xx-yy$). These restrictions
are lifted however if longer range spin interactions are considered
\cite{Shastry:1991}.

The collapse of the intermediate states allowed us to replace the
operators with projected spin operators confined to the restricted
Hilbert space of the $n=0,1$ manifolds. Thus, in this limited
Hilbert space, the formalism is similar to non-interacting
electrons in that the operators appearing in the scattering matrix
may be simplified. If the Hilbert space is enlarged to include
larger manifolds, then this would no longer be the case, and thus
including terms to higher order in $t/U$ becomes highly
non-trivial and is still one of the challenges to merge a weakly
interacting picture into a strongly interacting one.

We note that Equation~(\ref{Eq:ELF}) was derived effectively as an
expansion in $t/(U-\hbar\omega_{i})$. Therefore, the scattering
Hamiltonian is limited to cases when both the number of holes and
double occupied sites are restricted and off-resonance
conditions apply, with the incident photon energy
$\hbar\omega_{i}$ far away from $U$. Efforts to extend the
treatment to more general conditions involve understanding the
motion of holes or doubly occupied sites in an arbitrary spin
background. This has proved to be a hard task.

\subsection{Electronic Charge Relaxation}
\label{Section:charge_relaxation}

In Sections~\ref{SectionC}
and~\ref{Section:Formalism_correlations} we reviewed the general
formalism of the theory of Raman scattering for weakly and
strongly correlated systems. In this subsection we now specify the
electronic states from which light can be scattered and review the
various theoretical treatments for specific models of interacting
electrons. Emphasis is placed upon how symmetry can be used to
highlight electron dynamics on regions of the BZ, and general
features for each model system will be presented. We first
consider the case where the correlations among electrons are weak.

\subsubsection{Weakly-Interacting Electrons}
   \label{Section:weak}

In this subsection, we consider electrons as having very well
defined eigenstates labeled by energy and momentum and having
sharp spectral functions. Apart from the form of the energy
dispersion $\xi_{\bf k}$, the results are rather general and
governed largely by phase space considerations. We must, however,
consider the long-range Coulomb interaction in order to account
for charge backflow and screening, and we utilize the results
derived in Section~\ref{SectionC} and the general expression
Eq.~(\ref{Eq:gen_screened}).

\begin{figure}[b!]
\centerline{\epsfig{figure=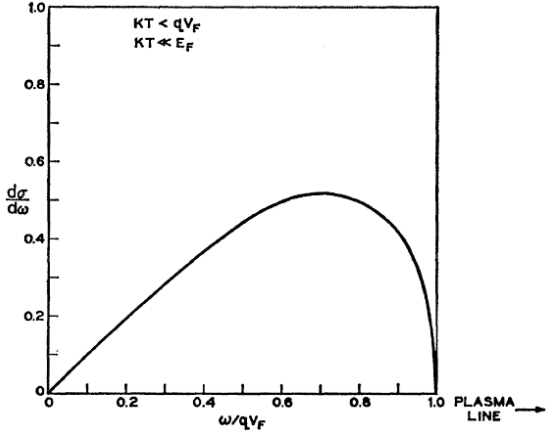,width=7.5cm,clip=}}
\vspace{0.1cm} \caption{Raman response of an electron gas. From
\textcite{Platzman:1965}.} \label{Fig:Wolff}
\end{figure}

The use of $\chi_{sc}$ rather than $\chi$ takes into account the
most drastic manifestation of the long-range Coulomb interaction,
viz. screening. For weakly interacting electrons, the Random Phase
Approximation (RPA) is acceptable, which replaces $\chi_{sc}$ by
the Lindhard function for non-interacting electrons. The response
functions are determined by the Lindhard kernel,
\begin{equation}
   \chi_{a,b}({\bf q},\Omega)=
   \frac{2}{V}\sum_{\bf k} a_{\bf k,q}b_{\bf k,q}
   \frac{f(\xi_{\bf k})-f(\xi_{\bf k+q})}
   {\xi_{\bf k}-\xi_{\bf k+q}+\hbar\Omega-i\delta},
\label{Eq:Lindhard}
\end{equation}
for general vertices $a,b$ appearing in
Eq.~(\ref{Eq:gen_screened}). In a free electron gas, the Raman
response is given by Eq.~(\ref{Eq:density}) or equivalently, the
last term in Eq.~(\ref{Eq:gen_screened}).  For large $q$,
collective excitations are unimportant and light scattering occurs
via creation of particle-hole excitations in the Landau continuum.
However, as the only phase space for creating particle-hole pairs
comes from finite $q$ transferred from the photons, the resulting
response is a continuum varying linearly with $\Omega$ at small
frequencies and extending up to a cut-off $\Omega_c=v_{\rm F}q$
from the borders of the continuum \cite{Mahan:2000}. The
low-energy intensity is proportional to $q^{2}$, and the only
excitation left at $q=0$ is the collective plasmon. The Raman
response for the free electron gas is shown in
Figure~\ref{Fig:Wolff}.

For electrons in a solid, however, the non-parabolicity of the
energy dispersion results in charge fluctuations which are
anisotropic in the small $q$ limit and thus can survive screening
and give more weight at low energy transfers. Formally, an
additional contribution to the response is given by the first two
terms in Eq.~(\ref{Eq:gen_screened}). Yet phase space restrictions
still produce an inescapable cut-off at $\Omega_c=v_{F}q$
\cite{Wolff:1968}, and the response resembles that shown in
Figure~\ref{Fig:Wolff}. This is also the case if scattering occurs
for the 2 dimensional electron-gas (2DEG) in the absence of a
magnetic field \cite{Jain:1987,Mishchenko:1999} or for complex
Fermi surfaces \cite{Ipatova:1983}. An RPA treatment for resonant
scattering has been given by \textcite{Wang:1999,Wang:2002}.

Recently, substantial progress has been made in understanding the
Raman response in the integer or fractional quantum Hall regimes
of the 2DEG. Space limitations do not allow us to review these
systems; so for brevity, we cite only a recent reference
\cite{Richards:2000}.

\subsubsection{Impurities}
   \label{sec:impurities}

Excitations at low energies in non-interacting electronic systems
can arise for small $q$ via electronic scattering from impurities,
where momentum contributed by impurity scattering can provide
phase space for electron-hole creation which is anisotropic in the
Brillouin zone. For example, if one considers a general
electron-impurity interaction of the form
\begin{equation}
   H_{\rm imp}=\sum_{{\bf k},{\bf k}^{\prime},\sigma}
   V_{{\bf k},{\bf k}^{\prime}}c^{\dagger}_{{\bf k},\sigma}c_{{\bf k}^{\prime},\sigma},
\end{equation}
with an anisotropic interaction $V_{{\bf k},{\bf k}^{\prime}}$,
the gauge invariant Raman response is given via the diagrams
presented in Figure~\ref{Fig:non_res}. If we make a symmetry
decomposition of the scattering amplitude $\gamma({\bf
k})=\sum_{L}\gamma_{L}\Phi_{L}({\bf k})$ in terms of basis
functions $\Phi_{L}$ of the Brillouin zone, the resulting response
in channel $L$ corresponding to a particular light polarization
orientation has a Drude Lorentzian form
\cite{Zawadowski:1990,Devereaux:1992,Falkovskii:1989}:
\begin{equation}
   \chi^{\prime\prime}({\bf q},\Omega)= N_{F}\gamma_{L}^{2}
   \frac{\Omega\tau_{L}^{*}}{1+(\Omega\tau_{L}^{*})^{2}},
\label{Eq:Drude-imp}
\end{equation}
with $N_{\rm F}$ the density of states at the Fermi level.
$1/\tau_{L}^{*}=1/\tau+Dq^{2}-1/\tau_{L}$ is the effective
scattering rate where
\begin{equation}
   1/\tau=1/\tau_{L=0}=n_{i} N_{F}\!\int \frac{dS_{{\bf k}}}{S}\int
   \frac{dS_{{\bf k^{\prime}}}}{S} \mid V_{\bf k,k^{\prime}}\!\mid^{2},
\end{equation}
involving an integration of the Fermi surface $S_{{\bf k}}$
normalized to the Fermi area $S$. Here $n_{i}$ is the impurity
concentration, $D=\frac{1}{3} v_{\rm F}^{2}\tau$ is the diffusion
constant. The anisotropy of the impurity scattering is
characterized via orthonormal basis functions $\Phi_L$
\begin{equation}
   V_{\bf k,k^{\prime}}=\sum_{L,L^{\prime}}
   \Phi_{L}^{\ast}({\bf k})\Phi_{L}({\bf k}^{\prime})V_{L,L^{\prime}},
\end{equation}
using an intelligent basis where the interaction is diagonal
$V_{L,L^{\prime}}=\delta_{L,L^{\prime}}V_{L}$. Then,
$1/\tau_{L}=2\pi n_{i} N_{F} V_{L}$ and $1/\tau_{L=0}$ is the
dominant contribution.

\begin{figure}[b!]
\centerline{\epsfig{figure=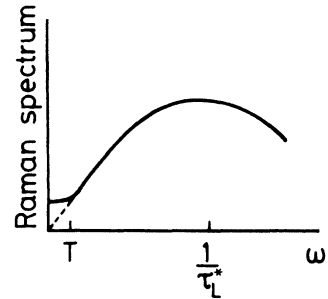,width=5.0cm,clip=}}
\vspace{0.1cm} \caption{Raman response from impurity scattering in
an otherwise non-interacting system. From
\textcite{Zawadowski:1990}.} \label{Fig:impurity}\end{figure}

The resulting Raman spectrum (Figure~\ref{Fig:impurity}) grows
linearly with frequency $\Omega$, decays as $1/\Omega$, and has a
peak when $\Omega\tau_{L}^{\ast}$ equals 1. The width of the
Lorentzian reflects the rate for which charge density excitations
having symmetry $L$ decay into all other channels. The light
polarizations select the type of excitation $L$ created, and thus
allow a way to probe the anisotropy of the impurity-electron
interaction. The decay of the charge density fluctuations can
occur via finite $q$ through the diffusion term and all
contributions other than $V_{L}$ which relax electrons out of the
state $L$. Put another way, $Dq^{2}+1/\tau$ are ``scattering out''
processes, while $1/\tau_{L}$ is a ``scattering in'' process,
giving an effective scattering rate $1/\tau_{L}^{\ast}$. This is a
consequence of a gauge invariant treatment including charge
backflow (Coulomb interaction) as well as density preserving
scattering (impurity vertex corrections).

In the limit of weak scattering, the response collapses into a
delta-function, reflecting momentum conservation. One can note the
obvious connection of the response plotted in
Figure~\ref{Fig:impurity} to the Drude conductivity, although even
for simple impurity scattering the two response functions are not
related by a power of frequency as soon as the impurity potential
has any momentum anisotropy. Otherwise, for purely isotropic
impurity scattering the conductivity and the Raman response are
related by a power of frequency - the so-called Shraiman-Shastry
relation given in Eq.~(\ref{Eq:SS})
\cite{Shastry:1990,Freericks:2001a}.

\subsubsection{Interacting Electrons - Non-Resonant Response}

However, the most important application of light scattering is for
systems where the electronic correlations are strong and cannot be
treated in standard RPA. Thus, while long-range Coulomb screening
is still important in order to maintain gauge invariance, the
interactions introduce generally complex dynamics in specific
regions of the BZ. In this case, the electron self energy
$\Sigma$, as well as the vertex corrections to the light
scattering amplitude $\gamma$, depend normally on both momentum
and energy, making the light scattering evaluation more difficult.
On the other hand, anisotropies of the electron dynamics can be
explored.

Here we start by considering non-resonant scattering, since this
is an area in which by far most theoretical treatments lie, as it
is simpler to evaluate than the mixed or resonant terms. We note
that many calculations of $Im(1/\epsilon)$ have been performed
from ab-initio approaches in the context of inelastic X-ray
scattering (see \textcite{Gurtubay:2004} and references therein
for recent work). There, the focus is largely on the ${\bf
q}-$dependence of the response, and electron-electron interactions
have been treated in various ways \cite{Ku:2002}. Yet, to our
knowledge, no calculation exists for Raman scattering in a simple
Fermi liquid in which inelastic scattering processes via the
Coulomb interaction are incorporated exactly, although recently
dynamical mean field theory (DMFT) in correlated metals has been
used \cite{Freericks:2001a}. This is because the irreducible
charge vertex is not generally known in models with strong
correlations, with the exception of the Falicov-Kimball model.
Thus we focus more on the polarization dependence and investigate
contributions to Raman scattering from non-conserved charge
fluctuations.

The general expression for the two-particle correlation function
describing the non-resonant Raman response reads
\begin{widetext}
\begin{equation}
   \chi_{\gamma,\gamma}({\bf q}=0,i\Omega)=
   -\frac{2}{V\beta}\sum_{i\omega}\sum_{\bf k}
   \gamma({\bf k}) G({\bf k},i\omega) G({\bf k},i\omega+i\Omega)\Gamma({\bf
   k};i\omega;i\Omega).
\label{Eq:full}
\end{equation}
Similar expressions are obtained for $\chi_{\gamma,1}$ and
$\chi_{1,1}$ where the vertices $\gamma$ are successively replaced by
1 to be inserted into Eq.~(\ref{Eq:gen_screened}), or may be
generally represented in terms of the reducible Raman vertex as
shown in Figure~\ref{Fig:non_res}. In the ladder approximation, the
renormalized vertex is given by a Bethe-Salpeter equation:
\begin{equation}
   \Gamma({\bf k};i\omega;i\Omega)=\gamma({\bf k})+\frac{1}{V\beta}
   \sum_{i\omega^{\prime}}\sum_{\bf k^{\prime}}
   V({\bf k}-{\bf k^{\prime}},i\omega-i\omega^{\prime})
   G({\bf k^{\prime}},i\omega^{\prime})
   G({\bf k^{\prime}},i\omega^{\prime}+i\Omega)
   \Gamma({\bf k^{\prime}};i\omega^{\prime};i\Omega).
\label{Eq:Bethe}
\end{equation}
\end{widetext}
Here, $V({\bf k},\omega)$ is the generalized electron-electron
interaction, and we have suppressed spin notation. If one neglects
vertex corrections such that the theory is not gauge invariant,
the Raman response has a particularly simple form given by
Eq.~(\ref{Eq:Kubo}). The effect of the long-range Coulomb
interaction is treated formally in the same way as in
Eq.~(\ref{Eq:gen_screened}), with the vertices replaced by the
normalized vertex as a solution to the Bethe-Salpeter
Eq.~(\ref{Eq:Bethe}).

Eqs.~(\ref{Eq:full}) and~(\ref{Eq:Bethe}) have been the starting
point for many studies of light scattering treating
electron-electron interactions in effective models. These include
systems which have nearly nested Fermi surface segments
\cite{Virosztek:1991,Virosztek:1992} or antiferromagnetic spin
fluctuations \cite{Kampf:1992,Devereaux:1999}. Similarly, a slave
boson approach to the $t-J$ model \cite{Bang:1993},
electron-phonon interactions
\cite{Kostur:1991,Kostur:1992,Itai:1992,Rashkeev:1993} and
fluctuation-exchange (FLEX) treatments of the Hubbard model
\cite{Dahm:1999} have been considered. While these studies involve
approximate solutions, more recently the use of DMFT has provided exact results in the limit of strictly local correlations in the Hubbard
\cite{Freericks:2003a,Freericks:2001b} and Falicov-Kimball models
\cite{Freericks:2001a}.

Two aspects of the Raman response are generally in the main focus:
the frequency dependence of the broad continuum extending well
past $qv_{\rm F}$, and the polarization dependence. We discuss
first the spectral response.

\begin{figure}[b!]
\centerline{\epsfig{figure=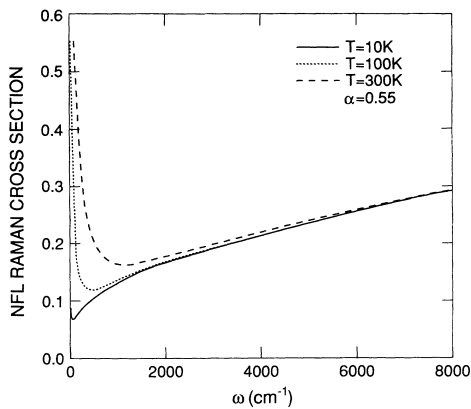,width=6.8cm,clip=}}
\vspace{0.1cm} \caption{Raman response as a function of
temperature obtained by \textcite{Virosztek:1992} for a system
with a nested Fermi surface. The upturn at low frequencies is to
the ($\log(\Omega)$) frequency dependence of the effective mass in
this model.} \label{Fig:viro}
\end{figure}

In the context of the cuprates, Varma and coworkers pointed out
that a flat, nearly frequency independent response could be
obtained if the imaginary part of the electron self-energy
depended linearly on frequency \cite{Varma:1989a,Varma:1989b}. The
response is then given in terms of the scale-invariant ratio of
$\hbar\omega/k_{\rm B}T$ and approaches a constant at large
frequency transfers. This can be understood phenomenologically by
replacing $1/\tau_{L}^{\ast}$ with
$1/\tau_{L}^{\ast}(\Omega,T)\propto \max(k_{\rm B}T,\hbar\Omega)$
in Eq.~(\ref{Eq:Drude-imp}).\footnote{This is only an
approximation since $1/\tau_{L}^{\ast}(\Omega,T)$ depends now on
energy. Causality requires that the relaxation function has real
and imaginary part, $M(\Omega,T)=$
$\Omega\lambda+i/\tau_{L}^{\ast}$ \cite{Goetze:1972,Opel:2000}.} A
scale-invariant response at low frequencies is a general
consequence of systems in proximity to a quantum-critical point,
but this scale invariance is broken outside the quantum critical
regime. ``Marginal Fermi liquid'' behavior emerges for instance
when scattering is considered in a nested Fermi liquid
\cite{Virosztek:1991,Virosztek:1992}, low Fermi-energy systems
\cite{Devereaux:1999,Dahm:1999}, and slave-boson systems
\cite{Bang:1993}. A broad background very similar to ``marginal''
behavior is also found for strongly coupled electron-phonon
systems \cite{Itai:1992,Kostur:1991,Kostur:1992}. As a
representative example, we show the response calculated by
\textcite{Virosztek:1992} for a nested Fermi liquid in
Figure~\ref{Fig:viro}.

Low energy electron dynamics can be extracted by studying the
Raman response in the limit $\Omega \rightarrow 0$. Neglecting
vertex corrections, the low frequency response reads
\cite{Devereaux:1999,Venturini:2002a}
\begin{eqnarray}
   \chi_{\mu}^{\prime\prime}(&\Omega& \rightarrow  0 ) = \Omega N_F
   \times {} \nonumber\\
   & &\times~\left<\gamma_{\mu}^{2}({\bf k}) \int d\xi
   \left(-{\partial f^{0}\over{\partial \xi}}\right) {Z_{\bf k}^{2}
   (\xi,T)}\over{2\Sigma^{\prime \prime}_{\bf k}(\xi,T)}\right>.
\label{chi0}
\end{eqnarray}
Here, $N_{\rm F}$ is the density of electronic levels at the Fermi
energy $E_{\rm F}$, $\Sigma^{\prime \prime}_{\bf k}$ is the
imaginary part of the single-particle self energy related to the
electron lifetime as $\hbar/2\Sigma^{\prime \prime}_{\bf k}
(\omega,T) = \tau_{\bf k}(\omega,T)$, $Z_{\bf
k}(\omega,T)=(1-\partial\Sigma^{\prime }_{\bf
k}(\omega,T)/\partial\omega)^{-1}$ is the quasiparticle residue,
$f^{0}$ is the equilibrium Fermi distribution function, and
$\langle \cdots \rangle$ denotes an average over the Fermi
surface. Thus the inverse of the Raman slope
\begin{equation}
   \Gamma_{\mu}(T)=\left[\frac{\partial\chi^{\prime\prime}_{\mu}
   (\Omega)}{\partial \Omega}\right]^{-1}
\label{Eq:Raman_slope}
\end{equation}
measures the effective ``scattering rate'' of the quasiparticles
in a correlated metal, and can be best thought of as a ``Raman
resistivity''.

\begin{figure}[b!]
\centerline{\epsfig{figure=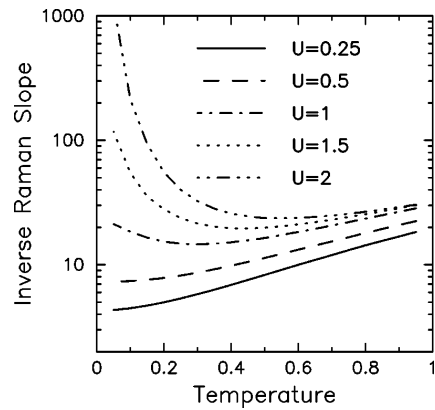,width=6.5cm,clip=}}
\vspace{0.1cm} \caption{Inverse Raman slope (see
Eq.~(\ref{Eq:Raman_slope})) close to a metal-insulator transition
at a value $U=\sqrt{2}$ in the Falicov-Kimball model for
$D=\infty$ \cite{Freericks:2001a}. All energies and temperatures
are measured in terms of the hopping $t$.}
\label{Fig:MIT_slope}\end{figure}

In systems with isotropic interactions, the polarization dependence
drops out and the slope of the low-frequency Raman response is
given in terms the low energy quasiparticle scattering lifetime,
$\Gamma_{\mu}(T) \propto \hbar/\tau(T)$, as an extension of
Eq.~(\ref{Eq:Drude-imp}). Yet, in strongly correlated systems, the
quasiparticle residue $Z$ and, importantly, vertex corrections,
enter as well.  In a correlated or a strongly disordered metal
(near an Anderson transition, e.g.), however, a finite energy might be
necessary to move an electron from one site to another one. Thus,
in spite of a non-vanishing density of states at the Fermi level,
as observed in an ARPES experiment for instance, no current can be
transported and $\Gamma_{\mu}(T) \gg \hbar/\tau(T)$. This is an
important difference between single- and two-particle properties.

Figure~\ref{Fig:MIT_slope} displays the inverse Raman slope
defined in Eq.~(\ref{Eq:Raman_slope}), as determined via a DMFT
treatment of the Falicov-Kimball model in the vicinity of a
metal-insulator transition, as a function of the Coulomb repulsion
$U$ \cite{Freericks:2001a}. It provides an illustrative example of
how Fermi-liquid-like features evolve as the lifetime of putative
quasiparticles increases due to decreased role of correlations.
For small $U$, the correlated metal displays an inverse slope
$\propto T^{2}$ as a canonical Fermi liquid in the metallic state.
A pseudogap opening in the density of states with increasing $U$
drives the inverse slope into insulating behavior, increasing as
the temperature decreases.

\begin{figure}[b!]
\centerline{\epsfig{figure=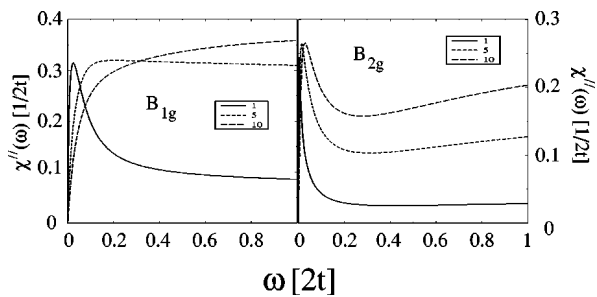,width=1\linewidth,clip=}}
\vspace{0.1cm} \caption{Polarization-dependent Raman response  in
a spin-fermion model for a fixed temperature for three different
values of the coupling constant. From \textcite{Devereaux:1999}.}
\label{Fig:NAFL}\end{figure}

As a second important application Eq.~(\ref{Eq:Raman_slope}) can
illuminate the anisotropy of electron dynamics due to the
momentum-dependent weighting factors of the polarization
orientations and self energies. The geometry of light scattering
orientations, as given by the form factors listed in
Table~\ref{Table_1}, project out the ratio of quasiparticle
residues and scattering rates in different regions of the BZ. As a
consequence, the Raman spectra show polarization dependent
behavior determined largely by the self energies and vertex
corrections near the regions projected by the scattering vertices
$\gamma$. Figure~\ref{Fig:NAFL} plots the $B_{1g}$ and $B_{2g}$
Raman response calculated in a spin-fermion model in which
electron scattering is most pronounced involving antiferromagnetic
reciprocal lattice momentum transfers ${\bf Q}=(\pi,\pi)$, leading
to ``hot'' quasiparticles near the BZ axes (projected by $B_{1g}$
form factors) and ``cold'' quasiparticles along the BZ diagonals
(projected by $B_{2g}$ form factors). Therefore, the Raman
response has a sharp quasiparticle peak for $B_{2g}$ scattering at
low energies due to the long quasiparticle lifetimes, while the
response in $B_{1g}$ is dominated by strong incoherent scattering
leading to a suppression of the quasiparticle peak at low energies
and an essentially structureless continuum.

Lastly, we note that the non-resonant Raman response has also been
calculated for exchange of fluctuation modes at wavevectors {\bf
Q} and -{\bf Q} for systems near a spin-density wave instability
\cite{Brenig:1992,Kampf:1992,Venturini:2000} and a charge-density
wave instability \cite{Caprara:2005}. Here, the Raman response is
sensitive to the light polarizations and has a peak centered at
twice the energy of the fluctuating mode.

\subsubsection{Interacting Electrons - Resonant Response}
\label{Section:resonant_theory}

In addition to the non-resonant response, one has the mixed and
resonant contributions to consider. Typically these diagrams are
neglected in the weak correlation limit, as they can be summed
into two-particle response functions as discussed in Section
\ref{SectionC}. In the insulating case, only the resonant terms
are kept, as the studies focus on excitations across a
charge-transfer or Hubbard gap. This has been calculated in
systems exhibiting 1D-Luttinger behavior
\cite{Sassetti:1998,Sassetti:1999,Kramer:2000,Wang:2004} where
bosonization techniques can be applied. Yet, generally treating
all diagrams on equal footing is technically demanding. Only
recently, an exact evaluation in DMFT has been performed
\cite{Shvaika:2004,Shvaika:2005}.

It is well known that many of the Raman signals in correlated
metals and insulators display complicated dependences on incoming
photon frequency $\omega_{i}$. For example, the $B_{1g}$
two-magnon feature at roughly $350$~meV in the thoroughly studied
insulating parent cuprates has a resonance for incident photon
energies near 3~eV. As a reaction to the experimental results in
the cuprates, much theoretical work has been devoted to Raman
scattering in a two-dimensional Heisenberg antiferromagnet using
the Fleury-Loudon model Eq.~(\ref{Eq:ELF}).

\begin{figure}[b!]
\centerline{\epsfig{figure=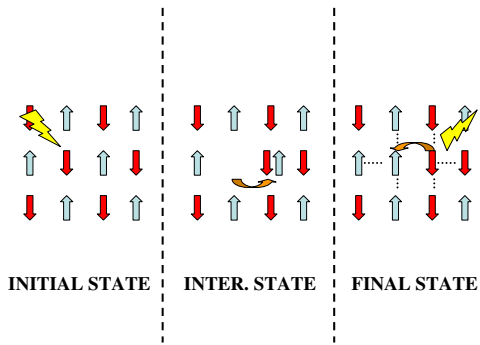,width=6.5cm,clip=}}
\vspace{0.1cm} \caption{Cartoon of the two-magnon scattering
process in a 2D Heisenberg antiferromagnet. An incident photon
causes an electron with spin $\sigma$ to hop leaving a hole and
creating a double occupancy in the intermediate state with energy
$\propto U$. One particle of the double with spin $-\sigma$ hops
back to the hole site liberating a photon with energy $\sim
U-zJ\sigma$, leaving behind a locally disturbed antiferromagnet
with $z$ exchange bonds broken in the final state as indicated by
the dotted lines.} \label{Fig:2_magnon}\end{figure}

In the nearest neighbor Heisenberg antiferromagnet, one treats the
spin operators using a Dyson-Maleev representation of magnons
with dispersion $2J$. Due to the $B_{1g}$ form factor, in the
absence of magnon-magnon interactions at T=0, a sharp peak at $4J$
appears as the top of the magnon dispersion is projected out, and
the response abruptly drops to zero for large Raman shifts \cite{Sandvik:1998}.
However, since the light scattering is localized to neighboring
spins, magnon-magnon interactions must be included, and the peak
becomes more symmetric and shifts to $\sim 3J$ via breaking
exchange bonds between local neighbors, as shown in
Figure~\ref{Fig:2_magnon}.

There have been many developments on the Fleury-Loudon model,
which has been addressed via critical fluctuation analysis
\cite{Halley:1978}, series expansions \cite{Singh:1989}, lower
\cite{Parkinson:1969,Morr:1997} and higher \cite{Canali:1992,Chubukov:1995b} order spin wave
theories, $t-J$ studies at finite doping \cite{Prelovsek:1996},
exact diagonalizations of small clusters \cite{Tohyama:2002},
excitonic cluster approaches \cite{Hanamura:2000}, finite
temperature Quantum Monte Carlo methods \cite{Sandvik:1998}, and
studies of bilayer effects \cite{Morr:1996}, 2-leg spin ladders
\cite{Jurecka:2001} and ring exchange \cite{Katanin:2003}, giving
a thorough treatment of two magnon scattering from spin degrees of
freedom in the non-resonant regime. Scattering from channels other
than $B_{1g}$, and describing the anisotropic lineshape of the
response, have been addressed via longer range spin exchange
interactions and by exact diagonalizations of magnons coupled to
phonons \cite{Freitas:2000}, using an earlier approach of
\textcite{Lorenzana:1995}. Lastly, recent developments concern
scattering from orbiton degrees of freedom \cite{Okamoto:2002} and
scattering within a resonant valence bond picture
\cite{Ho:2001}.\footnote{The reader is also referred to the review
article by \textcite{Lemmens:2003} for reviews on the theoretical
treatments of magnetic light scattering in low-dimensional quantum
spin systems.}

\begin{figure}[b!]
\centerline{\epsfig{figure=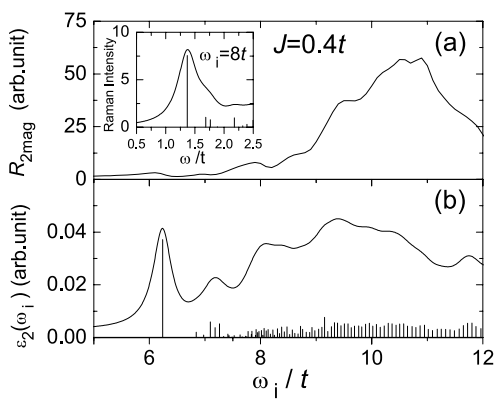,width=7.0cm,clip=}}
\vspace{0.1cm} \caption{The dependence of (a) two-magnon $B_{1g}$
Raman intensity (shown in the inset for $\omega_{i}=8t$) and (b)
absorption spectrum on the incoming photon energy $\omega_{i}$ in
a 20-site cluster of the Hubbard model with $U= 10t~ (J=0.4t)$.
The solid line in (b) is obtained by performing a Lorentzian
broadening with a width of $0.4t$ on the delta functions denoted
by vertical bars. From \textcite{Tohyama:2002}.}
\label{Fig:tohyama}\end{figure}

These approaches fail when the laser frequency is tuned
near an optical transition. In this regime, based on a spin
density wave approach, \textcite{Chubukov:1995a,Chubukov:1995b}
have formulated a so called ``triple-resonance'' theory from which
important features of the spectra can be derived.  Using a SDW approach
to the Hubbard model, they found that additional resonant diagrams of the type shown in Figure~\ref{Fig:resonant} (h)-(j) contribute
to the usual Loudon-Fleury terms, and derived a resonant profile in good agreement
with experiments. In addition, recent results by \textcite{Tohyama:2002} have been obtained for
the Raman response in the resonant limit from both spin and charge
degrees of freedom. In Figure~\ref{Fig:tohyama} we show their
results from exact diagonalization of the Hubbard model with a
20-site cluster \cite{Tohyama:2002}. The two-magnon response at
roughly $\hbar\Omega=2.7J$ is resonantly enhanced when the
incident photon frequency is tuned to the Mott gap scale $U$, in
qualitative agreement with the results of
\textcite{Chubukov:1995a,Chubukov:1995b}. Both approaches predict a resonant profile
for two-magnon Raman which differs from the absorption profile, as shown in Figure
\ref{Fig:tohyama}. \textcite{Tohyama:2002} have pointed out
that the resonance energies for the absorption spectrum and the two-magnon response are
not the same, due to differences in SDW coherence factors.

\subsubsection{Interacting Electrons - Full Response}

An approach treating the full fermionic degrees of freedom and,
simultaneously, treating non-resonant, mixed, and resonant
scattering on equal footing, is still in its infancy. The
theoretical challenge in calculating the full inelastic light
scattering response function is that the mixed diagrams involve
three-particle susceptibilities and the resonant diagrams involve
four-particle susceptibilities.  Only in the infinite-dimensional
limit, where most of the many-particle vertex renormalizations
vanish (all three-particle and four-particle vertices do not
contribute; only the two-particle vertices enter), one can imagine
arriving at exact results. As an exception, the full Raman
response function can be calculated in the Falicov-Kimball model,
because the two-particle irreducible charge vertex is known
exactly in the limit of large dimensions
\cite{Shvaika:2000,Freericks:2000}.

\begin{figure}[floatfix]
\centerline{\epsfig{figure=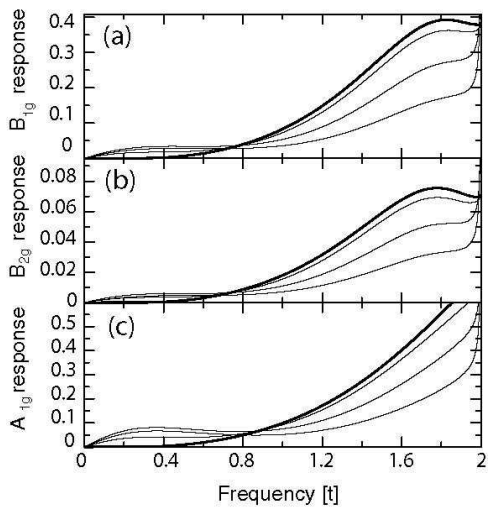,width=7.0cm,clip=}}
\vspace{0.1cm} \caption{Raman response in the 1/2-filled
Falicov-Kimball model  ($U=2t$) as a function of  transfered
frequency for various temperatures for fixed incident photon
frequency $\omega_{i}=2t$. The thickest curve is $T/t=0.05$, and
the temperature increases to 0.2, 0.5, and 1 as the curves are
made thinner. From \textcite{Shvaika:2004}.}
\label{Fig:isos}\end{figure}

Recently,  \textcite{Shvaika:2004,Shvaika:2005} obtained the full
electronic Raman response function, including contributions from
the non-resonant, mixed, and resonant processes within a
single-band model. In general, the resonance effects can create
orders of magnitude enhancement over the non-resonant response,
especially when the incident photon frequency is slightly larger
than the frequency of the non-resonant feature. The resulting
Raman response is a complicated function of the correlations, the
temperature, the incident photon energy, and the transfered
energy. It was found that resonance effects are different in
different scattering geometries, corresponding to different
symmetries of the charge excitations scattered by the light.

Resonance effects were found as a function of both the incoming
and the outgoing photon frequencies $\omega_{i,s}$. A double
resonance - occurring when the energy denominators of two pairs of
the Green's functions, appearing in the bare response shown in
Figure~\ref{Fig:resonant}, approach zero - gives the strongest
resonant enhancement of the response\cite{Shvaika:2004}. In
addition, an interesting resonance effect on both the
charge-transfer peak and the low-energy peak was found when the
incident photon frequency is of the order of the interaction
strength, showing that in general the total response cannot be
well described as a uniform resonance enhancement of the separable
non-resonant response. In agreement with the results of
\textcite{Tohyama:2002}, for an antiferromagnetic system this is a
direct consequence of the inseparability of energy scales in the
correlated electron problem, in contrast to non-interacting
electrons.

Shown in Figure~\ref{Fig:isos} is the temperature and
symmetry-dependent Raman response, including non-resonant,
resonant, and mixed terms in the Falikov-Kimball model. In the
insulating phase, spectral weight is depleted for small energy
transfers and piles up into the excitations at energies of order
$U$ as the temperature is lowered. The transfer of spectral weight
from lower to higher energies occurs across a temperature
independent so-called isosbestic point. An isosbestic point also
appears in studies of the Hubbard model
\cite{Freericks:2001b,Freericks:2003a}, implying that it is a
generic feature of the insulating phase, regardless of the
microscopic origin of the phase. We note that isosbestic behavior
already appears in the non-resonant contributions for $B_{1g}$
scattering. In the $A_{1g}$ and $B_{2g}$ symmetries, it emerges
only if resonant terms are included.

The local treatment of self energies in the single-site DMFT
approach imposes limitations on the theory of light scattering in
correlated systems. In particular, the full polarization
dependence of the Raman spectra would uncover the way in which
correlations affect electron dynamics in regions of the BZ,
providing a two-particle complement to ARPES, for example.
Progress here most likely will come from cluster dynamical mean
field theory able to treat nonlocal and anisotropic interactions
in a coarse-grained manner.

\subsubsection{Superconductivity}
\label{Section:sc_theory}

As discussed in Section \ref{Section:weak}, in the absence of
interactions, there is no phase-space for low energy Raman
scattering for ${\bf q}=0$ momentum transfers. In the
superconducting state, phase space restrictions are lifted since
light can break ${\bf q}=0$ Cooper pairs if the energy of the
light is greater than $2\Delta$. Thus the Raman response becomes
non-trivial, yet easily formulated, in BCS theory. As a consequence,
there has been an enormous amount of theoretical work devoted to
light scattering for temperatures below $T_{c}$ as an extension of
the theory for non-interacting electrons in the normal state. We
review that work here.

In the superconducting state, focus has been traditionally placed
on the two-particle non-resonant response in BCS theory. Formally
the Raman response is given by generalizing
Eqs.~(\ref{Eq:full}-\ref{Eq:Bethe}) in particle-hole space using
Pauli matrices $\tau_{i=0..3}$ in Nambu notation
\cite{Nambu:1960}:
\begin{widetext}
\begin{equation}
   \chi({\bf q}=0,i\Omega)=-\frac{2}{V\beta}\sum_{i\omega}\sum_{\bf k}
   Tr\left[\hat\gamma({\bf k}) \hat G({\bf k},i\omega)
   \hat\Gamma({\bf k};i\omega;i\Omega) \hat G({\bf k},i\omega+i\Omega)\right],
\label{Eq:full_s}
\end{equation}
where $Tr$ denotes the trace, and
\begin{equation}
   \hat\Gamma({\bf k};i\omega;i\Omega)=\hat\gamma({\bf k})
   +\frac{1}{V\beta}\sum_{i\omega^{\prime}}\sum_{\bf k^{\prime}}
   V_{i}({\bf k}-{\bf k^{\prime}},i\omega-i\omega^{\prime})\hat\tau_{i}
   \hat G({\bf k^{\prime}},i\omega^{\prime})
   \hat \Gamma({\bf k^{\prime}};i\omega^{\prime};i\Omega)
   \hat G({\bf k^{\prime}},i\omega^{\prime}+i\Omega)\hat\tau_{i}.
\label{Eq:Bethe_s}
\end{equation}
\end{widetext}
Here the bare Raman vertex of coupling to charge is
$\hat\gamma=\hat\tau_{3}\gamma$ and the interaction $V_{i}$
determines the channel of the vertex corrections. For example
$V_{i=3}$ corresponds to interactions coupling electronic charge,
while $V_{i=0}$ corresponds to spin interactions.

For the case of weak correlations, the Green's functions appearing
in Eqs.~(\ref{Eq:full_s}) and~(\ref{Eq:Bethe_s}) are given by the
BCS expression
\begin{equation}
   \hat G({\bf k},i\omega) = \frac{i\omega\hat\tau_{0}+\epsilon({\bf k})
   \hat\tau_{3}+\Delta({\bf k})\hat\tau_{1}}{(i\omega)^{2}-E^{2}({\bf k})},
\label{Eq:BCS}
\end{equation}
with $E^{2}({\bf k})=\xi^{2}({\bf k})+\Delta^{2}({\bf k})$ the
quasiparticle energies. In the weak coupling limit for the BCS
approximation, $V_{i=3}=-V$ for phonon-mediated pairing.

By far, the superconducting state has received the largest amount
of attention from theory, starting from the seminal contribution
of \textcite{Abrikosov:1961}, which predated the observation of
the effect by 19 years. The main focus in the early years was to
study the $2\Delta$ features in conventional $s-$wave
superconductors with small and large coherence lengths
\cite{Abrikosov:1961,Abrikosov:1987,Abrikosov:1988,Klein:1984},
including the effects of Coulomb screening \cite{Abrikosov:1973}
and examining the temperature dependence \cite{Tilley:1972}. If one
neglects vertex corrections, the ${\bf q}$-dependent Raman response in a
superconductor is given by a projected Maki-Tsuneto function
\cite{Maki:1962},
\begin{widetext}
\begin{eqnarray}
   &&\chi_{a,b}^{\prime\prime}({\bf q},\Omega+i\delta)=
   \frac{1}{N}\sum_{\bf k}a_{\bf k,q}b_{\bf k,q}\nonumber\\
   &&\times\biggl\{A_{+}({\bf k,q})\left[f(E({\bf k}))-f(E({\bf k+q}))\right]
   \left(\frac{1}{\Omega+i\delta-E({\bf k})+E({\bf k+q})}
   -\frac{1}{\Omega+i\delta+E({\bf k})-E({\bf k+q})}\right)\nonumber\\
   &&+A_{-}({\bf k,q})\left[1-f(E({\bf k}))-f(E({\bf k+q}))\right]
   \left(\frac{1}{\Omega+i\delta+E({\bf k})+E({\bf k+q})}
   -\frac{1}{\Omega+i\delta-E({\bf k})-E({\bf k+q})}\right)\biggr\},
\label{Eq:Tsuneto_long}
\end{eqnarray}
with the coherence factors $A_{\pm}({\bf k,q})=1\pm\frac{\xi({\bf
k})\xi({\bf k+q}) -\Delta({\bf k})\Delta({\bf k+q})}{E({\bf
k})E({\bf k+q})}$. A more common expression is the Raman response
for ${\bf q}=0$ which simplifies to
\begin{equation}
   \chi_{a,b}^{\prime\prime}({\bf q}=0,\Omega+i\delta)=
   \frac{2}{N}\sum_{\bf k}a_{\bf k}b_{\bf k}
   \left[\frac{\mid\Delta({\bf k})\mid}{E({\bf k})}\right]^{2}
   \tanh\left(\frac{E({\bf k})}{2T}\right)
   \left(\frac{1}{2E({\bf k})+\Omega+i\delta}
   +\frac{1}{2E({\bf k})-\Omega-i\delta}\right),
\label{Eq:Tsuneto}
\end{equation}
\end{widetext}
The full Raman response, including charge screening, is once again
given by Eq.~(\ref{Eq:gen_screened}), in which the vertices $a,b$
are replaced by the Raman ($a,b=\gamma$) and pure charge ($a,b=1$)
vertices \cite{Abrikosov:1973}.

\begin{figure}[b!]
\centerline{\epsfig{figure=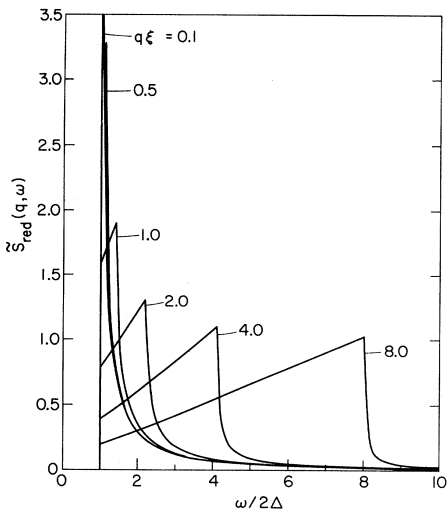,width=6.5cm,clip=}}
\vspace{0.1cm}\caption{Raman response for an $s-$wave
superconductor for $q\xi\pi/2=0.1, 0.5, 1.0, 2.0, 4.0,$ and $8.0$
with $\xi=\hbar v_{F}/\pi\Delta$ the Pippard-BCS coherence length
$\xi$ and $q \simeq 1/\delta$ the momentum transfer in a metal
with skin depth $\delta$. From \textcite{Klein:1984}.}
\label{Fig:KleinDierker}\end{figure}

For the case of an isotropic gap ($\Delta({\bf k})=\Delta$) and
momentum transfers ${\bf q}=0$, a threshold and a square-root
discontinuity appears at twice the gap edge $\Delta$, reflecting
the two-particle density of states. For finite ${\bf q}$, the
singularity is cut-off due to breaking Cooper pairs with finite
momentum, and the peak is shifted out to frequencies of roughly
$v_{\rm F}q$ as in the normal state
(Figure~\ref{Fig:KleinDierker}). Qualitatively similar behavior is
obtained for disordered $s-$wave superconductors
\cite{Devereaux:1992} in which $1/\tau_{L}^{*}$ (see
Eq.~(\ref{Eq:Drude-imp})) assumes the role of $v_{\rm F}q$  .

Further advances in the theory for conventional $s-$wave
superconductors were made for energy gaps with small anisotropy
\cite{Klein:1984}, coexistence with charge-density wave order
\cite{Balseiro:1980,Littlewood:1981,Littlewood:1982,Tutto:1992},
layered superconductors \cite{Abrikosov:1991}, impurities
\cite{Devereaux:1992,Devereaux:1993}, and final-state interactions
\cite{Klein:1984,Monien:1990,Devereaux:1993,Devereaux:1995a}.

We note in particular that the variation with ${\bf k}$ of the
Raman vertices $\gamma({\bf k})$ is coupled to the ${\bf
k}$-dependence of the energy gap $\Delta({\bf k})$ (see, e.g.,
Eq.~(\ref{Eq:Tsuneto})), leading to a strong polarization
dependence of the spectra. For isotropic $s$-wave superconductors,
the vertex does not affect the lineshape, and thus the spectrum is
polarization independent, apart from an overall prefactor. For this
case, a polarization dependence can be generated in BCS theory by
taking into account channel-dependent final-state interactions
\cite{Bardasis:1961} and/or impurity scattering. However, for the
most part this only produces a channel dependence in the vicinity
of the gap edge, and thus the main feature of the response is the
uniform gap existing for all polarizations. For anisotropic energy
gaps, the symmetry dependence of the spectra is a direct
consequence of the {\bf k}-summation (angular averaging), which
couples gap and Raman vertex and leads to constructive
(destructive) interference if the vertex and the gap have the same
(different) symmetry.

Generally, in superconductors with nodes of the energy gap,
power-laws in the low frequency and/or temperature variation of
transport and thermodynamic quantities emerge, replacing threshold
or Arrhenius behavior ubiquitous in isotropic superconductors.
However, due to the averaging over the entire Fermi surface, the
power-laws themselves do not uniquely identify the ground state
symmetry of the order parameter, but only can give the topology of
the gap nodes along the Fermi surface, e.g., whether the gap
vanishes on points and/or lines. Thus, one cannot distinguish
between different representations of the energy gap which have the
same topology. For instance, for the case of $d$-wave tetragonal
superconductors, there are five pure representations which have
line nodes on the Fermi surface. Two-particle correlation
functions, determining the density, spin or current responses, do
not have the freedom to probe various portions of the gap or its
phase around the Fermi surface.

With the advent of high-$T_c$ cuprates, a flurry of activity
ensued on theory of Raman scattering in $d-$wave superconductors
\cite{Falkovsky:1990,Monien:1989}, specifically including
polarization dependences \cite{Devereaux:1994a,Devereaux:1995a},
collective modes
\cite{Devereaux:1995a,Wu:1995a,Wu:1995b,Dahm:1998}, impurities
\cite{Devereaux:1995b,Devereaux:1997,Wu:1998,Devereaux:2003b},
temperature dependences
\cite{Devereaux:1995b,Devereaux:1997,Branch:1995,Devereaux:2003b},
screening
\cite{Devereaux:1995a,Manske:1997,Strohm:1998a,Branch:1996}, band
structure and bi-layer effects
\cite{Branch:1996,Strohm:1997,Devereaux:1996,Krantz:1994}, surface
and c-axis contributions \cite{Wu:1996,Wu:1997}, resonant effects
\cite{Sherman:2002}, and mixed-state pairing
\cite{Devereaux:1995a,Nemetschek:1998,Lee:2002}. The focus was
largely on how the symmetry selection rules could locate the
positions of the gap maxima and nodal points around the Fermi
surface.

\begin{figure}[b!]
\centerline{\epsfig{figure=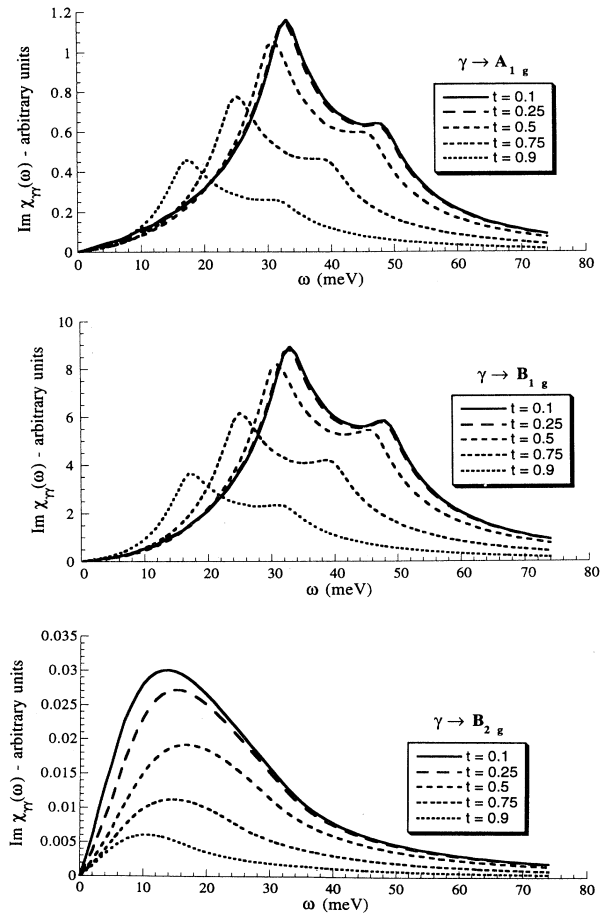,width=1\linewidth,clip=}}
\vspace{0.1cm} \caption{Raman spectra for unscreened $A_{1g}$
(top), $B_{1g}$ (middle) and $B_{2g}$ (bottom) response as a
function of reduced temperature $t=T/T_{c}$. Here $2\Delta\sim
37$meV, and the higher peak in $A_{1g}$ and $B_{1g}$ channels is a
van Hove feature. From \textcite{Branch:1995}.}
\label{Fig:Branch}\end{figure}

For $d_{x^{2}-y^{2}}$ superconductors, the interplay of
polarizations and gap anisotropy can be simply drawn. Referring to
Figure~\ref{Fig:project}, $B_{1g}$ orientations project out
excitations around the principle directions ($M$-points or
anti-nodal regions) of the BZ where the superconducting gap is
maximal and where the van Hove singularity is located, while
$B_{2g}$ orientations project the nodal regions along the
diagonals. As a consequence, the Raman response has a peak at
$2\Delta_{\rm max}$ for $B_{1g}$ and at slightly lower energy for
$B_{2g}$. The polarization dependence also enters the low
frequency behavior. Since line nodes yield a linear dependence on
energy of the density of states, the $B_{2g}$ response depends
linearly on $\hbar\Omega$ in the limit $\Omega \rightarrow 0$ for
a gap vanishing on the diagonals. For $B_{1g}$ orientations, in
contrast, the Raman vertex vanishes along with the energy gap at
the same points in the BZ. This yields an additional $\Omega^{2}$
contribution from the line nodes of the vertex, and the resulting
response varies as $\Omega^{3}$. The unscreened $A_{1g}$ response
measures an overall average throughout the BZ and thus picks up
the gap maxima as well as the linear density of states. This is
shown quantitatively in Figure~\ref{Fig:Branch}. The frequency
power laws also translate in low temperature power laws of the
response in the small frequency limit \cite{Devereaux:1995a}.

Disorder effects generally smear peak features at larger energies
and change the $B_{1g}$ exponent to 1, similar to the change in
the low temperature NMR rate for d-wave superconductors
\cite{Devereaux:1995b}. Moreover, similarities between the in-plane
conductivity and $B_{2g}$ Raman follows from the BZ weighting
around the nodes, while the $B_{1g}$ response is qualitatively
similar to the c-axis conductivity due to the weighting around the
anti-nodes \cite{Devereaux:2003b}. For example the residual
in-plane conductivity as $T\rightarrow 0$ is universal and given
by $\sigma(T=0)=ne^{2}/m\pi\Delta_{0}$, the slope of the $B_{2g}$
response $2N_{F}/\pi\Delta_{0}$ is also universal and insensitive
to impurity effects, while the $B_{1g}$ channel and c-axis
conductivity are non-universal, having additional impurity
dependent prefactors
\cite{Devereaux:1995b,Devereaux:1997,Devereaux:2003b}. The slope
of the $B_{2g}$ response follows the temperature dependence of the
in-plane conductivity, and both possess a peak at intermediate
temperatures due to a balance of DOS and lifetime effects as
temperatures are lowered from $T_{c}$. Yet, both the out-of-plane
conductivity and the $B_{1g}$ response do not show a peak due to
the more rapid variation of the projected DOS coming from
antinodal portions of the BZ.

We note that for a $d_{xy}$ energy gap, the above discussion
applies accordingly, with the role of $B_{1g}$ and $B_{2g}$
symmetry reversed. It was indeed an important development to
show that Raman scattering is unique in determining two-particle
electron dynamics independently in {\em different} regions of the
BZ in the superconducting state.

\begin{figure}[b!]
\centerline{\epsfig{figure=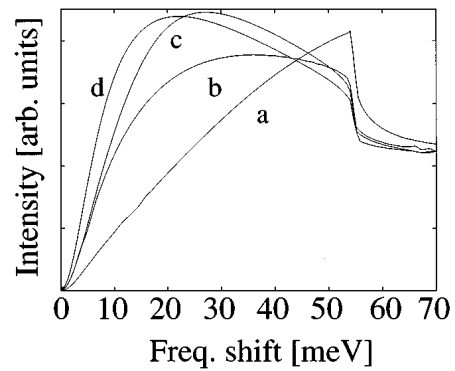,width=6.5cm,clip=}}
\vspace{0.1cm} \caption{Comparison of the screened $A_{1g}$
response obtained for different number of $d_{x^{2}-y^{2}}$ gap
harmonics $\Delta({\bf k})= \Delta_{0}
\{[\cos(k_{x}a)-\cos(k_{y}a)]/2 +
\Delta_{1}[\cos(k_{x}a)-\cos(k_{y}a)]^{3}/8 +$
$\Delta_{2}[\cos(k_{x}a)-\cos(k_{y}a)]^{5}/32\}$. Here $a-d$
correspond to the set $\Delta_{1,2}=(0,0), (1,0), (0,1),$ and
$(1,1)$, respectively, and $\Delta_{0}$ has been rescaled to give
the same value for the maximum gap. Generally the large peak at
$2\Delta$ (as shown in Figure~\ref{Fig:Branch}) is suppressed by
the backflow terms. From \textcite{Devereaux:1996}.}
\label{Fig:a1gsense}\end{figure}

While the low frequency power-laws are insensitive to band
structure (such as the shape of the Fermi surface), the
polarization selection rules can select features of the band
structure at higher energies. For example, in the cuprates, the van
Hove singularities from $(\pi,0)$ and related points yields a peak
at twice the quasiparticle energy $E({\bf k})$ in $A_{1g}$ and
$B_{1g}$ channels, as shown in Figure (\ref{Fig:Branch}). For
multiple bands, including the case of several Fermi surface
sheets, the responses for crossed polarizations are simply
additive, yet for $A_{1g}$ channels due to backflow effects an
additional interference term can be present if the charge
fluctuations are different for the different sheets
\cite{Krantz:1994,Devereaux:1996}.

Quite generally the backflow yields substantial reorganization of
spectral weight around $2\Delta_{\rm max}$ compared to the bare
response\cite{Branch:1995,Dahm:1999}. This is because the term
$\chi_{\gamma,1}$, which contributes in channels having the
symmetry of the lattice, is peaked and large at the same position
as the unscreened term $\chi_{\gamma,\gamma}$. Not surprisingly,
the reorganization depends delicately on the relative momentum
dependences of the Raman vertices and energy gap (number of BZ
harmonics for example, as shown in Figure~\ref{Fig:a1gsense}), as
well as on details of the band structure
\cite{Krantz:1994,Devereaux:1994a,Strohm:1997,Branch:1996}.

For the cuprates, Raman vertices have been calculated using LDA
\cite{Strohm:1997}, but limited progress has been made in
including the contributions of substantial electronic
correlations. In most other cases, either the effective mass
approximation has been used in calculations or simply a symmetry
classification has been made. While a detailed lineshape analysis
can be applied based purely on symmetry as explained above, it
must be kept in mind that even a comparison of overall intensities
between different geometries can at best be qualitative.

Yet, the continuation of these treatments from the superconducting
to the normal state is not straightforward. As can be seen
directly from Eq.~(\ref{Eq:Tsuneto}), the intensity vanishes
proportional to $\Delta^{2}$ as T approaches $T_c$. Hence, to
avoid phase space limitations in the absence of Cooper pairs, an
additional source for electronic scattering, such as the one mediating
the formation of Cooper pairs, must be included. While strong
coupling extensions of Raman scattering in $d-$wave
superconductors have recently been presented \cite{Dahm:1999,Devereaux:2000,Jiang:1996}, a merging of the
normal and the superconducting states is poorly understood. This
would require a not yet existing microscopic description of the
formation of $d-$wave superconductivity from the normal state.

\subsubsection{Collective Modes}
\label{Section:collective}

Raman scattering
has the almost unique ability to sort out collective modes of the
two-particle response in different symmetry channels, owing to the
freedom to independently adjust the two polarization vectors. The
collective mode spectrum one obtains depends upon which
interactions are included in Eq.~(\ref{Eq:Bethe_s}). We first
discuss the general consequences based on gauge-invariance and
focus on exciton-like modes.

In order to form a fully gauge-invariant theory, the interactions
responsible for superconductivity appear not only in $\hat G$, but
must also be included as vertex renormalizations $\hat \Gamma$. In
this way, the Raman response from pure charge density fluctuations
in the superconducting state yields the Goldstone mode from the
broken gauge symmetry - the phase or Anderson-Bogoliubov mode
\cite{Anderson:1958,Bogoliubov:1959,Nambu:1960}. In the absence of
the long-range Coulomb interaction, this mode is a soft sound mode,
yet the Coulomb interactions - inescapable for {\bf q}=0 - push
the sound mode up to the plasma frequency via the Higgs mechanism.
As a result, particle-number conservation is satisfied in the
superconducting state and $\chi_{sc}({\bf q}=0,\Omega)=0$,
independent of whether one considers Bloch states
\cite{Abrikosov:1973,Klein:1984,Monien:1990} or Anderson exact
eigenstates of the disordered problem \cite{Devereaux:1993}.

However, additional modes of excitonic origin may appear if one
considers further interactions between electrons in clean
\cite{Bardasis:1961} and disordered \cite{Maki:1962,Fulde:1965}
conventional superconductors. These excitons appear split off from
the continuum at $\hbar\Omega < 2\Delta$ if the interaction occurs
in higher momentum channels orthogonal to the BCS condensate.

Since Raman scattering couples to anisotropic charge density
fluctuations with symmetry selectivity to different channels $L$,
the light polarizations can be used to determine the exact nature
of bound states. \textcite{Balseiro:1980} considered the formation
of a phonon-Cooper pair bound state due to electron-phonon
coupling though neglecting channels higher than $L=0$. However,
this mode is canceled by the backflow applying generically to all
systems. Finite $L$ exciton formation in clean and disordered
superconductors, and the resulting appearance in Raman scattering,
have been considered explicitly by \textcite{Monien:1990} and
\textcite{Devereaux:1993}, respectively, bringing the symmetry of
the exciton and the polarization dependence to light.

We show in
Section~\ref{sec:A15} below that the effect of final state
interactions can be substantial in strongly coupled conventional
superconductors. This demonstrates the strength of the
electron-phonon coupling, not only in general, but also
specifically in channels orthogonal to the ground state.
Interestingly, the lattice instability found in some of these
materials has the same symmetry as the collective mode and the
electronic states which apparently drive the transition
\cite{Weber:1984}.

For $d_{x^{2}-y^{2}}$ superconductors, the collective mode spectra
have been investigated thoroughly by \textcite{Devereaux:1995a}
and others
\cite{Wu:1995a,Wu:1995b,Dahm:1998,Manske:1997,Strohm:1998a,Manske:1998}.
It was shown that the Anderson-Bogoliubov mode appears in $A_{1g}$
channels and massive modes can appear in other channels. Since the
pair state has only one representation in the $D_{\rm 4h}$ group,
massive collective modes arise when one considers interactions in
orthogonal channels. Recently, it has been suggested that the
presence of collective modes may allow one to distinguish charge-
or spin-mediated $d-$wave pairing
\cite{Chubukov:1999,Chubukov:2006}, highlighting the possible
importance in the context of the cuprates.

Generally, the collective mode spectrum can be quite diverse in
unconventional superconductors. In principle, additional broken
continuous symmetries can exist, such as $SO_{3}^{S}$ spin
rotational symmetry in spin-triplet systems and $SO_{3}^{L}$
orbital rotational symmetry in spin-singlet systems, if the gap
does not possess the full symmetry of the lattice. Furthermore,
massive collective modes can arise if the energy gap is degenerate
or has an admixture of different representations of the point
group. The massive modes can in principle lie below the gap edge,
and can thus be relevant for the low frequency dynamics of
correlation functions. In fact, Raman-active modes in spin-triplet
superconductors such as Sr$_{2}$RuO$_{4}$ have drawn recent
theoretical interest \cite{Kee:2003}, although the experimental
challenges are not negligible because of the low $T_c$ and the
related small energy gap in these materials.

Though very interesting, spin-triplet pairing or spin-orbit
effects are rare and more on the exotic side in superconductivity.
Competing ground states, however, are quite common whenever
correlation effects come into play. This is not at all confined to
the cuprates, but occurs also in, e.g., spin (SDW) and charge
(CDW) density wave systems. Usually, density-wave formation with
long-range order at least partially suppresses superconductivity
such as in $2H-{\rm NbSe_2}$ (see below). Then, additional modes
appear as a result of the competition between CDW ordering and
superconductivity, and collective modes appear as one modulates
either one or both order parameters.

\textcite{Littlewood:1981,Littlewood:1982} and
\textcite{Browne:1983} considered a direct coupling between charge
density and superconducting gap amplitudes, modulated for example
by a CDW phonon, although this was not specified. They obtained an
additional ``gap'' mode below $2\Delta$. Yet this mode was only
considered in the $L=0$ channel, and Coulomb interactions once
again remove this mode. \textcite{Lei:1985} considered an
effective CDW-SC coupling via a phonon and realized that in
anisotropic systems such collective modes in $L\ne 0$ channels may
appear. Finally, \textcite{Tutto:1992} treated electron-phonon and
CDW amplitude-phonon coupling on equal footing in finite angular
momentum channels, showing generally that the collective modes in
these channels are unaffected by Coulomb screening. The modes
obtained split off from the gap edge and appear as
excitations below the quasiparticle spectrum, much like excitons in semiconductors. Evidence of mixed CDW-SC pairing may be seen in Raman experiments via the presence
or absence of these modes. This has recently been extended by \textcite{Zeyher:2003} to MgB$_{2}$, having multiple energy gaps on different electron bands.

\begin{figure}[b!]
\centerline{\epsfig{figure=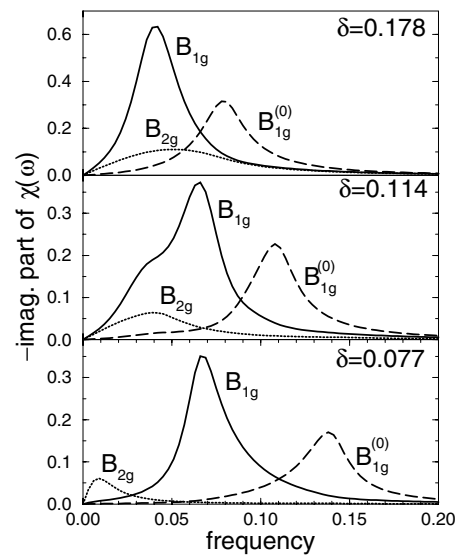,width=6.5cm,clip=}}
\vspace{0.1cm} \caption{Electronic Raman spectra of $B$ symmetries
for three different doping levels including coupling of $d$-CDW
and $d$-superconducting amplitudes. The superscript $0$ denotes
the spectra in the absence of collective modes. From
\textcite{Zeyher:2002}.} \label{Fig:Zeyher}
\end{figure}

The collective mode spectrum of coupled $d-$wave charge density
and superconductivity was investigated by \textcite{Zeyher:2002}
along the lines developed by \textcite{Tutto:1992} for
conventional CDW and superconducting systems. As for $s-$wave CDW
superconductors, collective modes split off from the maximum of
the gap edge. As an important difference in $d-$wave systems, the
modes distinctly affect the various symmetry channels. Besides the
reorganization of $A_{1g}$ spectral weight, additional modes alter
the $B_{1g}$ spectrum, as shown in Figure~\ref{Fig:Zeyher}.

Density waves instabilities need not necessarily compete with
superconductivity but rather can provide an effective coupling
mechanism \cite{Castellani:1995,Perali:1996} as long as quantum
and thermal fluctuations suppress long range order. Then,
collective modes may appear as fluctuation-induced modes. The
Raman scattering is usually determined from Aslamazov-Larkin
fluctuation diagrams considered for the conductivity
\cite{Aslamazov:1968}. To overcome the ${\bf q}=0$ phase space
limitations, the Raman response is given by the exchange of two
fluctuations modes at wavevectors ${\bf Q}_{c}$ and $-{\bf
Q}_{c}$, yielding generally a mode at energies of twice the mass
of the fluctuation propagator. Once again the polarization
dependence can select different fluctuation modes corresponding to
different ordering wavevectors coupling to either charge or spin
density modes. This was investigated for spin \cite{Brenig:1992}
and charge \cite{Caprara:2005} fluctuations in the normal state,
and for novel spin resonances in the superconducting state of the
cuprates \cite{Chubukov:1999,Venturini:2000,Chubukov:2006}.

Lastly, we remark that many other types of collective modes are
possible if one considers more exotic ground states with different
symmetry classifications. For example, a chiral spin liquid has
been investigated by \textcite{Khveshchenko:1994} in which helical
excitations were conjectured to exist and are in principle
measurable in $A_{2g}$ orientations which can be projected out via
proper sums of spectra taken with both linearly and circularly
polarized light. Other examples are modes induced by magnetic
fields or optical modes resulting from Dzyaloshinskii-Moriya
interactions in Heisenberg antiferromagnets as observed recently
in lightly doped $\rm La_{2-x}Sr_xCuO_4$, $0 \leq x \leq 0.03$
\cite{Gozar:2005b} and discussed by \textcite{SilvaNeto:2005},
directly demonstrating the importance of spin coupling to the
local environment.

\section{FROM WEAKLY TO STRONGLY INTERACTING ELECTRONS}
\label{Section:III}

In this section we review experimental results in systems other
than doped semiconductors (see reviews by, e.g.,
\textcite{Abstreiter:1984} and \textcite{Pinczuk:1989}) and
cuprates (see section \ref{sec:HTSC}) with a view towards
signatures in the Raman spectra arising from the development of
strong electronic correlations. We discuss various types of
superconductors and summarize results on correlated metals and
other strongly interacting systems.

The light scattering cross section in absorbing media, such as
systems with free carriers, is generally weak since the
interaction volume is small for the short penetration depth of
visible light, $\delta \ll \lambda_{i}=2\pi c/\omega_{i}$. As a
consequence, the momentum perpendicular to the surface is not
conserved, and the transfer ${\bf q}$ is not given any more by the
difference of the vacuum momenta of the involved photons ${\bf
k}_{i}-{\bf k}_{s}$ but essentially by $\delta=\lambda/(4\pi k)$
with $k$ the imaginary part of the index of refraction
\cite{Abrikosov:1961,Mills:1970}. Even in strongly absorbing
materials with $k > 1$, $1/\delta \ll \pi/a$ holds where $a$ is
the lattice constant, and the limit of small momentum transfer is
still effective. This introduces a new energy scale $\hbar v_Fq
\approx \hbar v_F/\delta$ with $v_F$ and $q$ being the magnitudes
of the Fermi velocity and the momentum transfer, respectively. In
all considerations, this scale must be put into relation to the
other relevant energies, such as the electron scattering rate
$\Gamma=\hbar/\tau$ in the normal and the gap $\Delta$ in the
superconducting state. These apparent academic considerations have
major impact on both the observability and the interpretation of
electronic spectra.

\subsection{Elemental Metals and Semiconductors}

In addition to the small scattering volume due to the absorption
of light by free carriers, a parabolic dispersion and a spherical
Fermi surface reduce the cross section of single-electron
excitations in metals and degenerate semiconductors strongly,
since in such systems the associated density fluctuations are
screened by the long-range Coulomb interaction. The few spectra we
are aware of have been taken on elements with a more complex band
structure such as Nb \cite{Klein:1982a,Klein:1984} or Dy
\cite{Klein:1991}. In Dy a broad continuum similar to that in the
high-$T_c$ cuprates \cite{Bozovic:1987} is found. In Nb the
superconducting state was studied. Due to the low transition
temperature $T_c$, the correspondingly small energy gap
$\Delta(T)$ and the small ratio $\delta/\xi$ with $\delta$ the
penetration depth of the light and $\xi$ the superconducting
coherence length (for details see section~\ref{sec:A15}) the
characteristic redistribution of scattering intensity is very hard
to observe. The peaks found at 1.8~K in the expected energy range
close to $2\Delta(T)$ are very weak, and no normal state spectra
have been measured for comparison \cite{Klein:1982a,Klein:1984}.

In fact, superconductors rather than normal metals were the
main focus in the early days of electronic Raman scattering. Only
after the discovery of the cuprates \cite{Bednorz:1986}, with
generally complicated and sometimes very surprising electronic
properties, did studies of the normal state become increasingly
attractive (see section \ref{sec:HTSC}).

\subsection{Conventional Superconducting Compounds} \label{sec:A15}

Among superconductors, intermetallic compounds like ${\rm Nb_3Sn}$
or ${\rm V_3Si}$ with A15 structure, can be considered conventional
both above and below $T_c$. They are strictly 3D,
superconductivity is mediated by phonons leading to an essentially
isotropic s-wave gap, and correlations
are believed to be of minor importance. This does not mean they
are simple. For instance, the Fermi velocity is very small,
close to the velocity of sound, and the Fermi surface is
multi-sheeted. Sufficiently perfect single crystals of ${\rm
Nb_3Sn}$ and ${\rm V_3Si}$ undergo a structural transformation
from a cubic to a tetragonal lattice at low temperature.
Nevertheless, A15 compounds are paradigms of strong-coupling
s-wave superconductors with a high density of electronic states at
$E_F$. Materials like the borocarbides, ${\rm MgB_2}$ or ${2H-\rm
NbSe_2}$ are certainly more complex and correlations or multiband
aspects come into play.

\subsubsection{A15 compounds}

\begin{figure}[b!]
\centerline{\epsfig{figure=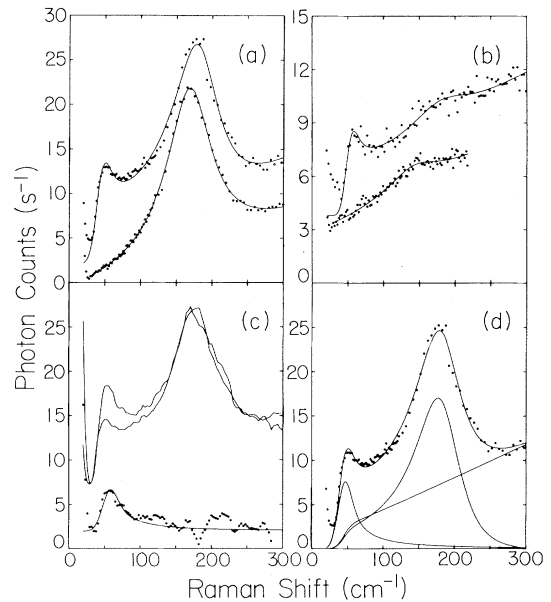,width=7.5cm,clip=}}
\vspace{0.1cm} \caption{Raman spectra of $\rm Nb_3Sn$. Lower
curves in (a) and (b) are at 40~K and upper curves are at 1.8~K.
Data in (c) and (d) are at 1.8~K. Below $T_c = 18$~K the intensity
at low energies is strongly suppressed with respect to the normal
state. Beyond a threshold of approximately $50~{\rm cm^{-1}}$ a
new peak appears. The symmetries are $E_g$ (a,~d), $T_{2g}$ (b),
$E_g + A_{1g}$ (c top), $E_g$ (c middle), and $A_{1g}$ (c bottom).
The $A_{1g}$ data in (c) are obtained by subtracting the middle
from the upper curve. All solid lines except for the upper two in
(c) are theoretical fits to a broadened Maki-Tsuneto function (see
section \ref{Section:sc_theory}). From \textcite{Dierker:1983}.}
\label{fig:Nb3Sn_1}\end{figure}

Superconductivity induced structures close to twice the gap edge
were found in mono-crystalline ${\rm Nb_3Sn}$
(Fig.~\ref{fig:Nb3Sn_1}) and ${\rm V_3Si}$
\cite{Klein:1982a,Dierker:1983,Hackl:1982,Hackl:1983} two years
after the discovery of gap modes in $2H-$NbSe$_{2}$ by
\textcite{Sooryakumar:1980} (section~\ref{sec:NbSe2}) and after an
early but unsuccessful attempt in polycrystalline ${\rm Nb_3Sn}$
by \textcite{Fraas:1970}. For $T < T_c$ the scattering intensity
is redistributed with a suppression below and a pile-up at
approximately $2\Delta \simeq 50~{\rm cm^{-1}}$. The well-defined
peak in $E_g$ symmetry follows the BCS prediction for the
temperature dependence of the gap up to approximately $0.85~T_c$
\cite{Hackl:1983,Hackl:1989}. Somewhat unexpectedly, the peak
frequencies of the superconductivity-induced features depend on
the selected symmetry (Table~\ref{table:A15}). Independent of
minor differences in the absolute numbers stemming from the data
analysis the $E_g$ peaks are significantly below those having
$A_{1g}$ and $T_{2g}$ symmetries. At first glance one could
think of a gap anisotropy to manifest itself. However, there is no
support from the tunneling results which rather indicate the
possible gap anisotropy to be opposite in $\rm V_3Si$ and $\rm
Nb_3Sn$ and very large or from calorimetric studies which should
track the smallest gap (Table~\ref{table:A15}). In addition, the
shapes of the Raman spectra are strongly symmetry dependent in
that the $E_{g}$ peak is much narrower than the others. The
meaning of this anisotropy was a matter of intense discussion.

The results in $A_{1g}$ scattering symmetry in ${\rm Nb_3Sn}$
\cite{Klein:1982a,Dierker:1983} and later in ${\rm V_3Si}$
\cite{Hackl:1988a} demonstrate clearly that the structures below
$T_c$ originate in light scattering from Cooper pairs
(Fig.~\ref{fig:Nb3Sn_1}), since there exist no Raman active
excitations at this symmetry, such as phonons or other bosonic
modes in the A15 structure from which the electrons can borrow
intensity. There is not even an electronic continuum above $T_c$
(see Fig.~\ref{fig:Nb3Sn_1}~(c) and \textcite{Hackl:1988a}). In
spite of similar band structures and densities of states at the
Fermi level, $N(E_{\rm F})$, \cite{Klein:1978} the intensities of
the modes are quite different in the two compounds as is the
overall scattering cross section. For this reason, the weak
$A_{1g}$ mode in ${\rm V_3Si}$ escaped detection for a while
\cite{Hackl:1988a}. Since there is nothing to interact with the
peak frequencies, the $A_{1g}$ structures should be close to the
energy gap in the respective material.  One actually observes
coincidence of both the $A_{1g}$ and $T_{2g}$ Raman energies with
those of bulk methods such as calorimetry and neutrons, while the
$E_{g}$ energies are substantially lower (Table~\ref{table:A15}).

\begin{table}
\begin{tabular}{|c|cccr||ccr|}
\hline

{Sample} & \multicolumn{4}{|c||}{Raman Energy ($\rm cm^{-1}$)} &
\multicolumn{3}{c|}{Reference Data ($\rm cm^{-1}$)} \\
& {$A_{1g}$} & {$E_{g}$} & {$T_{2g}$} & & & & \\
\hline
& & & & & 50 & tunneling & e \\
& & & & & 35--13 & tunneling & f   \\

${\rm Nb_3Sn}$ & \raisebox{1.5ex}[-1.5ex]{~~52~~} &
\raisebox{1.5ex}[-1.5ex]{~~41~~} &
\raisebox{1.5ex}[-1.5ex]{~~50~~} &
\raisebox{1.5ex}[-1.5ex]{a}  &53 & tunneling & g   \\
& \raisebox{1.5ex}[-1.5ex]{67} & \raisebox{1.5ex}[-1.5ex]{48} &
\raisebox{1.5ex}[-1.5ex]{70} &
\raisebox{1.5ex}[-1.5ex]{b}  & 62 & calorimetric & h \\
& & & & & 56 & neutrons & i \\
\hline

& & & & & 37 & tunneling & j \\
& & & & & 40--50 & tunneling & k \\

${\rm V_3Si}$ & \raisebox{1.5ex}[-1.5ex]{-} &
\raisebox{1.5ex}[-1.5ex]{40} & \raisebox{1.5ex}[-1.5ex]{-} &
\raisebox{1.5ex}[-1.5ex]{c} & 46 & IR & l \\

& \raisebox{1.5ex}[-1.5ex]{55} & \raisebox{1.5ex}[-1.5ex]{42} &
\raisebox{1.5ex}[-1.5ex]{52} &
\raisebox{1.5ex}[-1.5ex]{d} & 41 & IR & m \\
& & & & & 49 & calorimetric & h \\
\hline

\end{tabular}
\caption{Gap energies in A15 compounds as measured by Raman
scattering and other methods. a and c refer to results from fits
(see Fig.~\ref{fig:Nb3Sn_1} and section \ref{Section:sc_theory}),
b and d are peak frequencies. In two cases an anisotropy was found
by tunneling being indicated by a range (f and k). The first and
the second numbers are for [100] and [111] directions,
respectively. Results of the following publications are used: a
\cite{Dierker:1983}, b \cite{Hackl:1989}, c \cite{Klein:1984}, d
\cite{Hackl:1988a}, e \cite{Rudman:1984}, f \cite{Hoffstein:1969},
g \cite{Geerk:1984}, h \cite{Junod:1983}, i \cite{Axe:1973}, j
\cite{Moore:1979}, k \cite{Morita:1984}, l \cite{Tanner:1973}, m
\cite{Perkovitz:1976}.} \label{table:A15}
\end{table}

\begin{figure}[b!]
\centerline{\epsfig{figure=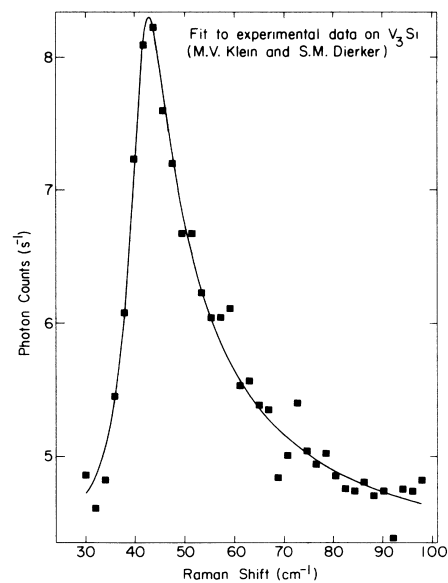,width=6.5cm}} \vspace{0.1cm}
\caption{Raman spectra in $E_g$ symmetry of ${\rm V_3Si}$. From
\textcite{Monien:1990}.} \label{Fig:V3Si_1}\end{figure}

We first note that surface sensitive methods such as tunneling
return somewhat smaller gap energies than bulk methods. Optical
spectroscopy results are also smaller most likely due to surface
treatment. Strain or disorder can indeed reduce $T_c$ in A15
materials since $N(E_F)$ decreases rapidly \cite{Mattheiss:1982}.
Similar reasons might apply for the Raman data in $\rm Nb_3Sn$ of
\textcite{Dierker:1983} although the fits (see
Figure~\ref{fig:Nb3Sn_1}) reveal gap values slightly below
(5--10~\%) the peak positions. Spectra of cleaved surfaces, such as
those of $\rm V_3Si$ and $\rm Nb_3Sn$ taken by
\textcite{Hackl:1988a} and \textcite{Hackl:1989}, respectively,
apparently give gaps closer to the bulk values.

\begin{figure}[b!]
\centerline{\epsfig{figure=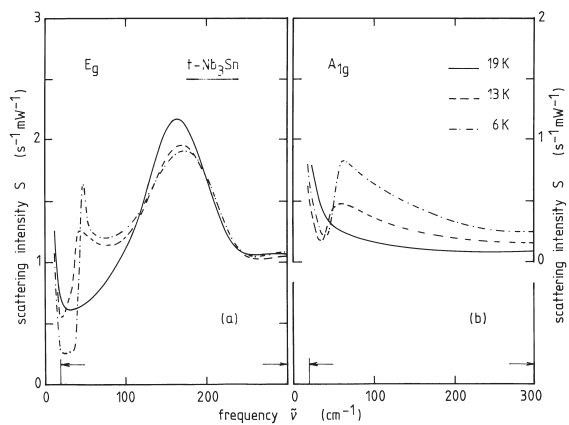,width=7.5cm,clip=}}
\vspace{0.1cm} \caption{Raman spectra of $\rm Nb_3Sn$ at $E_g$ (a)
and $A_{1g}$ (b) symmetries. The integrated spectral weight
(limits indicated by arrows) stays constant to within 3~\% in
$E_g$ symmetry while increasing by a factor of 3 in $A_{1g}$
symmetry on cooling from $T_c$ to 6~K. For clarity the data points
are omitted and only the results of a smoothing procedure are
displayed. The scatter of the data is smaller than 0.1 units
around the lines. From \textcite{Hackl:1989}, reproduced with permission from
Elsevier,~\copyright{~1989}.}
\label{Fig:Nb3Sn_2}\end{figure}

For these reasons, it seems worthwhile to look for other sources of
the anisotropy, and we consider an interpretation in terms of
final state interactions
\cite{Bardasis:1961,Zawadowski:1972,Klein:1984}. This means that
the two single electrons of a broken Cooper pair can still
interact in channels orthogonal to the pairing channel. The
strongly coupled $E_g$ phonon
\cite{Wipf:1978,Schicktanz:1980,Schicktanz:1982,Weber:1984} is in
fact orthogonal to the fully symmetric ($s$-wave) pairing channel.
Hence, it is capable of forming a bound state below the pair-breaking
threshold, explaining both the reduced energy and the linewidth of
the $E_g$ gap mode \cite{Monien:1990}. Fits to the results in $\rm
V_3Si$ are substantially improved by including the bound state
(Fig.~\ref{Fig:V3Si_1}) in comparison to those neglecting it
\cite{Klein:1984}. Additional experimental support comes from the
evolution with temperature of the spectra in $\rm V_3Si$ and $\rm
Nb_3Sn$ \cite{Hackl:1983,Hackl:1989}. In either compound, the
integrated spectral weight in $A_{1g}$ symmetry increases
significantly because a new scattering channel opens up below
$T_c$ due to the formation of Cooper pairs while staying
essentially constant in $E_g$ symmetry because the weight is being
transferred from the phonon to the bound state
(Fig.~\ref{Fig:Nb3Sn_2}).

In contrast to $E_g$ symmetry, the pair-breaking features in
$T_{2g}$ symmetry are weak and essentially at the $A_{1g}$
position. The question arises as to why there is no bound state
although there exists a phonon.  Clearly, the $T_{2g}$ phonon
intensity is weak and the line width is small, reflecting the
moderate coupling as opposed to $E_g$ symmetry where the complete
line width and the asymmetric Fano shape stem from the coupling to
conduction electrons \cite{Wipf:1978,Weber:1984}. The bound
state's energy split-off by approximately 30~\% indicates that the
very strong interaction drives the system close to an instability
of the $s$-wave ground state. On the other hand, the $T_{2g}$ mode
is only weakly coupled and the interaction with the conduction
electrons is not strong enough to substantially renormalize the
spectrum.

Symmetry arguments, the unique line shape and the intensity
transfer in $E_g$ symmetry, as well as the comparison to
calorimetric results, make us believe that the formation of a bound
state is more likely an interpretation of the $E_g$ results in A15
compounds than the manifestation of a gap anisotropy or a two-gap
scenario. In contrast, both effects may cooperate in 2D MgB$_{2}$
discovered to be a superconductor just recently by
\textcite{Nagamatsu:2001}.

\subsubsection{MgB$_{2}$ and the Borocarbides}

\begin{figure}[b!]
\centerline{\epsfig{figure=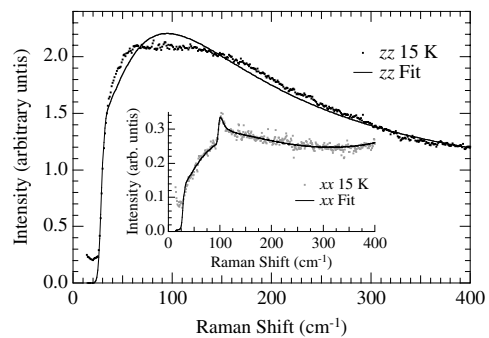,width=7cm}} \vspace{0.1cm}
\caption{Raman spectra for $xx$ and $zz$ polarization geometries
in the superconducting state of MgB$_{2}$ taken by
\textcite{Quilty:2003}. The solid lines are fits to the data using
the theory of \textcite{Devereaux:1992} for disordered $s-$wave
superconductors using a single gap and two gap model for $zz$ and
$xx$ polarizations. The $xx$ spectrum has also been interpreted in
terms of a collective bound state by \textcite{Zeyher:2003}.}
\label{Fig:MgB2}\end{figure}

Electronic Raman studies on MgB$_{2}$ have explored the
superconducting energy gap and changes in phonon lineshapes
occurring below T$_{c}$, starting with the work of
\textcite{Chen:2001} and followed thereafter by
\textcite{Quilty:2002,Quilty:2003}. Here the symmetry dependence
of the response allowed a direct observation of the pairing gap on
the two-dimensional $\sigma$ bands and the 3D $\pi$ bands. By
orienting the light polarizations along the c-axis of MgB$_{2}$
(perpendicular to the hexagonal planes) the $\sigma$ bands cannot
be probed for having little dispersion and thus the $\pi$ bands
are projected out, giving a value $2\Delta_{\pi}=29$ cm$^{-1}$.
Other polarizations are able to detect a larger pairing gap
$2\Delta_{\sigma}=100$ cm$^{-1}$. $zz$-polarized spectra in the
superconducting state are shown in Figure~\ref{Fig:MgB2} along
with fits from the theory for disordered $s-$wave superconductors.

The gap values are consistent with those from other techniques,
yet the fit yields values of the disorder-related scattering rate
different from those of the resistivity by a factor of 2
\cite{Quilty:2003}. \textcite{Zeyher:2003} has reanalyzed the fit
where the direct coupling of light to the $\sigma$ band is zero
and the $\sigma$ gap appears as a result of a coupling to the
Raman active $E_{2g}$ phonon, believed to be largely responsible for pairing. The $xx$ spectrum in the superconducting
state can be understood then as a superposition of a phonon line,
a background, and a collective bound state due to residual
interactions between electrons, similar to that observed in A15
compounds (see Figures \ref{Fig:V3Si_1} and
\ref{Fig:Nb3Sn_2}~(a)).

The superconducting energy gap has also been studied in some
detail in the borocarbide superconductors $R{\rm Ni_{2}B_{2}}C$
($R$=Y, Lu) by \textcite{Yang:2000a,Yang:2000b}. Sharp $2\Delta$
peaks were observed in $A_{1g}$ and $B_{2g}$ symmetries, while the
maximum in $B_{1g}$ symmetry is less pronounced and 20\% higher in
energy. All peaks showed a typical BCS-type temperature dependence
and disappeared above the upper critical field $H_{c2}$. Due to
the high surface quality and improved instrumentation, the residual
scattering intensity below the gap edge is much smaller than,
e.g., in the A15 compounds but finite with an approximately linear
variation with energy.

Since a direct coupling to a Raman active mode was not found in
the borocarbides, it is more complicated than in the A15 compounds
or in $\rm MgB_2$ to sort out whether a gap anisotropy, multi-gap
superconductivity or collective modes are responsible for the
variations in energy and line shape at the different symmetries.
Of course, the existence of bound states is not related to
Raman-active modes; only the experimental verification is more
indirect, e.g., via the line shape (see Figure~\ref{Fig:V3Si_1}).
Similarly, the linear low-energy scattering can suggest either the
presence of strong inelastic scattering due to large coupling
constants $\lambda$ \cite{Allen:1991,Yang:2000b} or gap nodes in
pairing states with lower symmetry, such as $s+g$-wave
superconductivity \cite{Lee:2002}. A quantitative analysis on the
basis of a realistic band structure could possibly help clarify
these, at present, open issues.

While superconductivity was the dominant correlation in the A15
compounds, $\rm MgB_2$ was more complex due to the interplay between
2D and 3D behavior. In the borocarbides magnetic order as a second
instability competing with superconductivity comes into play
\cite{Canfield:1998}. Although the Raman studies were performed on
non-magnetic compounds \cite{Yang:2000a,Yang:2000b} the vicinity
of different types of order is characteristic for this and the
following classes of systems.

\subsection{Charge Density Wave Systems} \label{sec:NbSe2}

The competition or coexistence of different ground states was
studied intensively in layered dichalcogenides in the 1970s and 1980s.
The interest in these charge density wave (CDW)
systems was revived after the discovery of superconductivity in
the cuprates for two reasons: in both compound classes
superconductivity competes with one or more other instabilities,
and, secondly, the dramatically improved instrumentation allowed
qualitatively new and unexpected insights into materials like
$2H-$NbSe$_{2}$. As an example, the electronic scattering rate
$\tau^{-1}$ exhibits marginal \cite{Varma:1989a} rather than Fermi
liquid like temperature and energy dependences (for a discussion
and for references see, e.g., \textcite{CastroNeto:2001}).

\subsubsection{$2H-$NbSe$_{2}$}

$2H-$NbSe$_{2}$ is a layered, though 3D, superconductor with an
in-plane coherence length $\xi_{\parallel}$ of approximately
70~\AA ~and ~$\xi_{\perp} = 25$~\AA ~\cite{Trey:1973}. The
penetration depth for visible light $\delta$ is of the order of
200~\AA, hence $\xi_{\perp} \ll \delta$. The discontinuity at
$2\Delta$ is expected to increase with $2\Delta/(\hbar v_F q)
\approx \delta/\xi$ \cite{Klein:1984}. In addition, the material
can be cleaved easily, facilitating the preparation of atomically
flat surfaces from which diffuse scattering of laser light is
minimized. These are favorable (though not easy!) conditions for
observing gap structures close to the elastic line.

\begin{figure}[b!]
\centerline{\epsfig{figure=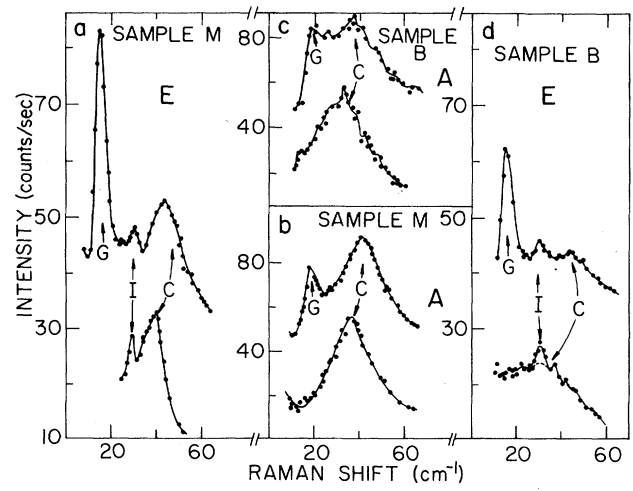,width=1\linewidth,clip=}}
\vspace{0.1cm} \caption{Raman spectra of $2H-$NbSe$_{2}$ above and
below $T_c$. The lower curves of each panel are taken in the
normal state (9~K) the upper ones at 2~K well below $T_c$ with the
sample immersed in superfluid He. At the $A$  ($\parallel -
\perp$) and the $E$ ($\perp$) symmetry the superconducting spectra
are offset by 40 and by 20 counts, respectively. According to the
strength of the CDW mode (labeled by C) Sample B and M have
slightly different impurity concentrations. From
\textcite{Sooryakumar:1980}.} \label{fig:NbSe2_1}\end{figure}

In Fig.~{\ref{fig:NbSe2_1}}, the first observation of the
redistribution of scattering intensity in the superconducting
state of $2H-$NbSe$_{2}$ with the sample immersed in superfluid He
is reproduced. The effect is measured for two samples with
slightly different impurity concentrations. In either case, the
fully symmetric $A$ and the $E$ responses are shown.\footnote{In
the (incommensurate) CDW phase the symmetry representations
$A_{1g}$ etc. of the $D_{6h}$ point group do not apply any more.}
The peaks have slightly different energies, and are located at 18
and 15~{\rm cm$^{-1}$}, respectively, close to the essentially
{\bf k}~independent leading edge gaps found in recent
photoemission experiments \cite{Valla:2004}. In the normal state
at 9~K the new low energy modes are absent, while the maximum
related to the CDW state is still present. As can be seen in all
panels, the CDW mode hardens below the superconducting transition.

\begin{figure}[floatfix]
\centerline{\epsfig{figure=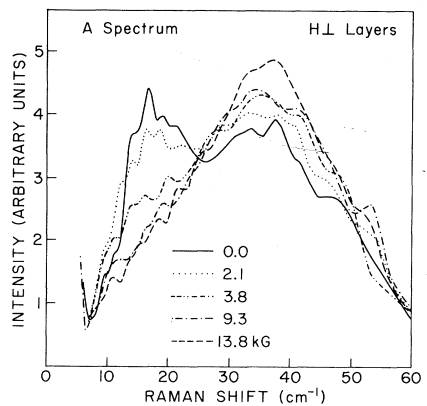,width=6.0cm,clip=}}
\vspace{0.1cm} \caption{Raman spectra in $A$ symmetry of
$2H-$NbSe$_{2}$ (sample B) in various magnetic fields at 2~K. From
\textcite{Sooryakumar:1980}.} \label{fig:NbSe2_2}\end{figure}

The difference between the two samples is apparently the impurity
concentration affecting the strength of both the CDW and the gap
mode. In fact, the CDW transition can be suppressed by either
pressure or an increasing number of defects, which may be
quantified by the residual resistance ratio \cite{Huntley:1974}.
In a systematic study of impurity effects, \textcite{Sooryakumar:1981b} showed that the gap excitations go away along with the CDW mode while the
superconducting transition temperature is essentially unchanged.
It is tempting to assume that the gap modes are directly coupled
to the CDW mode and exist only along with it. This interpretation
is supported by results obtained in a magnetic field
(Fig.~{\ref{fig:NbSe2_2}}). Upon increasing the field the gap
feature in $A$ symmetry is gradually suppressed while the CDW mode
gains intensity leaving the energy integral over the Raman
response $\chi^{\prime\prime}(\omega)$ constant to within 7\%
\cite{Sooryakumar:1981a}. In $E$ symmetry no clear sum rule could
be found \cite{Sooryakumar:1981a}, and it is possible that some of
the gap intensity appears independent of the CDW.

Particularly the result in $A$ symmetry
(Fig.~{\ref{fig:NbSe2_2}}), where the gap mode gains intensity at
the expense of the CDW mode, triggered the theoretical work to
follow.\footnote{For convenience we again give the related
references being discussed thoroughly in the context of collective
modes and in the previous paragraph:
\cite{Balseiro:1980,Littlewood:1981,Littlewood:1982,Browne:1983,Lei:1985,Klein:1984,Monien:1990,Tutto:1992}.
We would like to draw the readers's attention also to the closely
related Raman experiments in superfluid $^{4}$He
\cite{Greytak:1969} and their theoretical description
\cite{Zawadowski:1972}.} The available data are not supportive of
a sum rule in $E$ symmetry (see Fig.~{\ref{fig:NbSe2_1}}~a),
demonstrating similarities with the A15 compounds where also both
electronic scattering and coupling to phonons was observed. In
order to bring some light into the rather involved discussion, it
is worthwhile to reconsider the influence of impurities
\cite{Sooryakumar:1981b}.

At first glance, the reaction to disorder points in the same
direction as the results in magnetic fields. However, defects not
only suppress the formation of the CDW \cite{Huntley:1974} but
also, independently, reduce the intensity close to $2\Delta$
\cite{Devereaux:1992,Devereaux:1993} while normally leaving the
transition temperature $T_c$ of a conventional $s$-wave
superconductor unchanged \cite{Anderson:1959}\footnote{In A15
compounds the high density of electronic states at $E_F$
(partially responsible for the high $T_c$) depends sensitively on
disorder \cite{Mattheiss:1982}. Hence, disorder reduces $T_c$ fast
as opposed to what one would expect from the Anderson theorem.}.
For $\hbar v_F/\delta \ll \Delta,~\hbar/\tau$ the intensity at
$2\Delta$ is proportional to $\tau\Delta$
\cite{Devereaux:1992,Devereaux:1993}. In this respect Raman is
just opposite to the optical conductivity where the gap can be
observed only if $\hbar/\tau$ is of the order of $\Delta$ or
larger \cite{Mattis:1958}. This implies that the Raman gap feature
can be wiped out by impurities while $T_c$ remains essentially
constant; at the same time, though independently, the CDW
transition is suppressed. Hence, it is possible that the gap
features in $2H-$NbSe$_{2}$ exist on their own as pair-breaking
effect but the interaction with the CDW leads to a bound state.

Most of the other CDW systems are not superconducting, but show
very interesting behavior around the transition to the
charge-ordered phase. Some of them have been studied earlier using
light scattering. Here we briefly discuss a recent study of the
temperature-pressure phase diagram of the CDW state.

\subsubsection{$1T$-TiSe$_{2}$}

In $1T$-TiSe$_{2}$ a commensurate CDW is established below $T_{\rm
CDW} \simeq 200$~K. The amplitude of the CDW couples to
zone-boundary acoustic phonons which are folded to the center
below $T_{\rm CDW}$ \cite{Snow:2003}. Pronounced soft-mode
behavior can be observed as a function of temperature. In the
limit $T \rightarrow 0$ two strong lines at 115 and 75~cm$^{-1}$
in $A_{1g}$ and $E_g$ symmetry, respectively, dominate the
low-energy spectra. By increasing the pressure the CDW state first
stiffens along with the lattice then disappears rapidly in the
pressure range of 5 to 25~kbar. Above 25~kbar a quantum disordered
(essentially isotropic) metallic or semi-metallic state is found
although the Raman continuum typical for a metal is not reported.
The quantum mechanical melting of the CDW order is in many ways
similar to classical 2D melting, with the appearance of
crystalline and disordered CDW regimes, as well as an intermediate
``soft'' CDW regime in which the CDW exhibits strong fluctuations
and loses stiffness. Here, measurements on the development and polarization
dependence of the electronic continuum raises the possibility of following
quantum critical behavior in other systems with competing orders.
This is a promising direction for future studies.

\subsection{Kondo or Mixed-Valent Insulators}

All correlations discussed so far are related to
electron-lattice interactions in systems with screening
lengths of the order of the interatomic spacing. With
decreasing electronic density, new phenomena develop originating
from the competition between kinetic and potential energy of the
conduction electrons. Well known examples are the Mott
metal-insulator transition or Wigner crystallization. In either
case, the material becomes an insulator at low temperature due to
``immobilization'' of electrons rather than an energy gap as in
band insulators.

Raman scattering on band insulators and semiconductors has been
well-documented, with a focus placed on high energy charge
transfer excitations. Yet the development of the Raman response at
low frequencies can in principle shed light on the development of
electronic correlations with temperature and/or doping.

\begin{figure}[floatfix]
\centerline{\epsfig{figure=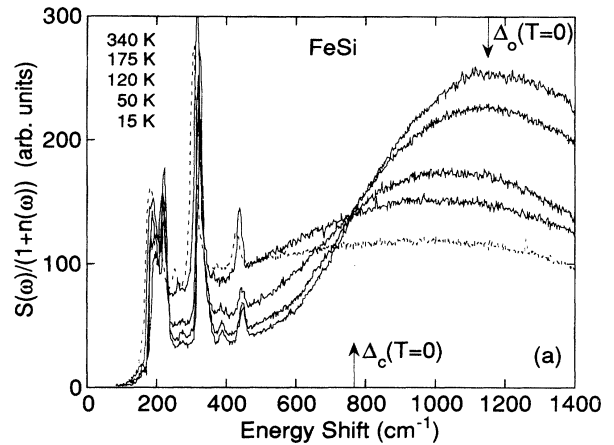,width=1\linewidth,clip=}}
\vspace{0.1cm} \caption{Temperature dependence of the Raman
response measured in FeSi by \textcite{Nyhus:1995a}. Here
$\Delta_{c}$ denote the position of energy gap developing in the
continuum at low temperatures, transferring spectral intensity into
the peak at energy $\Delta_{0}$ . The sharp low energy features
are phonons.} \label{Fig:FeSi}\end{figure}

Experiments by \textcite{Nyhus:1995a} on Kondo  insulating FeSi
shown in Fig.~\ref{Fig:FeSi} and
\textcite{Nyhus:1995b,Nyhus:1997a} on mixed-valent SmB$_{6}$ have
indeed shown the transfer of spectral weight from low to high
energies as the temperature is lowered into the insulating state.
This ``universality'' suggests that there is a common mechanism
governing the electronic transport in correlated insulators. As
these materials are cooled, they all show a pileup of spectral
weight for moderate photon energy losses with a simultaneous
reduction of the low-frequency spectral weight. This spectral
weight transfer is slow at high temperatures and then rapidly
increases as the temperature is lowered towards a putative
quantum-critical point corresponding to a metal-insulator
transition.

A characteristic energy appears which separates the region of
intensity depletion from intensity enhancement with temperature.
This characteristic frequency is essentially independent of
temperature and is thus called an isosbestic point in the spectra
(as shown in Figure \ref{Fig:FeSi}). Finally, it is often observed
that the range of frequency where the Raman response is reduced as
T is lowered is an order of magnitude or more larger than the
temperature at which the low-frequency spectral weight disappears.
As discussed in the Section \ref{Section:resonant_theory} these
findings are consistent with the loss of low-frequency scattering
due to the thermal depletion of excited states.  The
channel-dependence has not yet been a focus of interest. If it
were measured, important information concerning the evolution of
the potential energy with doping could be inferred \cite{Freericks:2005}.

\subsection{Magnetic, charge and orbital ordering:
Raman scattering in Eu-based compounds, Ruthenates, and the
Manganites.}\label{sec:manganites}

There has been a great deal of interest in the relationship
between diverse and exotic low-temperature phases of strongly
correlated systems \cite{Dagotto:2005}. In
particular, the manganites, ruthenates,
Eu-oxides, and hexaborides display charge ordered, paramagnetic insulating, and
ferromagnetic metallic phases as a function of doping, temperature, and/or
temperature. Due to the complex
interplay between spin, charge, and orbital degrees of
freedom, these systems present a
battleground where different ordered phases compete for primacy as
knobs of the Hamiltonian are changed \cite{Imada:1998}.

The phase diagram of strongly correlated materials is more complex
in systems having strong electron-lattice interactions as well as
orbital ordering tendencies. Raman spectroscopy in systems such as
the manganites and ruthenates have provided important information
on the evolution of lattice, charge, and spin dynamics across
phase boundaries. In many cases the transitions can be induced by
applying pressure.

While Raman scattering from phonons has traditionally provided important information
concerning the development of locally or globally symmetry-broken
states accompanied by the formation of static charge ordering,
electronic Raman spectroscopy can
be brought to bear on this problem as it can detect both
fluctuating or static charge and/or spin ordering, and may reflect on
the tendency toward orbital ordering as well. In addition, the
polarization dependence can shed light on the types of excitations
that are created in or near the ordered phases which may serve as
signatures that certain interactions are more prominent than
others. Thus Raman is a powerful spectroscopic method
by which the dynamics across quantum phase transitions can be
investigated in correlated systems.

Recent Raman studies on EuB$_{6}$ \cite{Nyhus:1997b}, Eu$_{1-x}$La$_{x}$B$_{6}$, EuO \cite{Snow:2001}, and Eu$_{1-x}$Gd$_{x}$O \cite{Rho:2002}, show that the
metal-semiconductor transition in these materials is accompanied
by distinct changes of the electronic continuum.  A
high-temperature paramagnetic semimetallic phase is well characterized by
scattering from diffusive charge excitations which become less
diffusive at lower temperatures when correlation effects have not yet set in.
However, the diffusive scattering rate, when fit with
Eq.~(\ref{Eq:Drude-imp}), increases with decreasing temperature and scales with the magnetic susceptibility as
the system begins to develop short-range magnetic order, typical
of insulating behavior, as shown in Figures~\ref{Fig:isos}
and~\ref{Fig:FeSi}. Finally, at low temperatures, a ferromagnetic
metallic phase occurs, showing a flat continuum characteristic of
a strongly correlated metal. The doping, polarization, and
magnetic field dependence of the spectra implies that the
metal-semiconductor transition is precipitated by the formation of
bound magnetic polarons above the ferromagnetic ordering
temperature.

Concerning the ruthenates, recently \textcite{Snow:2002} and \textcite{Rho:2003} have studied
the evolution of the spin and lattice dynamics through the
pressure-tuned collapse of the antiferromagnetic Mott-like phases
of Ca$_{2}$RuO$_{4}$, Ca$_{3}$Ru$_{2}$O$_{7}$, and
Ca$_{2-x}$Sr$_{x}$RuO$_{4}$ into a ferromagnetic and possibly
orbitally ordered metallic state at low temperatures. The studies
have shown many characteristic features resulting from the
interplay of strong electron-lattice and electron-electron
interactions. These include (i) evidence of an increase of the
electron-phonon interaction strength, (ii) an increased
temperature dependence of the two-magnon energy and linewidth in
the antiferromagnetic insulating phase, (iii) evidence of a charge
gap development significantly below the metal-insulator transition
(T$_{MI}$), and (iv) a hysteresis associated with the structural
phase change. The latter two effects are indicative of a
first-order metal-insulator transition and a coexistence of
metallic and insulating components for $T<T_{\rm MI}$. The
measurements have not yet been extended to probe the
unconventional superconducting state at low temperatures.

Raman measurements on cubic and layered manganites
have been used to explore the interplay of spin, charge and
orbital degrees of freedom. \textcite{Yamamoto:2000} and
\textcite{Romero:2001} observe a suppression of the low-energy
continuum at $B_{1g}$ symmetry upon entering the charge- and
orbital-ordered state. The interpretation is not yet settled.
Possible candidates are spin density or dynamical charge-orbital
fluctuations \cite{Yamamoto:2000} or a collective CDW excitation
\textcite{Romero:2001}. The controversy can probably not be solved
without a quantitative theoretical description.

\textcite{Bjornsson:2000} have measured cubic
La$_{1-x}$Sr$_{x}$MnO$_{3}$ and via phonon lineshape analysis have
shown strong electron-phonon interactions involving local lattice
distortions in the high temperature paramagnetic state which
gradually vanish below the ferromagnetic transition. A broad hump
in the electronic spectra develops at low temperatures in the
metallic state around 400 cm$^{-1}$ for x=0.2, and weakens in
intensity and shifts to higher energies for x=0.5. Although a
polaronic peak would weaken with increasing carrier density, the
shift towards higher energies with doping was interpreted rather
as a low-energy plasma excitations within the ferromagnetic
metallic phase. Since recent ARPES studies on the same compound
\cite{Mannella:2004} and bi-layer manganite \cite{Mannella:2005}
have revealed coexistence of quasiparticle and polaron features in
the metallic phase and do not show evidence for low energy plasma
excitations, more work is needed to clarify this issue.

A comprehensive study via reflectance and Raman measurements on
Pr$_{0.7}$Pb$_{0.21}$Ca$_{0.09}$MnO$_{3}$,
La$_{0.64}$Pb$_{0.36}$MnO$_{3}$,
La$_{0.66}$Pb$_{0.23}$Ca$_{0.11}$MnO$_{3}$, and
Pr$_{0.63}$Sr$_{0.37}$MnO$_{3}$, by \textcite{Yoon:1998}, have
shown that the electronic continuum displays a change from
diffusive polaronic peaks at high temperature to a flat
featureless continuum, similar to that observed in the cuprates,
in the low temperature ferromagnetic phase. A broad polaronic peak
around 1200 cm$^{-1}$ shifts to lower energies with increases
doping indicative of weakened polaron binding energies. This is
consistent with a crossover from a small-polaron-dominated regime
at high temperatures to a large-polaron-dominated low-temperature
regime. The low temperature phase also provided evidence for the
coexistence of large and small polarons, also in agreement with
the ARPES results.

\begin{figure}[b!]
\centerline{\epsfig{figure=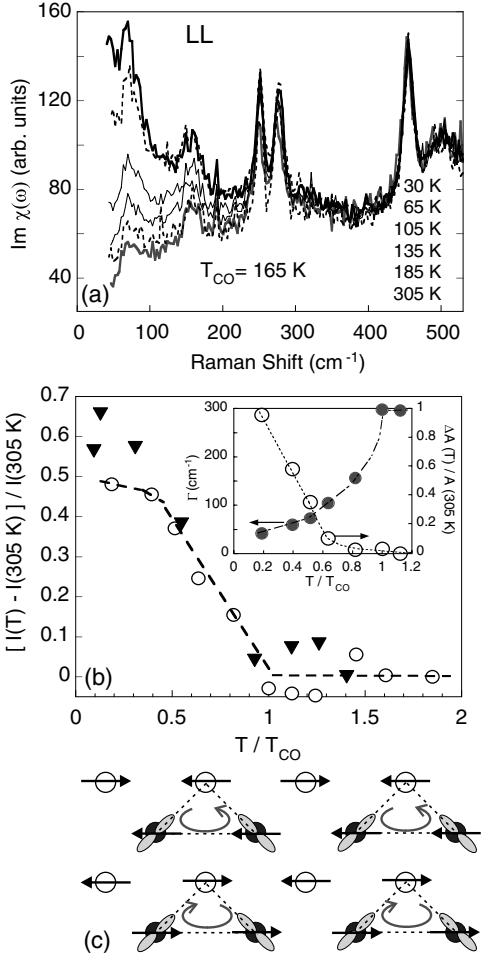,width=7.5cm,clip=}}
\vspace{0.1cm} \caption{(a) Temperature dependence of the Raman
spectra in the LL scattering geometry in
Bi$_{0.19}$Ca$_{0.81}$MnO$_{3}$ (b) Fractional change in the
integrated quasielastic Raman scattering intensity (50 -
350~cm$^{-1}$) as a function of $T/T_{\rm co}$ for samples with
charge-ordering temperatures $T_{\rm co} = 165$~K (circles) and
210~K (triangles). Inset: Fractional change in the quasielastic
scattering amplitude $A$ and fluctuation rate $\Gamma$ as a
function of $T/T_{\rm co}$ obtained via a fit of the data with
Eq.~(\ref{Eq:Drude-imp}). Lines are guides to the eye. (c) Example
of a closed-loop path for charge motion in the charge-ordered
phase ($x = 0.5$) which is not precluded by either the orbital
configuration or by the spin environment. Filled and empty circles
represent Mn$^{3+}$ and Mn$^{4+}$ sites, respectively. From
\textcite{Yoon:2000}.} \label{Fig:Yoon}\end{figure}

Recent work on Bi$_{1-x}$Ca$_{x}$MnO$_{3}$ $(x < 0.5)$ by \textcite{Yoon:2000}
is of particular interest in connection with polarization studies for the
evidence of charge ordering. As shown in Figure
\ref{Fig:Yoon}, Raman scattering offers a unique means of probing
the unconventional spin and/or charge dynamics that arise when
charge carriers are placed in the complex spin environment of a
charge-ordered systems. Using circularly (LL) in addition to
linearly ($xx$ and $yy$) polarized light anti-symmetric components
of the Raman tensor were isolated. In cubic crystals such as
Bi$_{1-x}$Ca$_{x}$MnO$_{3}$ they transform as the $T_{1g}$
irreducible representation (equivalent to $A_{2g}$ in tetragonal
materials like the cuprates). Upon entering the charge-ordered
phase a quasielastic scattering response appears with the $T_{1g}$
symmetry of the spin-chirality operator. Thus it was conjectured
that the chiral excitations were signatures of either a chiral
spin-liquid state associated with the Mn core spins, or of
closed-loop charge motion caused by the constraining environment
of the complex orbital and N\'eel textures. A possible path for
charge motion is shown in Figure \ref{Fig:Yoon}, emphasizing the
circular nature of charge transfer. It is remarkable, that the
spectral shape and the temperature variation of the characteristic
energy are quite similar to the low-energy response in the cuprates
(see section~\ref{sec:fluctuations}) although a state with static
order is entered in Bi$_{1-x}$Ca$_{x}$MnO$_{3}$. It is interesting
and perhaps important how the two types of response are related.

Recently \textcite{Saitoh:2001} have performed Raman measurements
on detwinned and orbitally ordered LaMnO$_{3}$ and have observed
multiple peak structures which they interpret as orbital
excitations or ``orbitons''. While this has been challenged by
\textcite{Grueninger:2002} on the basis of selection rules, more
recent measurements by \textcite{Krueger:2004} related the peak
features to second order phonon scattering activated via the
Franck-Condon mechanism \cite{Perebeinos:2001}. Even if we could
only scratch this interesting subject we hope we could
demonstrate, that Raman measurements continue to be of merit to
study the interplay of strong correlations and electron-phonon
coupling and the novel excitations which emerge in orbitally
ordered systems.

In this section we have shown how the Raman spectra evolve as the
degree of correlations increases in different materials. One
common aspect is the non-trivial polarization and temperature
dependence of the spectra which emerge in materials with
increasing complexity. Finally, it was shown that Raman scattering
can be applied to materials with varying degrees of competition
between ordered states. Nowhere is this more apparent than in the
results on the cuprates with high superconducting transition
temperature. Due to the large amount of work devoted to these
materials and the complexities of the issues raised, we split off
the discussion of cuprates and related compounds into the
following separate section.

\section{HIGH TEMPERATURE SUPERCONDUCTING CUPRATES}
\label{sec:HTSC}

\begin{figure*}[floatfix]
\centerline{\epsfig{figure=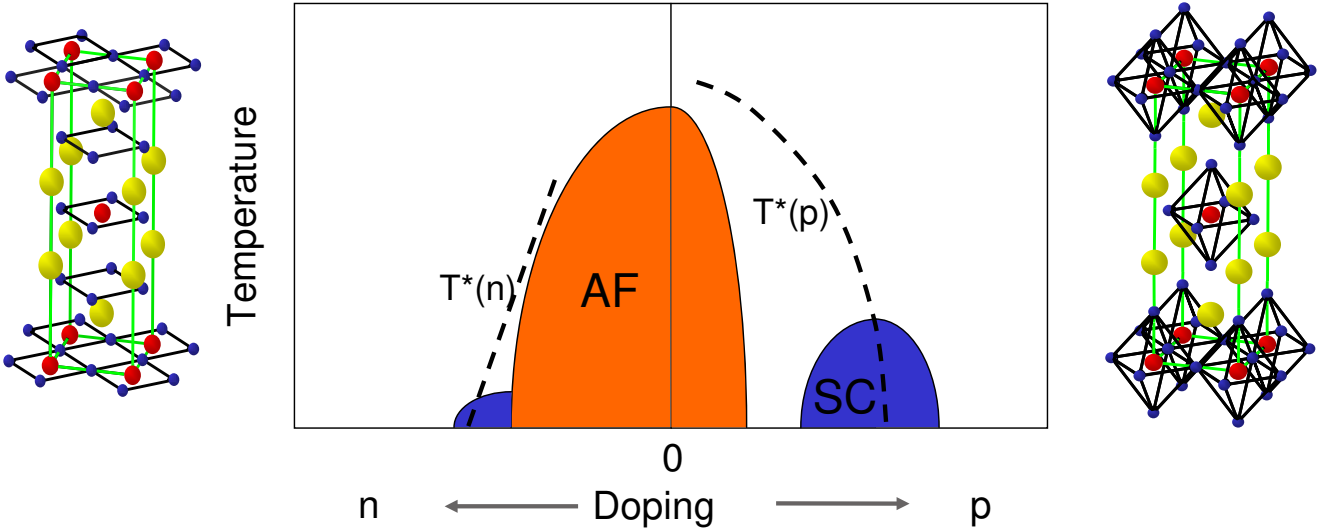,width=1\linewidth,clip=}}
\caption{Schematic phase diagram of the cuprates. On the
hole-doped ($p$) side long-range antiferromagnetic (AF) order
disappears rapidly. The maximal superconducting (SC) transition
temperature $T_c$ is strongly material dependent but always
observed at $p\simeq 0.16$. On the electron-doped ($n$) side the
AF phase is more extended. $T_c$ does not exceed 30~K at $n \simeq
0.14$. $T^{\ast}$ represents the approximate crossover temperature
to the pseudogap regime \cite{Timusk:1999}. On the l.h.s and the
r.h.s. the structures of prototypical $\rm Nd_{2-x}Ce_xCuO_4$ and
$\rm La_{2-x}Sr_{x}CuO_4$, respectively, are shown. The atoms are:
Cu - red, O - blue, La,Sr and Nd,Ce - yellow. All cuprates are
tetragonal or close to tetragonal with small material-dependent
deviations. $\rm Nd_{2-x}Ce_xCuO_4$ crystallizes in $T^{\prime}$
structure without O~octahedra and a slightly shorter $c$-axis.
$\rm La_{2-x}Sr_{x}CuO_4$ and all other hole-doped materials have
octahedra which are cut into half for materials with more than one
$\rm CuO_2$ plane.} \label{Fig:phase-diagram}\end{figure*}

The story of the cuprates began with the discovery of
superconductivity by \textcite{Bednorz:1986} in $\rm
La_{2-x}Ba_{x}CuO_4$ with  $x \simeq 0.1 \dots 0.2$. Soon after
the confirmation by \textcite{Cava:1987}, $\rm YBa_2Cu_3O_7$ with
a $T_c$ above 90~K was synthesized \cite{Wu:1987}. These
unexpected results led to further discoveries of superconductivity
in materials with a layered crystal structure with one or more
CuO$_{2}$ planes per unit cell (Fig.~\ref{Fig:phase-diagram}).
Materials synthesis has yielded compounds with increasingly higher
$T_{c}$, and major advances have been made in single-crystal growth
methods,\footnote{Recent work can be found in
\cite{Erb:1996a,Hardy:1993,Liang:2000,Liang:2002,Eisaki:2004,Onose:2001,Ando:2004}}
providing a diversity of samples. Independent of the material
family and its respective maximal $T_c$, superconductivity exists
for doping levels $0.05 < p < 0.27$ and $0.12 < n < 0.18$ with $p$
and $n$ the number of holes and electrons, respectively, per
plaquette ($\rm CuO_2$ formula unit) \cite{Loram:2001,Onose:2001}
as shown schematically in Figure~\ref{Fig:phase-diagram}. $n=0=p$
denotes half filling.

It became clear soon thereafter that the cuprates are doped Mott
insulators with strong electronic correlations dominating the
entire phase diagram (Fig.~\ref{Fig:phase-diagram}). The emergence
of high temperature superconducting phases in materials from which
strong correlations yield antiferromagnetism and large departures
from a canonical Fermi liquid theory has highlighted our
limited understanding of electronic correlations. While overdoped
systems seem to display a behavior of the resistivity close to
$T^{2}$ at low temperature and well-defined quasiparticles in
ARPES studies, strong deviations already occur for optimally doped
systems and become increasingly pronounced as the
antiferromagnetic phase is approached. At optimal doping, where
$T_c$ is maximal, the materials have already high normal state
resistivities and, hence, are canonical examples that bad metals
make good superconductors. Yet we still do not have a theoretical
framework to understand why.

Upon underdoping, the cuprates develop strong electronic
anisotropies in the $\rm CuO_2$ planes, which can be thought of as
a signature of correlations. For this reason, momentum resolution
is crucial for understanding the physical properties. ARPES
has been very important from the beginning, revealing, among many
other things, a strong {\bf k}~dependence of both the
superconducting energy gap and the pseudogap in the normal state
\cite{Damascelli:2003,Campuzano:2002}. As a more subtle effect,
the quasiparticle weight $Z_{\bf k}$ and the incoherent part of
the spectral function were observed to have a substantial
variation with {\bf k} and doping $p$, implying the importance of
correlation effects and the existence of strongly
momentum-dependent interactions \cite{Damascelli:2003}. More
recent studies of scanning tunneling microscopy (STM) have
indicated a presence of nanoscale disorder \cite{McElroy:2005} in
addition to the strong anisotropies identified from ARPES studies
\cite{Damascelli:2003}. Both tools have combined to give insight
of the tendencies these compounds have towards electronic ordering
before the antiferromagnetic phase is reached.\footnote{A
selection of references is
\cite{Howald:2003,Hoffman:2002,Vershinin:2004,Hanaguri:2004,Damascelli:2003}.}
This demonstrates directly that the simultaneous understanding of
both single- {\it and} two-particle response functions is
important.

In this context, electronic Raman scattering has played a major
role in characterizing the anisotropic dynamics of electrons
across the phase diagram. These include the intense study of
antiferromagnetism, where Raman measurements on the parent
insulating cuprate compounds were the first to yield an estimate
of the magnetic exchange $J$ from the energy of the two-magnon
scattering peak in the $B_{1g}$ channel \cite{Lyons:1988}. The
discovery of the broad and flat electronic continuum in the normal
state of the cuprates close to optimal doping \cite{Bozovic:1987}
became the signature of the anomalous and strange metallic phase
at high temperatures and spawned important ideas by
\textcite{Varma:1989a} which honed in on the physics at play in
these materials. The polarization dependence of this background
above \cite{Staufer:1990,Slakey:1991} as well as below $T_{c}$
\cite{Hackl:1988b,Cooper:1988b,Slakey:1990b} visualized strong
anisotropic interactions in these materials. Specifically, the
observation of a polarization-dependent gap opening in the spectra
below T$_{c}$ was instrumental in solidifying the symmetry and
orientation of the superconducting order parameter. Finally, more
recent work has focused on elucidating the behavior of competing
orders and complexity which come hand-in-hand with strong
correlations.

It is safe to say that no other system has been studied so
intensely via a bevy of experimental tools in recent years as the
cuprates. This is certainly true for Raman spectroscopy, where the
light scattering cross sections have been analyzed in a rich
number and quality of materials. In this section, we discuss Raman
results on the cuprates and related materials, with an overview of relating findings to
the physics uncovered in other systems. We show issues in which
consensus has been reached and other issues which are
controversial and require further analysis.

\subsection{From a Doped Mott Insulator to a Fermi Liquid}

\begin{figure*}[floatfix]
\centerline{\epsfig{figure=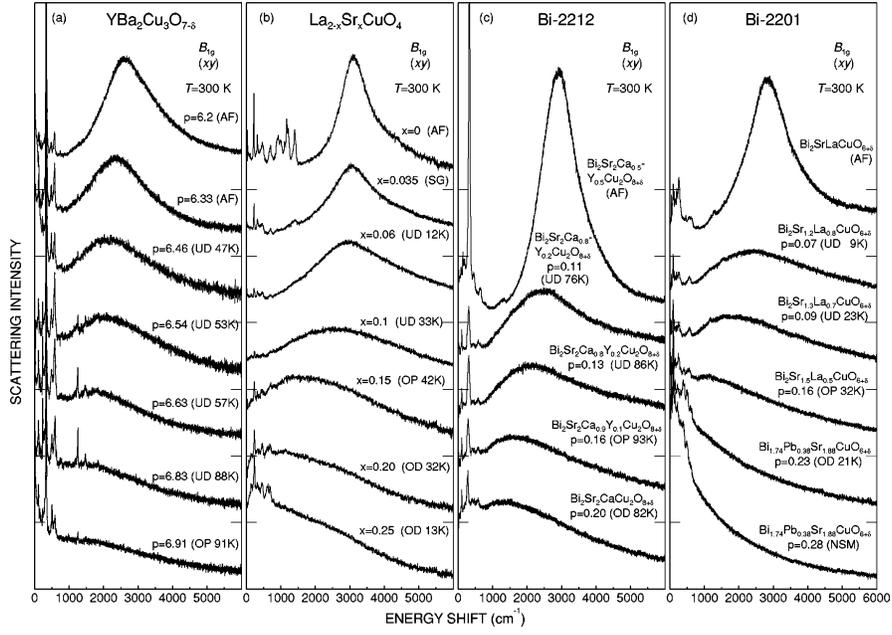,width=12cm,clip=}}
\vspace{0.1cm} \caption{Doping dependence of the two-magnon
$B_{1g}$ spectra in a number of compounds at 300 K. From
\textcite{Sugai:2003}.} \label{Fig:SugaiPRB2003}\end{figure*}

The so-called parent compounds of the cuprate superconductors are
antiferromagnetic Mott insulators. At half filling, charges are
completely localized and excited states are separated by the
Coulomb energy $U$ prohibiting double occupancy in the Hubbard
model.\footnote{The subject has been reviewed in detailed articles
and books \cite{Fulde:1995,Gebhard:1997}}  The ground state is a
Heisenberg antiferromagnet which develops 3D long-range order
below a N\'eel temperature of typically 300~K. The excitations are
spin waves with a very small (in-plane) anisotropy gap at ${\bf
q}=0$ and a nearest-neighbor exchange coupling $J$. The spin waves
have been studied right at the beginning by Raman scattering, and
$J$ was determined from the two-magnon peak to be of order
100~meV.\footnote{\cite{Lyons:1988,Lyons:1989,Sulewski:1990,Sulewski:1991,Knoll:1990,Sugai:1988,Ruebhausen:1997,Blumberg:1996}}
Very early the existence of chiral spin excitations, i.e.
excitations where the spins are rotated out of the easy ($\rm
CuO_2$) plane, was demonstrated \cite{Sulewski:1991}. The
resonance profile\footnote{A selection of references is
\cite{Lyons:1988,Knoll:1996,Blumberg:1996,Ruebhausen:1997}.} has
been described in terms of the triple resonance theory of
\textcite{Chubukov:1995a,Chubukov:1995b}. Very
recently, one magnon excitations at ${\bf q}=0$ were observed in
the N\'eel state of $\rm La_{2-x}Sr_{x}CuO_4$ \cite{Gozar:2004}
and related to Dzyaloshinskii-Moriya and $XY$ optical modes
resulting from a spin gap by \textcite{SilvaNeto:2005}. The
experiments are the ${\bf q}=0$ and $\hbar\Omega \rightarrow 0$
manifestation of canted spins which give also rise to scattering
in the two-spin-flip channel at large energies in $A_{2g}$
symmetry \cite{Sulewski:1991}. Here, a lot more information, such
as the Dzyaloshinskii-Moriya vector, the size of the anisotropy
gap, and the spin-lattice interaction strength, could be derived
\cite{Gozar:2005b}.

If the insulator is doped off half filling, antiferromagnetic long
range order disappears quickly with hole doping but survives up to
higher electron doping levels (Figure~\ref{Fig:phase-diagram}).
Yet short ranged antiferromagnetic correlations are not quenched
even at high doing levels, as seen by the persistence of the
two-magnon peak in $B_{1g}$ Raman scattering experiments on both
the hole-doped
\cite{Sugai:2003,Reznik:1993,Ruebhausen:1999,Blumberg:1994} and
the electron-doped side \cite{Onose:2004} of the phase diagram.
This is summarized in Figure~\ref{Fig:SugaiPRB2003} for a number
of compounds and shows how the two-magnon peak softens and
broadens with doping, eventually merging into the continuum at
higher doping levels.  Since the two-magnon intensity is dominated
by a double spin flip of nearest neighbors, it can be observed even
for small magnetic correlation lengths $\xi_{\rm m}$ of the order of
few lattice constants (see Figure~\ref{Fig:2_magnon}). The
two-magnon peak persists at least up to optimal doping for
hole-doped systems (Fig.~\ref{Fig:SugaiPRB2003}). The position of
the maximum and the peak intensity decrease by factors of roughly
2 and 20, respectively, for $0 \leq p \leq 0.16$. Consistent with
the Raman results, neutron scattering experiments in $\rm
La_{2-x}Sr_xCuO_4$ revealed magnetic excitations in the complete
superconducting range \cite{Wakimoto:2004}.

\begin{figure}[b!]
\centerline{\epsfig{figure=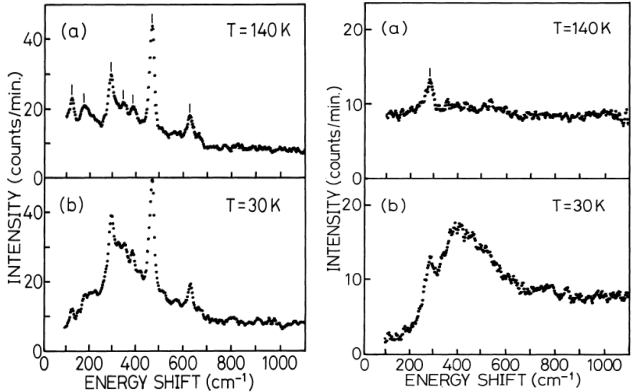,width=1\linewidth,clip=}}
\vspace{0.1cm} \caption{Raman spectra in the normal state (a)  and
superconducting state (b) for $A_{1g}+B_{2g}$ (left panel) and
$B_{1g}~(+A_{2g})$ (right panel) orientations in $\rm
Bi_2Sr_2CaCu_2O_{8+\delta}$. From \textcite{Yamanaka:1988}.}
\label{Fig:Yamanaka}\end{figure}

The broadening of the two-magnon response with doping and
temperature has been studied theoretically on clusters for the
$t-J$ model \cite{Prelovsek:1996} and by quantum Monte Carlo (QMC)
techniques \cite{Sandvik:1998}, respectively.
\textcite{Knoll:1990} analyzed their experiments at elevated
temperatures in terms of spin-lattice coupling. While many features of the data
at 1/2 filling for undoped cuprates
can be captured by studies of $H_{ELF}$ Eq. (\ref{Eq:ELF}), in general the
evolution of the magnon lineshape with doping and temperature and
its full polarization dependence is not very well understood. In
particular, it is an unsettled issue, how, when a more metallic
state develops for higher doping levels, the magnon line merges
into the relatively featureless continuum (shown in Figures
\ref{Fig:SugaiPRB2003} and \ref{Fig:Yamanaka}) that extends
well-beyond all relevant energy scales such as $\hbar qv_{\rm F}$,
the superconducting energy gap $\Delta$ or the maximal phonon
energy $\hbar \omega_{\rm D}$. In strongly overdoped yet
superconducting samples ($0.20 < p < 0.27$), the physical properties
in the normal state are still not those of a conventional metal.
The continuum itself displays significant polarization and doping
dependence, which will be addressed in section~\ref{sec:HTSC_NC}.

For $p>0.05$ ($n>0.12$), superconductivity emerges and reaches a
maximal $T_c$ at approximately $p=0.16$ ($n=0.14$). In the Raman
spectra, superconductivity induced peaks emerge out of the flat
continuum in the normal state, accompanied by spectral weight
reorganization for temperatures below T$_{c}$, as shown for $\rm
Bi_2Sr_2CaCu_2O_{8+\delta}$ in Figure~\ref{Fig:Yamanaka}.

In the following section we focus on the polarization dependence
of the Raman response below $T_c$ in both hole- and electron-doped
cuprates. We summarize how symmetry arguments can be used to
obtain the momentum dependence of electronic properties in general
and, specifically, that of the energy gap $\Delta({\bf k})$. The
doping dependence in hole-doped systems is postponed and will be
discussed in detail in section~\ref{sec:doping}.

\subsection{Superconducting Energy Gap and Symmetry}
\label{sec:HTSC_SC}

Early Raman results for the electronic continuum showed only
little difference between the normal and the superconducting
states \cite{Lyons:1987}. We know now that the relatively high
defect concentration in the first samples suppressed the
structures related to the gap. The synthesis of flux-grown $\rm
YBa_2Cu_3O_7$ with a sufficiently small number of defects
\cite{Kaiser:1987,Schneemeyer:1987} improved the situation
rapidly, and a clear indication of the pair-breaking effect was
obtained by \textcite{Cooper:1988a} on $\rm YBa_2Cu_3O_7$, showing
the emergence of a peak and reorganized spectral weight occurring
for temperatures below T$_{c}$ as expected from theory. Soon
thereafter a strong polarization dependence of the Raman spectra
was observed (Fig.~\ref{Fig:Yamanaka}), which can be considered the
first spectroscopic evidence of a gap anisotropy
\cite{Hackl:1988b,Cooper:1988b,Yamanaka:1988,Slakey:1990a}. In
contrast to conventional materials (see, e.g.,
Fig.~\ref{fig:Nb3Sn_1}), there is no sharp onset of the scattering
intensity at a threshold. As an explanation of the continuous
increase of the scattering intensity at small frequencies the
possibility of nodes was discussed early
\cite{Hackl:1988b,Monien:1989}. Originally shown in $\rm
Bi_2Sr_2CaCu_2O_{8+\delta}$ and $\rm YBa_2Cu_3O_7$, the studies
were quickly extended to other optimally hole doped cuprates. It
was shown by \textcite{Kang:1996,Kang:1997} that magnetic
fields suppress the peaks, independently indicating their relationship to superconductivity.

\subsubsection{Symmetry: $B_{1g}$ and $B_{2g}$}

The polarization dependence of both the peak frequencies and the
low-energy slopes of the superconducting spectra had been a vexing
problem for several years following the first observation of the
reorganized spectral weight and the polarization-dependent
response. As shown in Figure~\ref{Fig:Yamanaka}, the peak in the
$B_{1g}$ response developed at roughly 30 percent higher energies
than for $B_{2g}$ or $A_{1g}$ polarization geometries, and the low
frequency spectra rose as $\Omega^{3}$ for $B_{1g}$ and linearly
with $\Omega$ for other channels. In addition, the temperature
dependent depletion at low frequency shifts was faster in $B_{1g}$
symmetry than in other channels.

The satisfactory description emerged as soon as the momentum
dependences of the Raman vertices of different symmetries $\mu$,
$\gamma_{\mu}({\bf k})$, and of the energy gap $\Delta({\bf k})$
were properly taken into account, as outlined in Section
\ref{Section:sc_theory}. As indicated by a few experiments before
\cite{Hardy:1993,Shen:1993} and corroborated by many more later
\cite{Scalapino:1995}, a gap with $d_{x^{2}-y^{2}}$ symmetry is the
most compatible with the results in the cuprates
\cite{Devereaux:1994a}.

\begin{figure}[b!]
\centerline{\epsfig{figure=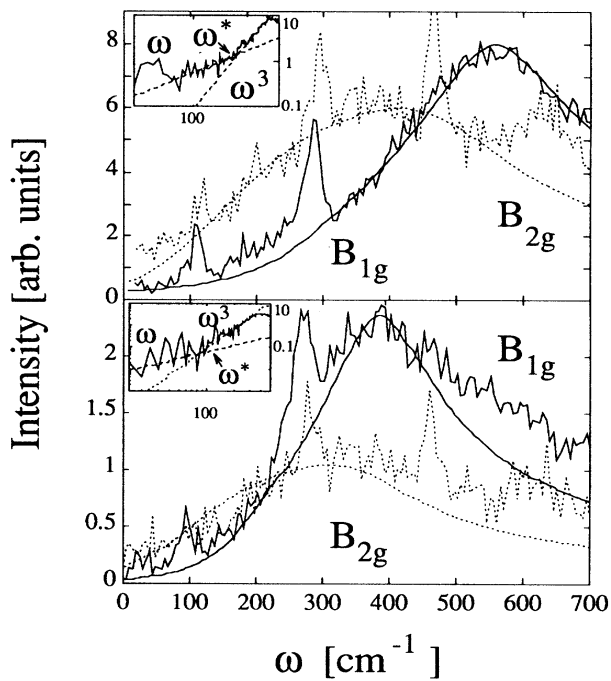,width=1\linewidth,clip=}}
\vspace{0.1cm} \caption{Fit to the $B_{1g}$ and $B_{2g}$ data on
as-grown (T$_{c}$=86~K, top panel) and oxygen-annealed
(T$_{c}$=79~K, bottom panel) $\rm Bi_2Sr_2CaCu_2O_{8+\delta}$ from
\textcite{Staufer:1992}. The inset is a log-log plot of the low
frequency $B_{1g}$ response, showing a cross-over from linear to
cubic behavior at a characteristic frequency
$\omega^{\ast}/\Delta_{0}$=0.38 and 0.45 for the top and bottom
panels, respectively. From \textcite{Devereaux:1995b}.}
\label{Fig:dirty-dwave}\end{figure}

The predictions of $d-$wave theory - such as frequency power-laws,
temperature dependences, and relative peak positions - were found
to be consistent with many optimally hole-doped compounds, such as
YBa$_{2}$Cu$_{3}$O$_{7}$ \cite{Devereaux:1995a,Chen:1994a}, $\rm
Bi_2Sr_2CaCu_2O_{8+\delta}$
\cite{Devereaux:1994a,Devereaux:1995a,Blumberg:1997a}, $\rm
La_{2-x}Sr_xCuO_4$ \cite{Chen:1994b}, $\rm Tl_2Ba_2CuO_6$
\cite{Devereaux:1995a,Kang:1996,Kang:1997,Blumberg:1997b}, $\rm
Bi_2Sr_2CuO_{6+\delta}$ \cite{Einzel:1996}, $\rm
Tl_2Ba_2CaCu_2O_8$ \cite{Maksimov:1990,Kang:1997}, $\rm
Tl_2Ba_2Ca_2Cu_3O_10$ \cite{Hoffmann:1994}, and later $\rm
HgBa_2Ca_2Cu_3O_{8+\delta}$ \cite{Sacuto:1998,Sacuto:2000}, $\rm
Bi_2Sr_2Ca_2Cu_3O_{10+\delta}$ \cite{Limonov:2002a} and $\rm
HgBa_2CuO_{4+\delta}$ \cite{Gallais:2004} The identification of
twice the maximal gap from the peak in $B_{1g}$ channels yielded
$2\Delta_{\rm max} \simeq 8~k_BT_c$ for all compounds, well above
the weak coupling values of $4.28~k_BT_c$ for $d$-wave pairing,
indicating the strong-coupling nature of the pair
state\footnote{Due to this high ratio, we note however that in certain cases
\cite{Zeyher:2002,Martinho:2004} the B$_{1g}$ peak is thought not
to be related directly to a $2\Delta$ feature. However we note that
$2\Delta/k_{B}T_{c} \sim 10$ emerges from strong coupling $d-$wave
treatments \cite{Monthoux:1994}, and is consistent
with the broadening of the $B_{1g}$ signal observed near 2$\Delta$.}.

An additional confirmation of $d-$wave pairing came from the
impurity effects on the Raman spectra in the superconducting state
\cite{Devereaux:1995b,Misochko:1999,Limonov:2002b}. Raman
scattering, like other types of responses (ARPES, infrared and
tunneling spectroscopy, NMR, etc.) couples only to the magnitude
and not to the phase of the order parameter and, hence, cannot
discriminate between a $d_{x^{2}-y^{2}}$ and $|d_{x^{2}-y^{2}}|$
gap, i.e. the sign change of the $d$-wave gap is not accessible.
Impurities however could distinguish between energy gap which were
conventional yet had accidental nodes from a pure $d-$wave energy
gap \cite{Borkowski:1994}. While peaks in the $B_{1g}$ channel are
generally suppressed by impurities, the theory for Raman scattering
in disordered $d-$wave superconductors predicted a crossover from
linear to cubic frequency dependence at a characteristic frequency
$\omega^{*}$ set by the impurity concentration, while a true
threshold would develop if the nodes were accidental
\cite{Devereaux:1995b}. Shown in Fig.~\ref{Fig:dirty-dwave}, along
with the improvement of sample quality and instrumentation,
the crossover from a linear to a cubic frequency dependence was
found to directly demonstrate the influence of impurities and
to further solidify the $d-$wave picture for hole-doped cuprates,
as cemented by SQUID measurements, reviewed in Refs.
\cite{Harlingen:1995,Tsuei:2000}\footnote{ We emphasize again that
disorder rapidly suppresses gap features in the Raman spectra (see
section~\ref{Section:sc_theory}). Even in a $d$-wave
superconductor where $T_c$ reacts sensitively to impurities the
peaks disappear long before $T_c$ vanishes. Hence, the
characterization of the samples is a central issue. Doping
generally introduces defects along with spins or carriers such as
Ni in the $\rm CuO_2$ plane or Sr in $\rm La_{2-x}Sr_xCuO_{4}$ and
clusters of oxygen in $\rm YBa_2Cu_3O_{6+x}$
\cite{Pekker:1991,Erb:1996b} suppressing the gap structures
rapidly.}.

Additional attributes have been intensely investigated, such as
the agreement of the effective mass approximation calculated
within LDA for $d-$wave superconductors in $\rm YBa_2Cu_3O_7$
\cite{Strohm:1997}, effects of orthorhombicity and mixing of
$s-$wave components allowed by symmetry
\cite{Strohm:1997,Nemetschek:1998}, as well as the issue of the
$A_{1g}$ peak seen in the superconducting state \cite{Krantz:1994},
which is among the complicated though eventually crucial problems
on the way to a better understanding of the cuprates.

\subsubsection{The $A_{1g}$ problem -- Zn, Ni and Pressure}
\label{Section:a1g}

While the $B_{1g}$ and the $B_{2g}$ spectra are in reasonable
agreement with the predictions on the basis of $d$-wave pairing
various problems arose in $A_{1g}$ symmetry which slowed
down the acceptance of the model in the beginning. One of the
objections was the strong intensity found in the $A_{1g}$ channel.
As discussed in Section \ref{Section:sc_theory}, the $A_{1g}$
response is complicated due to the backflow terms which are as
singular as the bare terms for frequencies at the gap edge, leading to
cancellations of diverging intensities. For
both a cylindrical Fermi surface and a tight-binding band
structure \cite{Krantz:1994,Devereaux:1994b}, the
effective mass approximation predicts a suppression of the
$A_{1g}$ intensity in comparison to other channels, in contrast to
what is found experimentally, as shown in Figures
\ref{Fig:Yamanaka} and \ref{Fig:Bi2223}. Since the effective mass
approximation is of questionable utility in highly correlated
systems like the cuprates, this objection was not as serious, as
an overall comparison of spectral intensities is not possible without
detailed knowledge of the Raman scattering matrix elements given
in Eq.~(\ref{M_oper}). Good fits to the data of the $A_{1g},
B_{2g}$ and $B_{1g}$ symmetries (with the overall intensities as
free parameters) could be obtained \cite{Devereaux:1995a}.
However, a sensitivity to band structure and higher harmonics of
the energy gap and Raman vertices found in the calculations
implied a similar sensitivity to details of the materials which
was not observed in the data.

Further, it was argued by
\textcite{Krantz:1994} that scattering in a multi-band system -
such as in the CuO$_{2}$ bilayer - would yield a sharp intensity
at twice the gap edge, which was again inconsistent with the data
on single- and double-layer systems available at that time.

The issue of multiband scattering was partially resolved by
\textcite{Devereaux:1996}.
It was found that if the energy gaps were equal on at
least two sheets of the Fermi surface split by bi-layer hopping, a
diverging intensity would be possible. However the intensity is
only proportional to the difference of the individual Raman
vertices of the bands. This is a qualitative reason why
the multi-band case would give peaks only under quite special
conditions, implying that the fits obtained from the single band case
are still valid. Yet, it does not satisfactorily explain the
sensitivity of the $A_{1g}$ response to band and energy gap
anisotropy factors.

\begin{figure}[b!]
\centerline{\epsfig{figure=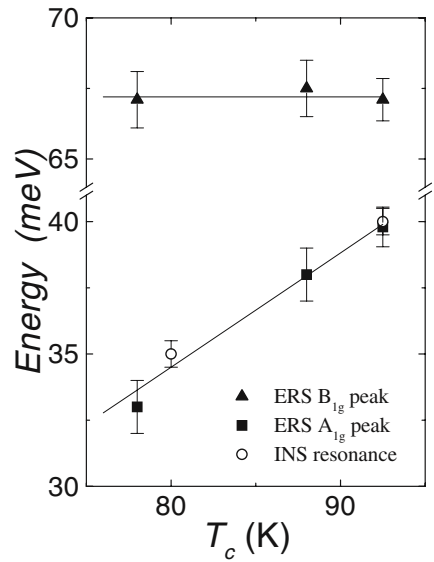,width=6.5cm,clip=}}
\vspace{0.1cm} \caption{Raman and neutron-resonance peak energies
as a function of the critical temperature T$_{c}$ in $\rm
YBa_2(Cu_{1-x}Zn_x)_3O_{6.95}$ for x = 0, 0.01, and 0.03. The
horizontal line for the $B_{1g}$ peak positions is just a guide to
the eye, while the $A_{1g}$ peak and the neutron resonance are
fitted by a straight line representing $\rm 5k_{B}T_{c}$. From
\textcite{Gallais:2002}.} \label{Fig:Gallais}\end{figure}

A recent undertaking to understand the origin of the $A_{1g}$
intensity has involved the response of the peak to partial
replacement (up to a few percent) of Cu by Zn and/or Ni in $\rm
YBa_2Cu_3O_{7-\delta}$
\cite{Martinho:2004,Gallais:2002,Limonov:2002b,LeTacon:2005}.
\textcite{Martinho:2004} found that the intensity of the $A_{1g}$
peak was insensitive to either Ni or Zn doping, while the peak in
$B_{1g}$ is suppressed by Zn, in agreement with earlier results
\cite{Limonov:2002b}. As shown in Figure~\ref{Fig:Gallais},
\textcite{Gallais:2002} found that the peak position in $A_{1g}$
was reduced by Ni impurities and made the important observation
that it followed that of the magnetic spin resonance mode
\cite{Sidis:2004}. \textcite{LeTacon:2005} found that Zn doping
increased the low frequency spectral weight and suppressed the
$B_{1g}$ peak without changing its position. It is not clear why
the position of the $B_{1g}$ peak does not change while $T_c$
decreases. Possible explanations are an inhomogeneous distribution
of the impurities or an accidental cancellation effect between the
decreasing $2\Delta(T_c)$ and an increase of the peak energy
$\hbar\Omega_{\rm peak}^{B_1g}$ because of defects
\cite{Devereaux:1995b}. The shift of the $A_{1g}$ peak with Zn
argued for a two-component picture including a contribution from a
collective mode such as the $\pi$~resonance in the spin wave
spectrum \cite{Hinkov:2004}.

Two approaches concerning the presence of a collective mode in the
spectra superimposed upon a well-screened background have been put
forward recently by \textcite{Venturini:2000} and
\textcite{Zeyher:2002}. \textcite{Venturini:2000} showed that a
collective mode, uniquely appearing in $A_{1g}$ and not other channels due
to symmetry, emerges
due to coupling to the 41 meV spin resonance mode seen in neutron
scattering measurements in a number of compounds
\cite{Sidis:2004}. \textcite{Zeyher:2002} argued that the peak in
the $B_{1g}$ channel may be identified as a collective mode split off
from $2\Delta$ \cite{Blumberg:1997a} as a consequence of the
simultaneous presence of long-range $d$-CDW and
$d$-superconducting order and that the $A_{1g}$ peak may be
related to a superconducting amplitude mode. In either case the
sensitivity to parameters characterizing the anisotropies of
energy gap, band structure, and Raman vertices were not present,
yet it is still unclear which, if either, are able to explain the
$A_{1g}$ peak in the superconducting state.

\begin{figure}[b!]
\centerline{\epsfig{figure=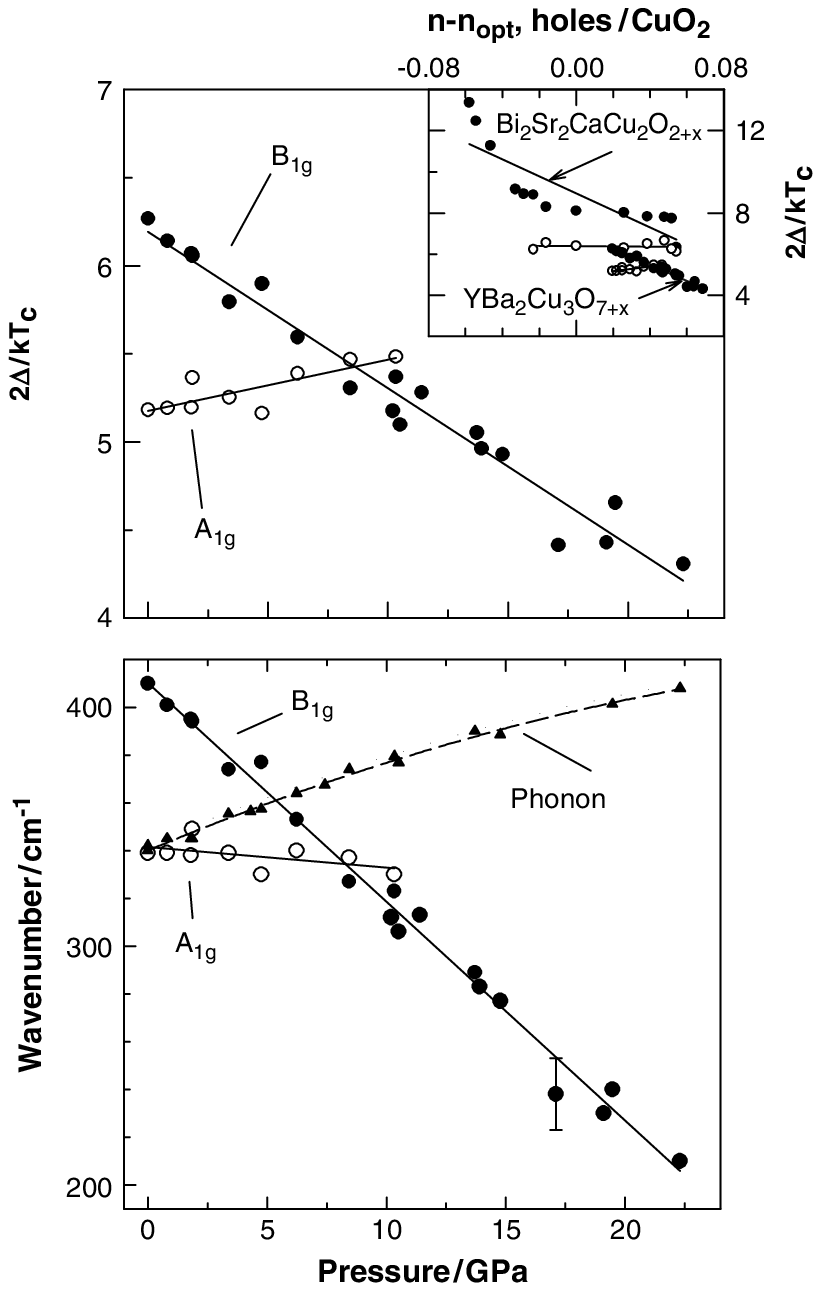,width=8.0cm,clip=}}
\vspace{0.1cm} \caption{Pressure dependences of the
superconducting peaks in $\rm YBa_2Cu_3O_{6.95}$ for $B_{1g}$
(full circles) and $A_{1g}$ (open circles) symmetries. The
position of the $B_{1g}$ phonon is also indicated (triangles). The
inset shows a comparison to the Raman data on
$\rm Bi_2Sr_2CaCu_2O_{8+\delta}$ samples at ambient pressure with different doping
levels \cite{Kendziora:1995}. From \textcite{Goncharov:2003}. Reproduced with permission from Wiley ~\copyright{~2003}.}
\label{Fig:Goncharov}\end{figure}

Yet a remarkable set of experiments by \textcite{Goncharov:2003} plotted in Figure~\ref{Fig:Goncharov} highlights some shortcomings in the above mentioned scenarios. Upon pressure the
peak in the B$_{1g}$ channel and $T_{c}$ move downward in a similar fashion as upon overdoping (discussed in detail in Section \ref{sec:doping}), and the $B_{1g}$ phonon hardens as expected, while the peak
position in $A_{1g}$ remains remarkably constant. This implies
that the $B_{1g}$ peak is intimately tied to superconductivity and
the $A_{1g}$ peak has a substantial contribution from a channel
which is relatively insensitive to pressure and changes in
superconducting properties. In the discussions of possible
candidates for the mode highlighted above, the pressure
sensitivity of spin, charge, and superconducting order would all be expected to be large. Thus the origin of this peak remains presently unclear. One can speculate
that it may be related to possible phonon modes involving mixed
Bi-O and Ba-O which have been indicated to be less sensitive to
pressure, but clearly further work is needed to clarify this
matter \cite{Goncharov:2003}.

In multi-layer compounds with more than 2 adjacent CuO$_{2}$
planes, a strong $A_{1g}$ peak in the superconducting state
occurs at roughly the peak frequency of the $B_{1g}$ channel, as
shown in Figure \ref{Fig:Bi2223} for $\rm
Bi_2Sr_2Ca_2Cu_3O_{10+\delta}$. In addition, strong phonon
resonances were found here and for multilayer Hg compounds \cite{Hadjiev:1998}, where some $A_{1g}$ c-axis phonons shift by as much as 20 cm$^{-1}$ at the
superconducting transition, more than in any other high T$_{c}$
compound, implying a strong renormalization of the $A_{1g}$
continuum below T$_{c}$.

Recently, \textcite{Munzar:2003} argued that a $c-$axis plasma mode
would be expected to be Raman active in cuprate materials with
more than 2 CuO$_{2}$ planes, giving rise to an additional
contribution in the $A_{1g}$ channel only due to mass fluctuations
with opposite sign on different CuO$_{2}$ layers. The position of
the plasma resonance was predicted to lie in a frequency range
close to the continuum peak, as shown in Figure~\ref{Fig:Bi2223}.

\subsubsection{Resonance Effects} \label{sec:resonance}

\begin{figure}[floatfix]
\centerline{\epsfig{figure=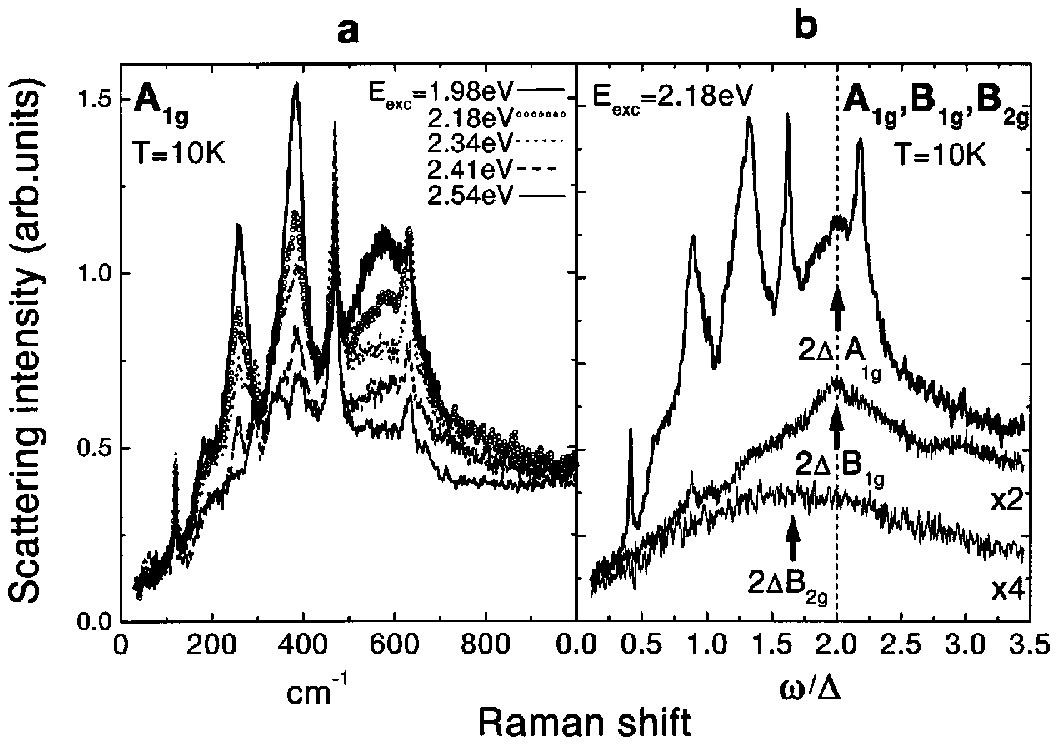,width=1\linewidth,clip=}}
\vspace{0.1cm} \caption{Resonance enhancement of the of the
$A_{1g}$ Raman response in $\rm Bi_2Sr_2Ca_2Cu_3O_{10+\delta}$ at
10~K. From \textcite{Limonov:2002a}.}
\label{Fig:Bi2223}\end{figure}

As another experimental knob to turn, resonance studies of the
peaks developing in the superconducting state provide information
on the character and the interactions of the quasiparticles
forming the condensate. For this reason, studies of the gap
feature as a function of the incoming photon energy were pursued.
Effects in the energy range of the gap were observed in various
compounds with one \cite{Kang:1996,Blumberg:2002} more than one
$\rm CuO_2$ layer
\cite{Hadjiev:1998,Ruebhausen:1999,Limonov:2002a,Sacuto:1998,Budelmann:2005}.
There are two distinct, though not necessarily independent,
effects: (i) the spectral weight and the shape of some phonons of
predominantly $A_{1g}$ symmetry change more or less dramatically
below $T_c$ \cite{Hadjiev:1998,Limonov:2002a}, and (ii) the
intensity of the electronic continuum is amplified in the vicinity
of $2\Delta_{\rm max}$ (Fig.~\ref{Fig:Bi2223})
\cite{Ruebhausen:1999,Limonov:2002a,Blumberg:2002}. The resonance
effects typically occur for excitation energies of $2\pm 0.2$~eV
and are particularly strong in compounds with 3 or 4 $\rm CuO_2$
layers \cite{Limonov:2002a}.\footnote{It is interesting to note
that the resonance of the two-magnon peak is close to the charge transfer gap of
2.5~eV or higher
\cite{Knoll:1996,Blumberg:1997a,Ruebhausen:1997}.} Importantly,
the normal state is only weakly affected. Hence, although the
resonance energy is close to the 1.6~eV absorption edge
\cite{Uchida:1991,Singley:2001}, constraints are imposed on
explanations in terms of interband transitions between the lower
and the upper Hubbard band (c.f., Eq.~(\ref{Eq:SS_matrix}) and
\textcite{Shastry:1990}), since the width of the resonance is much
larger than $2\Delta_{\rm max}$. At least in the triple-layer
compounds, an additional channel can be opened up below $T_c$ due to the c-axis plasma resonance \cite{Munzar:2003} which may interact with low-energy phonons and renormalize their shape and intensity substantially. The
existence of a resonating continuum in double-layer compounds is
still a matter of debate. Some authors find the pair-breaking
peaks to resonate \cite{Ruebhausen:1999,Blumberg:1997a,Budelmann:2005}, while
others do not \cite{Venturini:2002c,Limonov:2002a}.

Given this background, the strong resonance effects in
electron-doped $\rm Nd_{2-x}Ce_xCuO_4$ were a quite interesting
and somewhat unexpected feature \cite{Blumberg:2002}. While the
$B_{2g}$ response in the superconducting state is strongly
enhanced toward the red, neither the $B_{1g}$ spectra below $T_c$
nor the normal state spectra in general are particularly sensitive to the
energy of the exciting light. An explanation in terms of Hubbard
physics is certainly tempting but, for the symmetry
and temperature dependence, not completely exhaustive. In our
opinion, the subject needs further experimental and theoretical
clarification before arriving at a level of predictive power.

After the observation of resonance effects in $\rm
Nd_{2-x}Ce_xCuO_4$ and along with an improved material quality the
number of studies in electron-doped systems increased continuously
and facilitated several interesting insights which will be
summarized in the following paragraph.

\subsubsection{Electron versus Hole-Doped Materials}

Some cuprates such as $\rm Nd_{2-x}Ce_xCuO_4$ crystallize in the
$T^{\prime}$ structure which is characterized by missing oxygen
octahedra (see Figure~\ref{Fig:phase-diagram}) and, as a
consequence, a short $c$-axis \cite{Tokura:1989}. The maximal
$T_c$ of approximately 30~K is obtained in thin films of $\rm
La_{2-x}Ce_xCuO_4$ \cite{Naito:2000}. None of the electron-doped
cuprates is completely ordered, and oxygen appearing in an apex
position is the main defect \cite{Radaelli:1994} even in
superconducting samples.

\begin{figure}[b!]
\centerline{\epsfig{figure=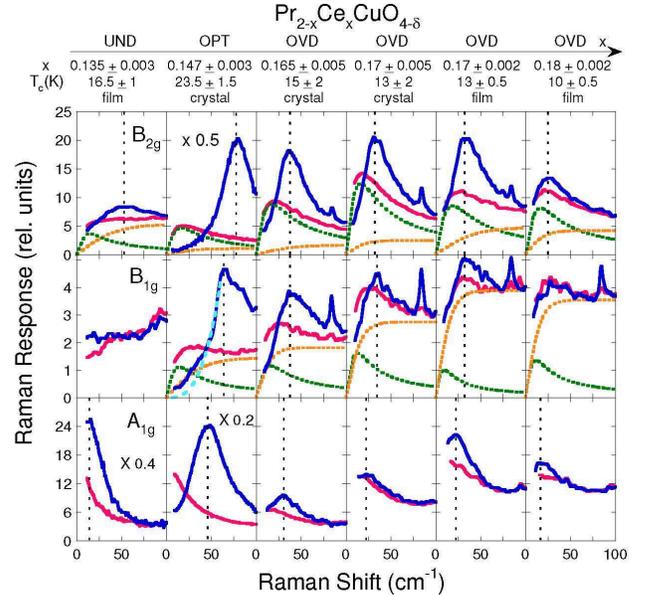,width=8.5cm,clip=}}
\vspace{0.1cm} \caption{Doping dependence of the low energy
electronic Raman response of $\rm Pr_{2-x}Ce_{x}CuO_4$  single
crystals and thin films for $B_{2g}$, $B_{1g}$ and $A_{1g}$
channels obtained with 647~nm excitation. The columns are arranged
from left to right in order of increasing cerium doping.
Abbreviations UND, OPT and OVD refer to under-doped, optimally
doped and over-doped samples, respectively. The normal state
response (light/red) measured just above the respective $T_c$ is
decomposed for the $B_{2g}$ and $B_{1g}$ channels into a
Drude-like component with a constant carrier lifetime (green
dotted line) and an extended continuum (yellow dotted line).
Superconducting spectra (dark/blue) are taken at $T \approx 4$~K.
For the OPT crystal a low-frequency $\omega^3$ power law is shown
at $B_{1g}$ symmetry. From \textcite{Qazilbash:2005}.}
\label{Fig:PCCO}\end{figure}

Phonons \cite{Heyen:1991} and crystal-field excitations
\cite{Jandl:1993} in $\rm Nd_{2-x}Ce_{x}CuO_4$ were studied soon
after the discovery \cite{Tokura:1989} and explained thoroughly.
An exception was the $A_{1g}$ line at $590~{\rm cm^{-1}}$,
assigned by \textcite{Heyen:1991} as a localized mode of
interstitial oxygen in apex position (cf.
Fig.~\ref{Fig:phase-diagram}). Later on, \textcite{Onose:1999}
could indeed demonstrate that the mode is suppressed after
annealing the samples in reducing atmosphere in order to induce or
enhance $T_c$. The first electronic Raman spectra in the
superconducting state of slightly overdoped $\rm
Nd_{1.84}Ce_{0.16}CuO_4$  showed a small gap anisotropy
\cite{Stadlober:1995}. The ratio $4.1 \leq 2\Delta/k_BT_c \leq
4.9$ is similar to that of strong-coupling conventional
superconductors like Pb, Nb or $\rm Nb_3Sn$ (see
Table~\ref{table:A15}). The temperature dependence of the gap is
relatively close to the BCS prediction
\cite{Stadlober:1995,Blumberg:2002}. This motivated an
interpretation in terms of an anisotropic $s$-wave gap, yielding a
reasonable description of both the shape and the positions of the
$B_{1g}$ and $B_{2g}$ pair-breaking peaks \cite{Stadlober:1995}.

Similar spectral shapes were also found at other doping levels for
both $\rm Nd_{2-x}Ce_{x}CuO_4$ and $\rm Pr_{2-x}Ce_{x}CuO_4$, as
shown in Figure~\ref{Fig:PCCO}. The low-energy sides of the peaks
were found to be almost doping independent and closer to those of
hole-doped cuprates than to those in conventional superconductors
when samples and instrumentation facilitated improved measurements
close to $\Omega = 0$.

Recent ARPES
\cite{Armitage:2001,Matsui:2005} and interferometric experiments
\cite{Tsuei:2000,Chesca:2003} provided evidence of a $d$-type gap
bringing antiferromagnetic spin fluctuations back into play as a
possible coupling mechanism. Since the diameter of the Fermi
surface encircling $(\pi,\pi)$ is smaller here than in hole-doped
systems, the antiferromagnetic ordering vector ${\bf
Q}_{AF}=(\pi,\pi)$ connects spots close to $(\pi/2,\pi/2)$ rather
than in the vicinity of $(\pi,0)$ for $p$~doping and the gap is
expected to have a maximum at the hot spots and to decrease
towards $(\pi,0)$.

On this basis \textcite{Blumberg:2002} proposed a novel type of non-monotonic
$d$-wave gap with maxima only approximately 15$^{\circ}$ away from
the diagonal. An explicit calculation using the suggested
$\Delta({\bf k})$ \cite{Blumberg:2002} in a one-band model
approximately reproduces the overall line shapes but reveals
discrepancies in the peak positions \cite{Venturini:2003} (see
also \textcite{Blumberg:2003}). Better agreement between
experiment and theory can be obtained with a monotonic $d-$wave
gap in a two band picture \cite{Liu:2005} for which evidence has
been found by magnetotransport \cite{Fournier:1997}.

Four remarks are to be considered: (i) In
ARPES, the gap maximum is found approximately at the hot spot
where a pseudogap is observed above $T_c$ \cite{Matsui:2005}. (ii) No quasiparticle peaks indicating coherence in the superconducting state are resolved in ARPES \cite{Armitage:2001,Matsui:2005}. (iii) Twice the gap
energy observed by ARPES is $\sim 40$\% smaller than that derived from the Raman spectra \cite{Matsui:2005}. (iv) Phase-sensitive
experiments \cite{Alff:1999} and results on the magnetic
penetration depth \cite{Kokales:2000,Skinta:2002a} are supportive of
an $s$-type gap.

Hence in spite of mounting evidence of $d$-wave pairing, explanations for the symmetry dependence of the Raman spectra and several other experiments are still missing. A possible explanation could lie in a crossover from
$d$ to $s$ pairing upon increasing doping \cite{Skinta:2002b}. However,
Raman data from differently doped $\rm
Pr_{2-x}Ce_xCuO_4$ (see Figure~\ref{Fig:PCCO}) do not show a
variation of the lineshape and sufficiently strong symmetry
dependence of the pair-breaking features in the proper doping
range \cite{Qazilbash:2005}.

In the hole-doped systems, on the other hand, strong variations of
the lineshapes at $A_{1g}$ and $B_{1g}$ symmetry are found
although there is little doubt about the persistence of the
(dominant) $d$-wave nature of the superconducting gap in the
entire phase diagram \cite{Tsuei:2000}. The doping and
polarization dependence of the Raman spectra in superconducting
$p$-type cuprates will be the subject of the next section.

\subsection{Superconducting Gap: Doping Dependence}\label{sec:doping}

\begin{figure}[b!]
\centerline{\epsfig{figure=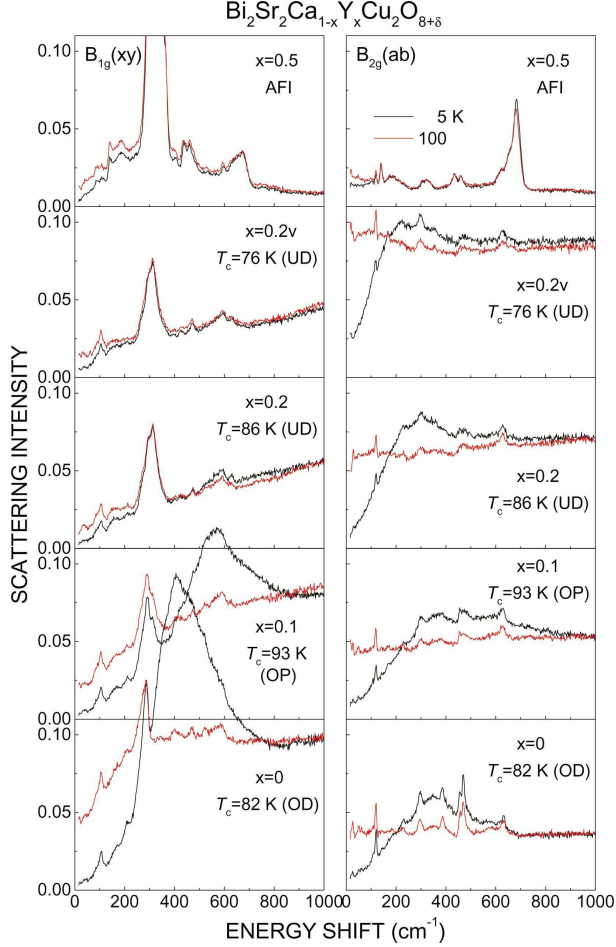,width=1\linewidth,clip=}}
\vspace{0.1cm} \caption{Doping dependence of the $B_{1g}$ and
$B_{2g}$ Raman spectra in Bi-2212 at 5 K and 100 K. From
\textcite{Sugai:2003}.} \label{Fig:Sugai}\end{figure}

One of the early Raman experiments on doping effects in overdoped
$\rm Bi_2Sr_2CaCu_2O_{8+\delta}$ showed the energy of the
pair-breaking peak in $B_{1g}$ symmetry to decrease much faster
than those at the other symmetries and $T_c$ \cite{Staufer:1992}.
The first systematic study was performed by
\textcite{Kendziora:1995} on $\rm Bi_2Sr_2CaCu_2O_{8+\delta}$, shown in the inset of Figure \ref{Fig:Goncharov}. The
doping level in this material can be varied continuously and
reliably between $0.15 \leq p \leq 0.23$ by changing the oxygen
concentration \cite{Triscone:1991,Kendziora:1993}. At lower oxygen doping,
the structure is possibly metastable or unstable, at higher doping
the oxygen diffuses out even at room temperature. Underdoping is better achieved by replacing Ca with Y. As an essential
result, the pair-breaking peaks at $A_{1g}$, $B_{1g}$, and $B_{2g}$
symmetry are found to depend in different ways on doping and,
consequently, on $T_c$. This doping dependence is shown in Figure \ref{Fig:Sugai} for
recent measurements on $\rm Bi_2Sr_2Ca_{1-x}Y_xCu_2O_{8+\delta}$ by \textcite{Sugai:2003}.
Most remarkably, the $B_{1g}$
pair-breaking peak is proportional to $(1-p)$ rather than $T_c$ in
the doping range indicated, but fades in intensity for underdoped materials.
The peak energies in $B_{2g}$ symmetry however scale more or less with the transition temperature. For
$p<0.15$ the $B_{1g}$ peak becomes very weak, yet a clear 2$\Delta$ peak
survives in the $B_{2g}$ channel for all doping levels.

\begin{figure}[b!]
\centerline{\epsfig{figure=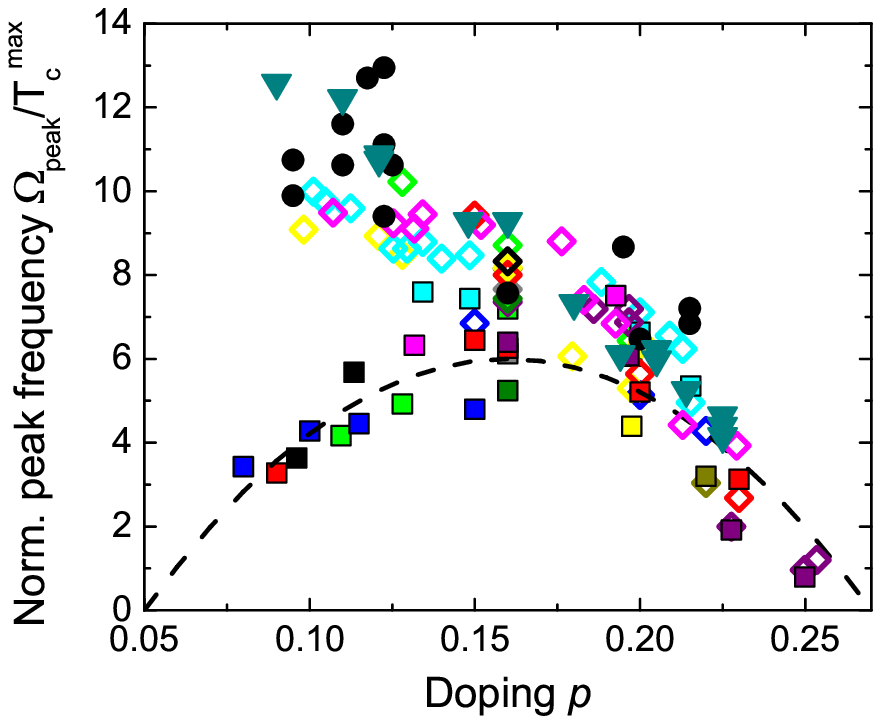,width=1\linewidth,clip=}}
\vspace{0.1cm} \caption{Compilation of the position $\Omega_{\rm
peak}$  of the peak in the Raman response in the superconducting
state normalized to $T_{c}^{\rm max}$. $B_{1g}$ (open diamonds)
and $B_{2g}$ (squares) orientations are shown for a variety of
compounds. The respective references are color coded as follows:
Bi-2212 [red \cite{Venturini:2002c}, green
\cite{Sugai:2000,Sugai:2003}, cyan \cite{Kendziora:1995}, yellow
\cite{Liu:1999,Blumberg:1997a}, pink \cite{Hewitt:2002}, and gold
\cite{Masui:2003}], LSCO [blue \cite{Sugai:2003}], Bi-2223 [grey
\cite{Masui:2003}], Tl-2201 [purple
\cite{Gasparov:1997,Gasparov:1998,Nemetschek:1993,Kang:1997}],
Tl-2223 [olive \cite{Hoffmann:1994,Stadlober:1994}], and Hg-1201
[black \cite{Gallais:2005}]. The dashed black line is the
interpolation formula $6T_{c}/T_{c}^{\rm
max}=6[1-82.6(p-0.16)^{2}]$. For comparison, results for twice the
maximal leading edge gap of ARPES (circles) \cite{Campuzano:2002} and
for the peak-to-peak energy in the tunneling density of states
(triangles) \cite{Zasadzinski:2002} are included.}
\label{Fig:Delta}\end{figure}

In the decade to follow there were numerous studies on differently
doped cuprates, including $\rm Bi_2Sr_2CaCu_2O_{8+\delta}$,\footnote{\cite{Hackl:1996,Blumberg:1997a,Liu:1999,Ruebhausen:1999,Opel:2000,Sugai:2000,Hewitt:2002,Venturini:2003,Sugai:2003,Budelmann:2005}}
$\rm YBa_2Cu_3O_{6+x}$,\footnote{\cite{Cooper:1988b,Nemetschek:1997,Chen:1993,Altendorf:1992,Reznik:1993,Opel:2000,Sugai:2003,Limonov:1998,Limonov:2000,Masui:2005}}
$\rm La_{2-x}Sr_xCuO_{4}$,\footnote{\cite{Chen:1994b,Naeini:1999,Venturini:2002a}} $\rm HgBa_2CuO_{4}$,\footnote{\cite{LeTacon:2005,Gallais:2005}} $\rm HgBa_2Ca_2Cu_3O_{8}$,\footnote{\cite{Sacuto:1998,Sacuto:2000}} $\rm Tl_2Ba_2CuO_{6}$,\footnote{\cite{Nemetschek:1993,Kang:1996,Gasparov:1997,Blumberg:1997b}}
$\rm Tl_2Ba_2CaCu_2O_{8}$,\footnote{\cite{Kang:1997}} and $\rm
Tl_2Ba_2Ca_2Cu_3O_{10}$~\footnote{\cite{Stadlober:1995,Gasparov:1997}}. A
compilation of experimental results is shown in
Fig.~\ref{Fig:Delta}. The scatter of the data points partially
reflects the development of the sample quality. However, there
are also discrepancies between the results which are related to
the interpretation of the experiments.

We first summarize the generally accepted features close to and above optimal doping:
\begin{itemize}
\item
at $p \simeq 0.16$ the peaks in the three Raman active symmetries
are in relative positions expected for $d$~wave pairing, the ratio
$2\Delta_{\rm max}/k_BT_c$ is approximately 8,\footnote{\cite{Hackl:1988b,Cooper:1988b,Yamanaka:1988,Devereaux:1994a,Chen:1994b,Gasparov:1997}}
\item
in the overdoped range, $p > 0.16$, the $B_{1g}$ peak frequency
decreases faster than $T_c$ obeying $\Omega_{\rm
peak}^{B_{1g}}/T_c^{\rm max} \approx 46(0.28-p)$ (with
$\Omega_{\rm peak}^{B_{1g}}$ and $T_c^{\rm max}$ in $\rm cm^{-1}$
and K,
respectively),\footnote{\cite{Kendziora:1995,Blumberg:1997a,Naeini:1999,Venturini:2002c,Sugai:2003,Masui:2003}}
\item
whenever the $B_{2g}$ peak can be observed its maximum follows
$T_c$, $\Omega_{\rm peak}^{B_{2g}} \propto
T_c,$\footnote{\cite{Kendziora:1995,Opel:2000,Venturini:2002b,Venturini:2002c,Gallais:2005,Misochko:1999}}
\item
the $A_{1g}$ peak frequency follows either the magnetic
$(\pi,\pi)$ mode\footnote{\cite{LeTacon:2005,Gallais:2002}} or,
in the case of resonantly enhanced light scattering, $\Omega_{\rm
peak}^{B_{1g}}$.\footnote{\cite{Limonov:2002a}}
\end{itemize}

The gap close to the node, which is projected out in $B_{2g}$
symmetry, is also found by penetration depth, low bias STM, and the Nernst effect
to more or less follow the transition temperature for all doping levels
\cite{Panagopoulos:1998,Deutscher:1999,Deutscher:2005,Xu:2000}. However it
is inconsistent with peak-to-peak measurements of tunneling density of
states\cite{Zasadzinski:2002}, the energy of the $(\pi,0)$ peak in
the spectral function\cite{Campuzano:2002}, and thermal conductivity
\cite{Sutherland:2005}, which when interpreted in terms of
$d-$wave quasiparticle picture \cite{Durst:2000} indicate that the
gap energy continues to increase with decreased doping. This
highlights one of the major issues in the cuprates of whether the
superconducting pairing energy rises with underdoping or follows
T$_{c}$, and this certainly requires further study.

In many experiments, a second energy scale is observed which
varies approximately as $p_0-p$, similar to $\Omega_{\rm
peak}^{B_{1g}}$. The majority of authors \cite{Timusk:1999} call
it a pseudogap $\Delta^{\ast}$ opening below the crossover
temperature $T^{\ast}$ (see Fig.~\ref{Fig:phase-diagram}) with
$\Delta^{\ast} \propto T^{\ast}$. The understanding of the
pseudogap is a matter of intense research at present. A possible
relation to the $B_{1g}$ Raman data will be discussed below in
section \ref{sec:doping}.

In the underdoped range, the interpretation is more controversial.
The main issues are whether or not the $B_{2g}$ pair-breaking peak
can be observed at all doping levels and how the $B_{1g}$ spectra
evolve for $p<0.16$.

The $B_{2g}$ problem seems to converge with the improvement of the
sample quality. The cleaner the samples, the clearer the $B_{2g}$
pair-breaking peak, as expected theoretically
\cite{Devereaux:1995b}. The problem can be visualized in $\rm
YBa_2Cu_3O_{6+x}$ where oxygen tends to cluster and to form
pinning centers if the Cu-O chains are not completely filled
\cite{Erb:1996b}. However, in addition to fully oxygenated $\rm
YBa_2Cu_3O_{7}$, partially ordered phases exist for $x=0.5$ with
$T_c\simeq 60$~K and $x=0.353$ with $T_c \rightarrow 0$
\cite{Liang:2000,Liang:2002} where every second and third chain is
filled, respectively. Therefore, the maxima are pronounced in $\rm
YBa_2Cu_3O_{6.50}$ and $\rm YBa_2Cu_3O_{6.98}$ \cite{Opel:2000},
while the peak is smeared out at optimal doping, $\rm
YBa_2Cu_3O_{6.93}$, and practically disappears slightly below
\cite{Sugai:2003}. The problem of oxygen clustering exists also in
$\rm Bi_2Sr_2CaCu_2O_{8+\delta}$, but the resulting potentials are
weaker and essentially doping independent, as can be seen directly
from the comparison of different overdoped samples
\cite{Opel:2000,Venturini:2002c} exhibiting nearly constant
intensity of the $B_{2g}$ pair-breaking features. In a similar
fashion, partial replacement of Ca by Y does not create a strong
impurity potential either. This was directly shown by electron
spin resonance (ESR) \cite{Janossy:2003} in $\rm
YBa_2Cu_3O_{6.1}$, where Ca in place of Y is a very weak impurity
which does not localize carriers even at low temperature and
doping. We conclude that Y and O doping can be used more or less
simultaneously in $\rm Bi_2Sr_2CaCu_2O_{8+\delta}$ as long as Y is
distributed statistically. The direct comparison of the data
presented in Fig.~\ref{Fig:Sugai} with those of
\textcite{Opel:2000} and \textcite{Venturini:2002c} indeed shows
that the $B_{2g}$ pair-breaking peaks have doping and dopant (O,
Y) independent intensities if a high sample quality can be
maintained. Under these conditions, the $B_{2g}$ maxima seem to
exist at all doping levels and apparently scale with $T_c$. Since
in $\rm YBa_2Cu_3O_{6+x}$ and $\rm La_{2-x}Sr_xCuO_{4}$ doping
changes both carrier concentration $p$ and mean free path $\ell$,
the pair breaking peaks appear and disappear depending on the
doping-dependent order.

It is important to remember these considerations for the analysis
of the $B_{1g}$ data. This is particularly important when
intensity issues are discussed. Therefore, we focus first on
Y-under- and O-overdoped $\rm Bi_2Sr_2CaCu_2O_{8+\delta}$
(Fig.~\ref{Fig:Sugai}) to determine the intensity evolution of the
$B_{1g}$ pair-breaking structure. It becomes very weak right below
optimal doping while no significant weakening of the $B_{2g}$
structures is observed. The gradual suppression of the $B_{1g}$
coherence peaks in Raman goes along with the disappearance of the
$2\Delta$ peaks in scanning tunneling microscopy
\cite{McElroy:2003}. This is corroborated by results in partially
ordered $\rm YBa_2Cu_3O_{6.5}$, in $\rm La_{1.9}Sr_{0.10}CuO_{4}$,
$\rm HgBa_2CuO_{4}$, and $\rm HgBa_2Ca_2Cu_3O_{8}$, where the
$B_{2g}$ structures are well resolved, while in $B_{1g}$ symmetry,
no pair-breaking effect could be found
\cite{Naeini:1999,Opel:2000,Venturini:2002b,Gallais:2005}.

When the peak is observed in $B_{1g}$ in the underdoped region, as shown
in Figure \ref{Fig:Delta}, its position does not track T$_{c}$ and
lies within the scatter of points or slightly
below the values of $2\Delta$ generally obtained via ARPES
\cite{Campuzano:2002} and peak-to-peak positions in tunneling
measurements \cite{Zasadzinski:2002}.\footnote{We point out
however that there is considerable uncertainty in determining the
gap from ARPES, as both the leading edge and the peak of the
spectrum have been used. In either case, the anti-nodal spectral
function is very broad in underdoped systems and identifying an
energy scale from a broad feature is not without uncertainty.} It
has been argued by \textcite{Chubukov:1999,Chubukov:2006} and
\textcite{Zeyher:2002} that the $B_{1g}$ peak in Raman
measurements in underdoped compounds may be due to a collective mode
appearing below T$_{c}$ from either spin-coupling or $d-$CDW
order, respectively, split off from twice the gap maximum
$2\Delta$. Yet this interpretation is inconsistent with the
doping behavior of the pairing gap determined from $B_{2g}$
symmetry, as discussed above. In addition, the findings of the
doping behavior of the anti-nodal quasiparticles in the normal
state (listed below in Section \ref{sec:doping}) indicate that
anti-nodal quasiparticles have become incoherent already above
optimal doping. Therefore it is not intuitively obvious how a
propagating mode could emerge deep in a region of incoherence. An
alternative scenario is that the peak observed in $B_{1g}$
channels is strongly altered by incoherence and is a measure of
the binding energy of localized anti-nodal electrons, such as via
the formation of resonance-valence bond singlets or small
polarons. Yet why a peak should appear remains to be understood.
In any case, this issue highlights another vexing problem in the
cuprates: how involved are the anti-nodal electrons in
superconductivity for underdoped systems.

There were reports about $B_{1g}$ pair-breaking peaks at low
doping $p \leq 0.1$ \cite{Slakey:1990b,Blumberg:1997a} appearing
already above $T_c$, at a doping independent position of
approximately 600~$\rm cm^{-1}$. In the discussion of the
existence of preformed pairs in the pseudogap state above $T_c$
and of collective modes inside the gap (see
Section~\ref{Section:collective}) that are indicative of the pairing
potential, this result has obviously some importance. However, in
the vast majority of the experiments, the observation could not be
confirmed. Further complication comes from the existence of an
oxygen impurity mode in close vicinity \cite{Hewitt:1999,Quilty:1998}. Future
studies have to clarify this issue.

In conclusion, Raman scattering experiments reveal two energy
scales in the superconducting state which have a different
dependence on doping: while the peaks in $B_{2g}$ symmetry follow
the transition temperature, the maximum energy of those in
$B_{1g}$ symmetry decreases as $(p_{0}-p)$. In the majority of the
experiments, the coherence peaks in $B_{1g}$ symmetry are found to
fade away rapidly for $p<0.16$. The origin of the features
observed at small doping remains controversial.

\subsection{Normal State: Dichotomy of Nodal and Anti-Nodal Electrons}
\label{sec:HTSC_NC}

The studies of the symmetry and doping dependence of the Raman
spectra in the superconducting state are suggestive of
interactions with a pronounced structure in momentum space which,
in addition, vary with doping (see, e.g., Fig.~\ref{Fig:Delta}).
One of the key questions is therefore how the interactions
renormalize the quasiparticles and their response functions in the
normal state.\footnote{The properties of the normal state have
been studied and reviewed extensively in the last two decades.
Recent reviews are by \textcite{Timusk:1999},
\textcite{Loram:2001} and \textcite{Basov:2005}. A thorough study
of the dc and Hall transport properties in various compounds
performed on high-quality samples of the last generation was
presented recently by \textcite{Ando:2004}} Owing to the crystal
structure of the cuprates, the anisotropy of the in-plane to the
out-of-plane transport translates into a momentum dependence of
the in-plane transport properties
\cite{Forro:1993,Turlakov:2001,Devereaux:2003b}. Hence, Raman
scattering can substantially supplement optical conductivity
measurements above $T_c$. An anisotropy of the normal-state Raman
spectra was actually observed soon after the gap
anisotropy.\footnote{\cite{Staufer:1990,Slakey:1991,Reznik:1993,Katsufuji:1994,Yamanaka:1996,Hackl:1996,Blumberg:1999,Naeini:1999,Opel:2000}}

\subsubsection{Unconventional Metal-Insulator Transition}\label{sec:ordering}

\begin{figure}[b!]
\centerline{\epsfig{figure=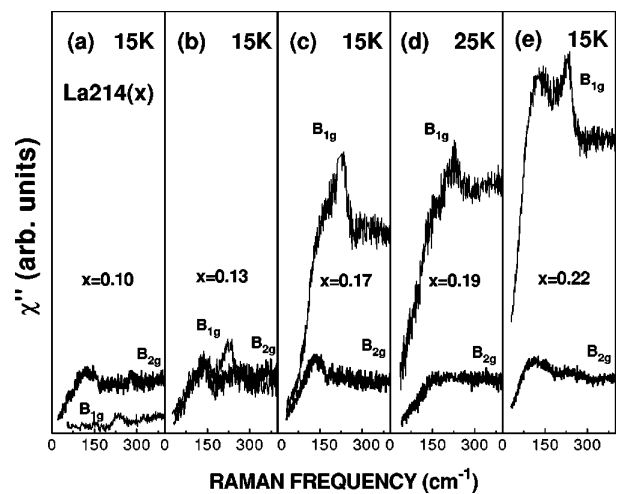,width=1\linewidth,clip=}}
\vspace{0.1cm} \caption{Direct comparison of the low-energy
$B_{1g}$ and $B_{2g}$ Raman response functions measured at
temperatures indicated (all below T$_{c}$) in
La$_{2-x}$Sr$_{x}$CuO$_{4}$ for different values of Sr doping. The
scale is the same for all frames. From \textcite{Naeini:1999}.}
\label{Fig:Naeini}\end{figure}

The momentum dependence of the transport as measured by Raman
scattering is indeed dramatic. Generally, the nodal region
along the diagonal of the Brillouin zone ($B_{2g}$ symmetry)
remain essentially unchanged at all doping levels, while the
$(\pi,0)$~regions ($B_{1g}$ symmetry) suffer an overall loss of
oscillator strength by approximately an order of magnitude if the
doping level is reduced from 0.22 to 0.10 (Fig.~\ref{Fig:Naeini}).
When this doping dependence was first observed,
\textcite{Katsufuji:1994} realized that the variation of the
$B_{1g}$ spectra with $p$ cannot be explained by only changing the
filling of a rigid band where the ratio of the $B_{1g}$ to the
$B_{2g}$ scattering intensities is essentially determined by
$(t/t^{\prime})^2$ with $t$ and $t^{\prime}$ the nearest and
next-nearest neighbor hopping matrix elements,
respectively \cite{Einzel:1996}. Changing the ratio $(t/t^{\prime})$
enough to account for the observed Raman intensities
leads to unrealistic band structures. The suggested importance of correlation effects on the Raman intensities was pointed out early on.

\begin{figure}[b!]
\centerline{\epsfig{figure=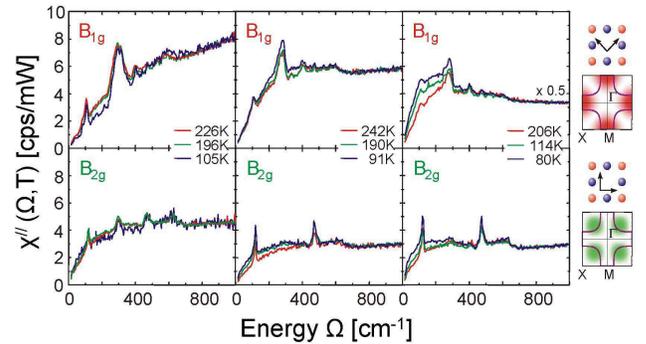,width=1\linewidth,clip=}}
\vspace{0.1cm} \caption{Direct comparison of the low-energy
$B_{1g}$ and $B_{2g}$ Raman response functions measured at
temperatures indicated (all above T$_{c}$) in $\rm
Bi_2Sr_2CaCu_2O_{8+\delta}$  for different doping levels $p=0.15$,
0.20, and 0.22 (from left to right). From
\textcite{Venturini:2002a}.}
\label{Fig:VenturiniPRL2002}\end{figure}

The intensity changes of the $B_{1g}$  spectra as a function of
doping (Fig.~\ref{Fig:Naeini}) are accompanied by a qualitative
change of the temperature dependence at a given doping level. For
$\rm Bi_2Sr_2CaCu_2O_{8+\delta}$, the effect is shown in the range
$0.15 \leq p \leq 0.23$ (Fig.~\ref{Fig:VenturiniPRL2002}). For
$B_{2g}$ symmetry, there is little change of both the overall
intensity and the initial slope which depends on temperature as
expected for a metal. As opposed to $B_{2g}$ symmetry, the
temperature dependence of the $B_{1g}$ spectra reverses sign
(Fig.~\ref{Fig:VenturiniPRL2002}~(a)), indicating non-metallic
behavior  at $p=0.15$. As outlined above
(Eq.~(\ref{Eq:Raman_slope})), the initial slope of the response
function is proportional to a transport lifetime $\tau$ or,
equivalently, a conductivity. Accordingly, the inverse of it,
$[\partial \chi_{\mu}^{\prime \prime}/\partial\Omega]^{-1}$ is a
resistivity which, for the Raman selection rules, is momentum
sensitive as indicated by the index $\mu$.

It has been shown by \textcite{Opel:2000} that both the energy
dependences and the magnitudes of $\Gamma_{\mu}(\Omega,T)=
\hbar/\tau_{\mu}(\Omega,T)$ and $m_{\mu}^{\ast}(\Omega,T)/m_{b} =
1+\lambda_{\mu}(\Omega,T)$ can be determined from the spectra
using a memory function analysis \cite{Goetze:1972} in combination
with an energy integral over $\chi_{\mu}^{\prime\prime}/\Omega$
(``sum rule''). By extrapolating the results to $\Omega=0$, very
reliable numbers for $\Gamma(\Omega \rightarrow 0,T)$ can be
obtained.\footnote{The memory function or extended Drude analysis
is particularly useful for a single-component response. Further
details and limitations are discussed in the appendix of reference
\cite{Opel:2000}. In Raman scattering the integral is as crucial
as the plasma frequency in IR spectroscopy if magnitudes {\it and}
energy dependences are to be derived in a similar way as in the
analysis of the reflectivity \cite{Basov:2005}. ``Sum rule'' might
be somewhat misleading and should be used with care since the
integral over $\chi_{\mu}^{\prime\prime}(\Omega)/\Omega \propto
\sigma^{\prime}(\Omega)$ (see Eq.~(\ref{Eq:SS})) is not a
conserved quantity like the number of carriers in the famous f-sum
rule. Without the integral the energy dependence of one of the
quantities must be dropped and/or a fit to model functions for
$\Gamma_{\mu}(\Omega)$ and $1+\lambda_{\mu}(\Omega)$ is required
\cite{Slakey:1991,Yamanaka:1996,Hackl:1996,Blumberg:1999,Naeini:1999}.}

\begin{figure}[b!]
\centerline{\epsfig{figure=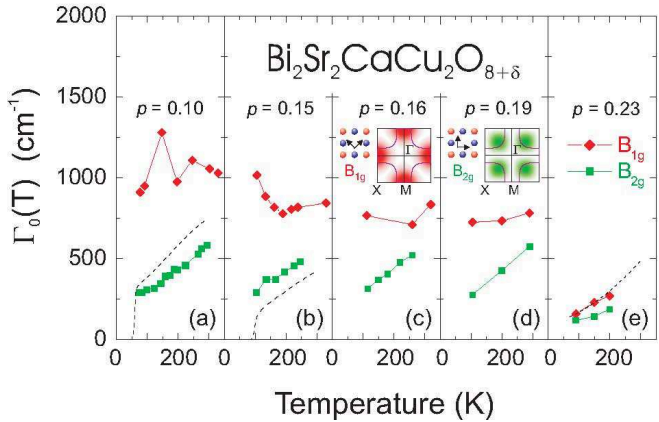,width=1\linewidth,clip=}}
\vspace{0.1cm} \caption{Raman relaxation rates (``resistivities'')
at $B_{1g}$ and $B_{2g}$ symmetries of $\rm
Bi_2Sr_2CaCu_2O_{8+\delta}$. $B_{1g}$ and $B_{2g}$ is sensitive to
anti-nodal and nodal regions of the Brillouin zone as indicated.
The dashed lines represent transport data. Using plasma frequencies from IR spectroscopies, the Drude formula is
used for the conversion from resistivities to relaxation rates.
From \textcite{Hackl:2005}.} \label{Fig:HacklAdvP2005}\end{figure}

In Fig.~\ref{Fig:HacklAdvP2005}, the ``Raman resistivities''
$\Gamma_{\mu}(\Omega,T)$ in the limit $\Omega=0$ are plotted for
$\mu = B_{1g},B_{2g}$ at various temperatures and doping levels.
The $B_{2g}$ results compare well to conventional resistivities
$\rho(T)$ using the Drude expression $\Gamma_{\mu}(\Omega=0,T) =
\varepsilon_0 (\tilde{\omega}_{\rm pl})^{2}\rho(T) $ with the
renormalized plasma frequency $\tilde{\omega}_{\rm pl}$. With the
plasma frequency in eV and the resistivity in $\mu\Omega$cm, one
finds $\Gamma_{\mu}(T) \simeq 1.08 (\tilde{E}_{\rm pl})^{2}\rho(T)
$. At $p=0.23$, the transport is essentially isotropic. Below
$p=0.22$, anisotropy develops quite abruptly, and for $p<0.16$,
the temperature dependence of the $\Gamma_{B_{1g}}(\Omega=0,T)$
becomes non-metallic. This crossover behavior has been interpreted
in terms of an unconventional metal-insulator transition at
$0.20<p<0.22$, where the antinodal transport is gradually quenched
while the nodal one remains essentially unaffected
\cite{Venturini:2002a}.

Since the current vertex has a similar {\bf k}~dependence as the
$B_{2g}$ Raman vertex, IR spectroscopy and, similarly, dc
transport project out mainly the nodal part of the Fermi surface
\cite{Devereaux:2003b}, and show therefore good qualitative
agreement with the $B_{2g}$ Raman results, which exhibit little
variation with doping for $0.1 < p <0.23$ beyond the change of
$\tilde{\omega}_{\rm pl}$, as shown in Figure \ref{Fig:HacklAdvP2005}.
The suppression of the antinodal transport goes along with a reduction of the quasiparticle
strength $Z_{\bf k}$ in the vicinity of $(\pi,0)$ \cite{Kim:2003},
and a collapse of the Korringa law $(T_1T)^{-1}= {\rm const}$ as
observed by NMR, where $T_1$ is the spin-lattice relaxation time \cite{Alloul:1989,Billinge:2003}.

Below $p \simeq 0.16$, insulating behavior can also be observed in
conventional transport at low temperature if superconductivity is
suppressed with high magnetic fields
\cite{Boebinger:1996,Ando:2004a}. The logarithmic divergence of
the resistivity in the limit $T \rightarrow 0$ indicates
localization of the carriers. The crossover from metallic to
non-metallic behavior occurs at temperatures  $T < T_c$, well below
those found in $B_{1g}$ Raman scattering. This is consistent with
the observation that the $B_{2g}$ spectra do not show any anomaly
in the normal state. The synopsis of NMR, conventional, and Raman
transport leads to the quite consistent conclusion that carriers
get gradually localized upon decreasing doping. According to the
Raman results, the suppression of free-carrier transport starts at
$p \simeq 0.21$ at the anti-nodal regions of the Fermi surface and
gradually moves towards the nodal points with decreasing $p$. The
conductivity apparently disappears only very close to or at zero
doping, making connection to recent ARPES results which show
quasiparticles along the nodal direction even at the lowest (yet
finite) doping level \cite{Yoshida:2003}.

\begin{figure}[b!]
\centerline{\epsfig{figure=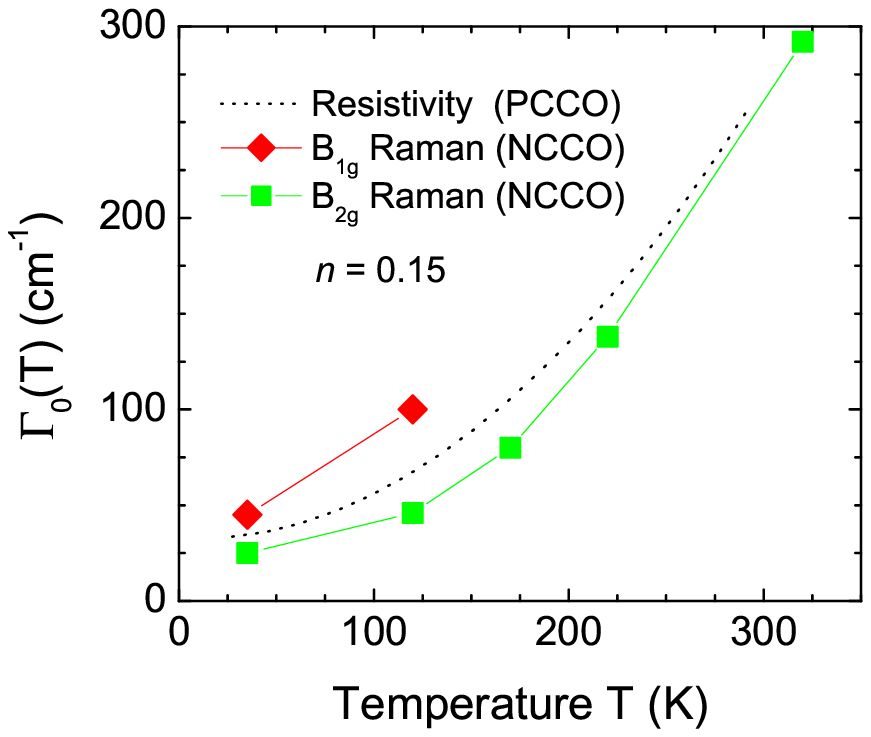,width=0.85\linewidth,clip=}}
\vspace{0.1cm} \caption{Raman relaxation rates at $B_{1g}$ and
$B_{2g}$ symmetries of $\rm Nd_{1.85}Ce_{0.15}Cu_2O_{4}$
\cite{Koitzsch:2003}. $B_{1g}$ and $B_{2g}$ are sensitive to
anti-nodal and nodal regions, respectively. The dashed line
represents recent transport data for $\rm
Pr_{1.85}Ce_{0.15}Cu_2O_{4}$ thin films \cite{Dagan:2004}. The
Drude model with a plasma energy of 1~eV \cite{Homes:1997} is used
for the conversion of the resistivity to relaxation rates.
Compiled for this review.} \label{Fig:NCCO-Gamma}
\end{figure}

Although there are no studies of both the doping and temperature
dependences in $n$-doped cuprates, the comparison of the
low-temperature data of $\rm Pr_{2-x}Ce_{x}Cu_2O_{4}$
(Fig.~\ref{Fig:PCCO}) to those in $\rm Bi_2Sr_2CaCu_2O_{8+\delta}$
(Fig.~\ref{Fig:Sugai}) or $\rm La_{2-x}Sr_{x}Cu_2O_{4}$
(Fig.~\ref{Fig:Naeini}) is worthwhile. In fact, a significant
decrease of the overall $B_{1g}$ intensity for small $n$ is
revealed whereas the continuum at $B_{2g}$ symmetry is not as
doping independent as on the $p$-doped side, but rather becomes
weaker towards lower doping. Raman relaxation rates have been
determined explicitly only for $\rm Nd_{1.85}Ce_{0.15}Cu_2O_{4}$
\cite{Koitzsch:2003}. Both in $B_{1g}$ and $B_{2g}$ symmetry, Raman
``resistivities'' are close to those found by conventional
transport, as shown in Figure~\ref{Fig:NCCO-Gamma}. This is
certainly at variance with the results in hole-doped systems at
comparable carrier concentrations. From the viewpoint of Raman
scattering, $n$-doped cuprates at optimal doping in the normal state look more like
over-doped $p$-type samples with isotropic transport properties.
The interpretation is not clear at the moment, but this similarity may be
perhaps related to the rather involved band structure on the $n$-doped side
\cite{Fournier:1997,Onose:2004}.

\subsubsection{Quantum Critical Point(s)}

The quite unusual transport properties described above are
accompanied by various crossover phenomena, such as the opening of
a pseudogap \cite{Timusk:1999}, the collapse of the Korringa law
\cite{Billinge:2003}, and several other intriguing
observations \cite{Loram:2001}. Recent studies show pseudogap
phenomena also in electron doped cuprates
\cite{Alff:2003,Onose:2001,Onose:2004}. All those anomalies fit
into the broader context of a quantum critical point (QCP) buried
below the superconducting phase.\footnote{For general references
see \textcite{Sachdev:1999,Vojta:2003}.} A QCP occurs in the phase diagram at
$T=0$ at a
critical value $x_c$ of a control parameter, such as doping $x$ and/or pressure.
The most general property of a QCP is
the existence of thermal and quantum fluctuations up to high
temperatures for $x \simeq x_c$, which prevent the transition into
an ordered phase at finite $T$. The end point of the N\'eel phase
at $p \simeq 0.02$ as well as $p=0.05$ and $p=0.27$ delimiting
superconductivity are examples on the hole-doped side. However,
an interesting putative QCP may be hidden below $T_c$ at
the zero temperature extrapolated value of $T^{\ast}(p,n)$ (see
Fig.~\ref{Fig:phase-diagram}), indicating competition between
different types of order and superconductivity
\cite{Castellani:1997,Sachdev:1999,Andergassen:2001,Chakravarty:2001,Kivelson:2003}.
Below $T^{\ast}$ partial or even long range order can be
established.

In a normal metal, the kinetic energy of the electrons $E_{\rm
kin}$ is much larger than the Coulomb energy $U$ because of
screening. If the ratio $U/E_{\rm kin}$ increases upon decreasing
carrier density and approaches 1 various types of instabilities
can arise which usually induce a transition to an insulator.
Classical examples are the Mott transition or the Wigner crystal.
The cuprates are close to this limit, and several additional
possibilities have been discussed. Prominent examples are
spontaneous orbital currents \cite{Varma:1997}, spin ordering
\cite{Zaanen:1989,Machida:1989,Tranquada:1995,Tranquada:2004,Kivelson:2003},
charge ordering such as stripes or density waves
\cite{Castellani:1995,Chakravarty:2001,Andergassen:2001,Kivelson:2003}
and Fermi surface deformation fluctuations \cite{Metzner:2003}. In
all cases, static order is found only, if at all, at very low
doping and/or in specifically modified structures such as $\rm
La_{2-y-x}{\it Re}_ySr_xCuO_4$ for $y\simeq 0.4$ and $Re={\rm
Nd,Eu}$ \cite{Tranquada:1995,Klauss:2000} or the nickelates. In
La$_{1.775}$Sr$_{0.225}$NiO$_{4}$ \cite{Pashkevich:2000} and
La$_{1.67}$Sr$_{0.33}$NiO$_{4}$ \cite{Blumberg:1998,Yamamoto:1998}
the formation of static distortions due to the formation of
stripes are big enough to split phonon and magnon lines due to
zone doubling and to suppress the $B_{1g}$ continuum at low
energies.

Of course, static order is much easier to verify than
fluctuating order. Nevertheless, fluctuating incommensurate spin order has been
observed in $\rm La_{2-x}Sr_xCuO_4$ for $x \ge 0.055$
\cite{Fujita:2002}, $\rm YBa_{2}Cu_{3}O_{6.85}$, and
$\rm YBa_{2}Cu_{3}O_{6.6}$
\cite{Hinkov:2004}, while commensurate spin order is seen at low energies in
$\rm YBa_{2}Cu_{3}O_{6.353}$ \cite{Stock:2006}. This means that in a temperature dependent
volume characterized by a coherence length $\xi_s(T)$ a
superstructure of the antiferromagnetically ordered spins is
established for higher doping levels. At the same time the charges which are expelled from
the antiferromagnetic regions becoming spatially organized as
well. This type of order has similarity with a liquid crystal and
is sometimes referred to as nematic
\cite{Kivelson:1998,Kivelson:2003}. From these experiments it
cannot be decided whether the fluctuating magnetic superstructure
is a property of the spins or of the charges. For statically
ordered $\rm La_{2-y-x}Nd_ySr_xCuO_4$ with doping $x=1/8$,
\textcite{Tranquada:1995} showed that charge ordering precedes
spin ordering when the temperature is reduced. It has indeed been shown that charge
ordering fluctuations are quite common phenomena in correlated
systems \cite{Castellani:1995,Metzner:2003}.

As long as there is no long range or static order which competes
with superconductivity, the fluctuations can establish Cooper
pairing \cite{Perali:1996,Chakravarty:2001}. For this reason, the
understanding of the dynamics of incipient charge and spin order
is a very interesting subject in the cuprates.

\subsubsection{Role of Fluctuations and Incipient Ordering Phenomena at small doping}
\label{sec:fluctuations}

\begin{figure}[b!]
\centerline{\epsfig{figure=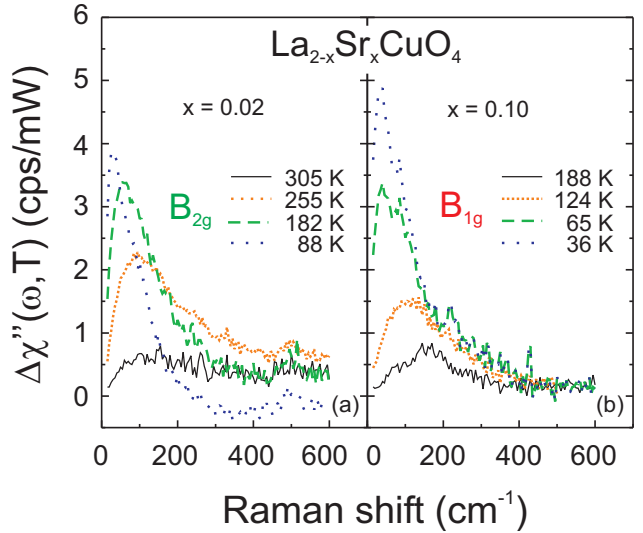,width=1\linewidth,clip=}}
\vspace{0.1cm} \caption{Low energy response of underdoped $\rm
La_{2-x}Sr_{x}CuO_4$. A Drude-like peak \cite{Zawadowski:1990}
with a characteristic energy $\Omega_c(x,T)$ is revealed after
subtraction of the 2D response of the ${\rm CuO_2}$ planes. At
$x=0.02$ (a) and 0.10 (b) the additional response is observed in
$B_{2g}$ and $B_{1g}$ symmetry, respectively. The styles of the
lines (colours) do {\em not} correspond to similar temperatures
but rather highlight the {\em scaling} of the response with
temperature: similar spectra are obtained if the temperatures
differ by approximately a factor of 2. From
\textcite{Tassini:2005}.} \label{Fig:TassiniPRL2005}\end{figure}

The study of dynamical order requires inelastic probes typically
sensitive at finite ${\bf q}$. In the case of charge ordering, resonant
X-ray scattering is the most promising as it
can provide direct evidence for the charge ordering at a particular
wavevector \cite{Abbamonte:2004}, while neutrons can be used only if the charges modulate the lattice or
the spin structure. While optical methods are confined to ${\bf q} =0$
two excitations with opposite momenta can be created, such as
multi-phonon or two magnon scattering. Therefore light can be
scattered from Cooper pairs or if an electron-hole pair couples to
two excitations. It depends on the individual context
how fluctuation phenomena are best studied.

\begin{figure}[b!]
\centerline{\epsfig{figure=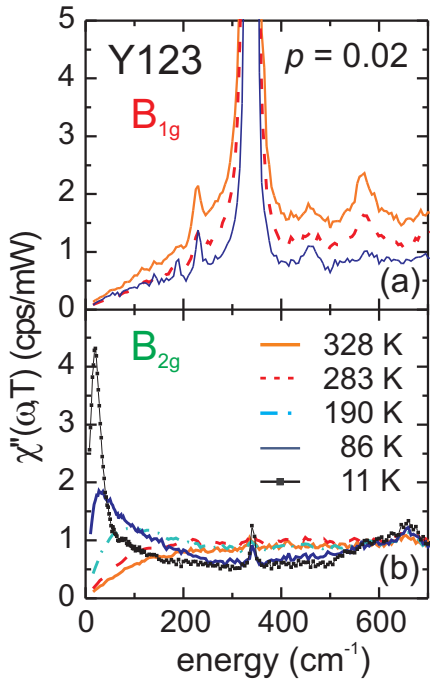,width=6.0cm,clip=}}
\vspace{0.1cm} \caption{Raman response $\chi_{\mu}^{\prime
\prime}(\omega,T)$ of ($\rm Y_{0.97}Ca_{0.03})Ba_{2}Cu_{3}O_{6.1}$
in $B_{1g}$ (a) and $B_{2g}$ (b) symmetry. The doping level is
close to $p=0.02$. From \textcite{Hackl:2005}.}
\label{Fig:HacklAdvP2005-2}\end{figure}

In the copper-oxygen systems, an unexpected additional component in
the $B_{1g}$ spectra was observed in $\rm La_{1.90}Sr_{0.10}CuO_4$: as shown in Figure~\ref{Fig:TassiniPRL2005}, a peak
is found in the 100~$\rm cm^{-1}$ range which gains intensity and
moves to very low but finite energy with decreasing temperature
\cite{Venturini:2002b,Tassini:2005}.\footnote{We note that the the
lines observed recently by \textcite{Gozar:2004} in $\rm
La_{2-x}Sr_{x}CuO_4$ ($x=0,~0.01$) have nothing in common with the
response discussed here and were clearly identified as one-magnon
excitations (see also \textcite{SilvaNeto:2005}).} At high energy
for all temperatures, and at high temperature for all energies, the
spectra of the cuprates with $p \simeq 0.10$ in general are
similar and show a strong anisotropy between nodal and anti-nodal
regions in the Brillouin zone (see Fig.~\ref{Fig:Naeini},
\ref{Fig:VenturiniPRL2002}, and~\ref{Fig:HacklAdvP2005}). At low
temperature the low energy feature is unique to $\rm
La_{1.90}Sr_{0.10}CuO_4$, while the spectra become flatter in $\rm
YBa_2Cu_3O_{6.5}$ and $\rm
Bi_2Sr_2(Y_{0.38}Ca_{0.62})Cu_2O_{8+\delta}$ (see
Fig.~\ref{Fig:VenturiniPRL2002}). To appreciate the differences,
one has to recall that the conventional transport properties of
all these compounds are very similar \cite{Ando:2004}. Before
we discuss possible interpretations we wish to compare results for
different doping levels and materials.

From the comparison of the low energy peak to infrared results, a
relationship to fluctuating stripe order was conjectured
\cite{Venturini:2002b}. For further support, it seems worthwhile to
exploit the selection rules Raman scattering offers. The effect in
$B_{1g}$ symmetry is indeed compatible with the orientation of
stripes along the Cu--O bonds at $p=x=0.10$ \cite{Fujita:2002}.
This is simply because order along the principal axes of an
essentially tetragonal lattice corresponds to an orthorhombic
distortion. Fluctuations correspond to micro-twinning, meaning that
$x$ and $y$ cannot be accessed individually. Hence, $B_{1g}$
symmetry measuring only the difference $xx-yy$ (on a microscopic
scale) projects out exactly this distortion. As a familiar example,
we recall that fully oxygenated $\rm YBa_2Cu_3O_{7}$ has Cu--O
chains along the crystallographic $b$~axis making $b>a$. Even if
the sample is twinned, the chain contributions are always
superimposed on the $B_{1g}$ spectra, while the $xx$ and $yy$
spectra are equal (as opposed to a single-domain crystal). For
$x<0.055$, a reorientation by $45^{\circ}$ is observed by neutron
scattering \cite{Fujita:2002}. Since $B_{1g}$ and $B_{2g}$ are
equivalent modulo a $\pi/4$ rotation in the basal plane of a tetragonal lattice, the
related peak should now appear in $B_{2g}$ rather than in $B_{1g}$
symmetry. In fact, this has be observed in $\rm
La_{1.98}Sr_{0.02}CuO_4$, as shown in Figure~\ref{Fig:TassiniPRL2005}
\cite{Tassini:2005}.

Spectra similar to those in $\rm La_{1.98}Sr_{0.02}CuO_4$
(Fig.~\ref{Fig:TassiniPRL2005}) are
found in $\rm (Y_{0.97}Ca_{0.03})Ba_2Cu_3O_{6.1}$ with $p \simeq
0.02$ (Fig.~\ref{Fig:HacklAdvP2005-2}~(b)). Owing to the
homogeneity of the sample, the peak is very narrow and well
defined. As the low energy peak emerges, the intensity of the continuum is
suppressed by roughly 30~\% over an energy range of approximately
550~$\rm cm^{-1}$ in a similar though much stronger way than at
higher doping \cite{Nemetschek:1997,Opel:2000}. In $B_{1g}$
symmetry, there is an overall loss of spectral weight up to much
higher energies, similar to $\rm YBa_2Cu_3O_{6+x}$ and
$\rm Bi_2Sr_2CaCu_2O_{8+\delta}$ at higher dopings,
demonstrating the transition to a correlated
insulator \cite{Freericks:2001a}.

\begin{figure}[b!]
\centerline{\epsfig{figure=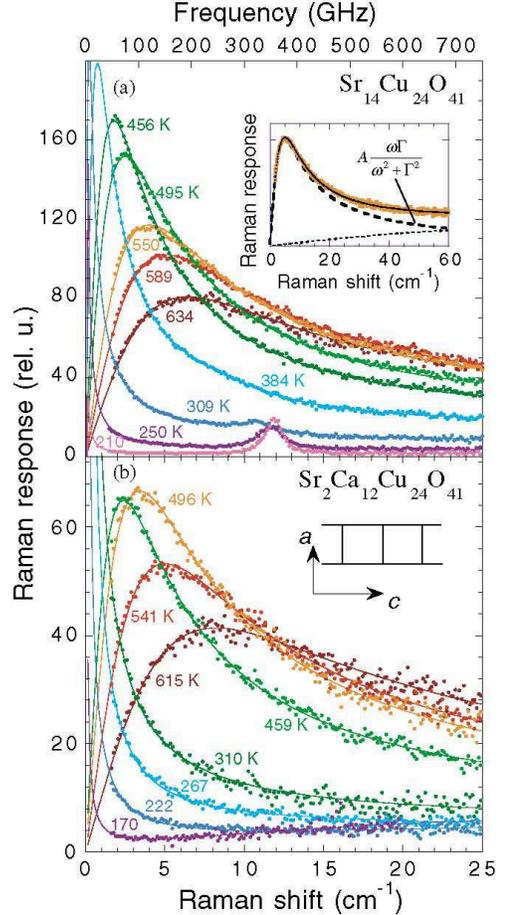,width=7.0cm,clip=}}
\vspace{0.1cm} \caption{Temperature dependent Raman response for
$cc$ polarization in ladder compounds. Upper inset: fit of the
data with Eq. \ref{Eq:Drude-imp} plus a small background. Lower
inset: Two-leg ladder structure. From \textcite{Gozar:2003}.}
\label{Fig:Gozar}
\end{figure}

Simultaneously with the studies in $\rm La_{1.90}Sr_{0.10}CuO_4$,
the quasi 1D ladder compound Sr$_{14-x}$Ca$_{x}$Cu$_{24}$O$_{41}$
was investigated \cite{Blumberg:2002,Gozar:2003} which is
considered a model system for the high-$T_c$ cuprates
\cite{Sigrist:1994,Dagotto:1996,Dagotto:1999}. At temperatures
above approximately 450~K low energy spectral peaks similar to those in $\rm
La_{1.90}Sr_{0.10}CuO_4$ are found (Fig.~\ref{Fig:Gozar}),
although their energies and widths are an order of magnitude
smaller. The width obeys an Arrhenius law with a doping
independent activation energy $\Delta$ of approximately 2100~K. In
the insulator ($x=0$), the conductivity (for $T<300$~K) reveals a
similar $\Delta$ while for $x=12$ the conductivity is metallic
above 70~K \cite{Gozar:2003,Gozar:2005a}. The position of the
maximum and the width decrease upon cooling. In both cases, for $T < 450$~K the
peaks move below the detection limit of 2~cm$^{-1}$. Below 250~K,
no indication of the peaks can be detected any more. It is
interesting to note that at least in undoped
Sr$_{14}$Cu$_{24}$O$_{41}$ a charge-ordered state develops below
$T_{CO} \approx 250$~K \cite{Abbamonte:2004}.

In the ordered state Sr$_{14}$Cu$_{24}$O$_{41}$ has an optical
response which is well described by that of a pinned charge
density wave \cite{Littlewood:1987,Blumberg:2002}. For $T > 250$~K,
the results are interpreted in terms of a damped plasma
oscillation above $T_{CO}$
\cite{Blumberg:2002,Gozar:2003,Gozar:2005a}. Since free carriers
damp the mode, the width is expected to increase proportional the
conductivity just opposite to what one would expect from free
carrier response. This complicates the interpretation of the
results in metallic Sr$_{2}$Ca$_{12}$Cu$_{24}$O$_{41}$ where the
width still increases with temperature while the conductivity
decreases. \textcite{Gozar:2003} argue that different regions of
the Fermi surface contribute to the transport and the damping in
essentially 2D Sr$_{2}$Ca$_{12}$Cu$_{24}$O$_{41}$.

A comparison of panels (a) and (b) of Figure~\ref{Fig:Gozar}
shows that the continuum surviving at low temperature is
significantly stronger in the metal (Fig.~\ref{Fig:Gozar}~(b)). We
therefore think that the low-energy mode found at $T>250$~K could
also originate from incipient charge order. In
Sr$_{14-x}$Ca$_{x}$Cu$_{24}$O$_{41}$ the mode is then superimposed
on a free electron response in a similar way as in the high-$T_c$
cuprates.

It is interesting that similar types of spectra are also found in Bi$_{1-x}$Ca$_{x}$MnO$_{3}$ (see
section~\ref{sec:manganites}), exhibiting comparable
temperature dependences.
While the results in ladders are interpreted in terms of an
overdamped plasma mode in a CDW system above the ordering temperature
\cite{Blumberg:2002,Gozar:2003,Gozar:2005a}, the response in the
cuprates was proposed to originate from fluctuations of the charge
density in the vicinity of a charge-ordering instability
\cite{Caprara:2005}. Since the charge modulation observed e.g. by
tunneling microscopy
\cite{Howald:2003,Hoffman:2002,Vershinin:2004} is at finite {\bf
q}, two fluctuations have to be exchanged to fulfill the $q=0$
selection rule in Raman scattering \cite{Caprara:2005,Venturini:2000}.
\textcite{Caprara:2005} have shown that charge ordering fluctuations at incommensurate wavevectors determined via neutron scattering \cite{Fujita:2002}
yield proper lineshapes and selection rules in
$\rm La_{1.90}Sr_{0.10}CuO_4$ and
$\rm La_{1.98}Sr_{0.02}CuO_4$. To which extent the spin channel is involved and whether similar considerations apply to other dopings and cuprate families
have not been explored yet and remain important future topics.

In contrast to Sr$_{14-x}$Ca$_{x}$Cu$_{24}$O$_{41}$, the
low energy peak is strongly doping dependent in $\rm
La_{2-x}Sr_{x}CuO_4$. The temperature scale is different
by a factor of 2 for the two doping levels studied
(Fig.~\ref{Fig:TassiniPRL2005}), implying that $T^{\ast}(0.02)
\approx 2T^{\ast}(0.10)$. The doping level $p_c$ determined from
extrapolating $T^{\ast}$ to zero, $T^{\ast}(p_c)=0$, yields $0.15
\leq p_c \leq 0.20$ in close vicinity of the QCP inferred from
other experiments. It is therefore possible that the low energy response
is related to the thermal and quantum fluctuations above a hidden
critical point.

In summary, the comparison of the cuprates studied suggests that
there is a superposition of two anomalies in $B_{1g}$ symmetry:
(i) below $p \simeq 0.22$ antinodal quasiparticles become
localized in all compounds as can be observed at sufficiently high
temperature; (ii) with decreasing temperature a well-defined
peak develops in $\rm La_{1.90}Sr_{0.10}CuO_4$ at low energy.
Although nothing comparable can be resolved in Y- and Bi-based
compounds at the same doping level, the additional response may
not necessarily be absent. It rather can mean that the low energy
features are broader and shifted to higher energies, hence becoming
very similar in shape to the response from the 2D planes. Since the relationship between superconductivity and spin
and/or charge fluctuations is a key issue in the cuprates further
work seems worthwhile here.

\section{CONCLUSIONS AND OPEN QUESTIONS}
\label{Section:conclusions}

The results discussed in the course of this review demonstrate
that Raman spectroscopy is and has been an invaluable tool to
investigate the dynamics of strongly correlated electrons.
Improvements in the experimental technique have opened up the
field to a variety of new systems. Light scattering has offered
unique insights into dynamics in different regions of the
Brillouin zone, showing the development of ordering and the
competition between various phases as the role of correlations
increases. Materials such as superconductors with charge-density
wave order, correlated insulators, ruthenates, manganites, and
finally the superconducting cuprates show the rich variety of
phenomena which have been unveiled via Raman investigations. Along
the way, light scattering has considerably deepened our knowledge
of dynamics and correlation effects, and has provided several key
ingredients towards the development of a comprehensive theoretical
description of these materials.

Summarizing the findings on correlated systems over the past
several years, the key ingredients stemming from Raman
investigations include: (i) The development of anisotropies as
correlations are increased. Symmetry-selective measurements
provide a tool to zoom in on the dynamics in different regions of
the Brillouin zone. This is evidenced by Raman scattering studies
on MgB$_{2}$ as well as the cuprates in the normal and
superconducting phases, and the ordered phases in the ruthenates
and the manganites. (ii) The existence of collective modes.  Using
Raman techniques the prevalence of certain types of order as a
consequence were identified. Evidence for the importance of
collective modes comes from CDW-superconductors and magnon
scattering in antiferromagnetic insulators, and implications for a
number of possible fluctuation or ordering modes exist in the
cuprates. (iii) A qualitative understanding of quantum critical
behavior. The symmetry projection of parts of the Brillouin zone
completes the picture of the battleground between competing orders
concomitant with an underlying quantum phase transition. Here,
spectral weight transfers as a function of temperature, doping,
and pressure on a number of materials combined with polarization
dependent studies have opened a new door in the area of quantum
criticality.

It is clear that a number of issues are still of primary interest
in correlated systems in general. These include (i) the origin of
the electronic continuum in a number of compounds which look
surprisingly similar at first pass. (ii) The origin of the
polarization dependence in materials which exhibit instabilities
towards ordered phases. (iii) The mapping of Brillouin
zone-projected electron dynamics close to a quantum critical
point.

The cuprates continue to provide a wealth of information and
puzzles in the area of superconductivity and strong electronic
correlations. Issues in which consensus has been reached include
(i) the presence antiferromagnetic correlations over a wide range
of doping levels evidenced from the two-magnon peak, (ii) the
broad continuum in the normal state as a common feature of all the
cuprates, (iii) the $d-$wave nature of the pair state below
T$_{c}$ in a number of hole-doped materials derived from the low
frequency power laws and polarization dependences, (iv) the
dichotomy between the dynamics of $B_{1g}$ and $B_{2g}$
quasiparticles at low dopings, where the low frequency anti-nodal
$B_{1g}$ behavior is governed by incoherence at the same time as
the nodal $B_{2g}$ quasiparticles show relatively
doping-independent metallic character, similar to the findings of
transport quantities, and (v) the disappearance of this dichotomy
for appreciably doped samples.

In the cuprates there are still several unsettled questions. (i)
What is the origin of the $A_{1g}$ and/or $B_{1g}$ peaks in the
superconducting state? Do they originate from the redistribution
of superconducting quasiparticles or are they collective modes or
of lattice origin? (ii) How does the nodal/anti-nodal dichotomy
picture of coherence/incoherence merge into metallic description
at high doping, BCS superconductivity for temperatures below
T$_{c}$, and antiferromagnetism near half filling? (iii) What is
the microscopic origin for low energy peaks at low doping, and how
from this can information be obtained on the competition between
ordered phases?

These are issues which we believe will form the plan of
development of Raman scattering in the cuprates as well as other
materials in the years to come. The continuation of investigations
on many new and well-characterized samples will further our
knowledge of materials and correlations, and applications of the
use of pressure and magnetic field will allow an exploration of
quantum criticality and evolution of anisotropic electron dynamics
in a variety of systems. The rapid development of Raman scattering
as we have outlined indicate that the study of the electronic
dynamics of complex materials will remain a vibrant and promising
field of research.

\section*{Acknowledgements}

We acknowledge support by the Alexander von Humboldt Foundation,
ONR Grant No. N00014-05-1-0127, and NSERC (T.P.D.) and by the
Deutsche Forschungsgemeinschaft under grant nos. Ha~2071/2 and
Ha~2071/3 (R.H.). The latter project is part of Research Unit 538.

We would first like to express our
gratitude to A. Zawadowski who kindled and fostered our interest
in the field of Raman spectroscopy and whose guidance and
scientific insight is immeasurably appreciated. We would also like
to express our thanks to the current and former members of our
groups for their cooperation, daily discussions, and many useful
comments: Ch. Hartinger, B. Moritz, R. Nemetschek, M. Opel, W. Prestel, A.
Seaman, B. Stadlober, L. Tassini, F. Venturini and F. Vernay. We
also would like to take this opportunity to acknowledge our
collaborators in the field: Y. Ando, H. Berger, R. Bulla, A. Chubukov, S.~L.
Cooper, D. Einzel, A. Erb, L. Forr\'o, J. Freericks, Y. Gallais,
K. Hewitt, J. C. Irwin, A. J\'anossy, A. Kampf, M.~V. Klein, K.
Maki, J. Naeini, A. Sacuto, A. Shvaika, I. T\"utt\H{o}, A.
Virosztek, and A. Zawadowski. Over the years, we have also
benefited tremendously from scientific discussions with many other
colleagues: Y. Ando, O.~K. Andersen, K. Andres, D. Belitz, L.
Benfatto, G. Blumberg, A. Bock, S.~V. Borisenko, N. Bontemps, I.
Bozovic, P. Calvani, M. Cardona, B.~S. Chandrasekhar,
T. Cuk, A. Damascelli, C. Di~Castro, J. Fink, P. Fulde,
M. Gingras, A. Goncharov, A. Griffin, R. Gross, G. G\"untherodt, W.
Hanke, J. Hill, R.~W. Hill, S. Ishihara, A. J\'anossy, M. Jarrell, K.
Kamar\'as, H.-Y. Kee, B. Keimer, C. Kendziora, P. Knoll, M.
Le~Tacon, P. Lemmens, S. Maekawa, N. Mannella, D. Manske, I. Mazin, W. Metzner,
G. Mih\'aly, L. Mih\'aly, N. Nagaosa, D. Pines, L. Pintschovius, D. Reznik, M. R\"ubhausen, T.
Ruf, G. Sawatzky, R. Scalettar, Z.-X. Shen, E. Sherman, R.~R.~P.
Singh, V. Struzhkin, S. Sugai, C. Thomsen, T. Tohyama, C.~M.
Varma, W. Weber, J. Zaanen and R. Zeyher. Lastly, it is a pleasure
to thank S. L. Cooper, J. Freericks, M. Opel, I. T\"utt\H{o}, A. Shvaika, and A.
Zawadowski for the critical reading of this review article.

T.P.D. and R.H. would like to acknowledge the hospitality and
support of the Walther Meissner Institut and the Laboratoire de
Physique du Solide at the Ecole Sup\'erieure de Physique et de
Chimie Industrielles and, respectively, and the University of
Waterloo, where parts of this review were completed.

\bibliographystyle{apsrmp}

\end{document}